\documentclass{article}
\usepackage[latin9]{inputenc}
\usepackage{verbatim}
\usepackage{bm}
\usepackage{amsmath}
\usepackage{amssymb}
\usepackage{graphicx}
\usepackage{color,soul}
\usepackage{cancel}
\usepackage{multirow}
\usepackage{amsthm}
\usepackage{enumitem}
\usepackage{relsize}

\makeatletter

\providecommand{\tabularnewline}{\\}

\usepackage{hyperref}
\newtheorem{thm}{Theorem}[section]
\newtheorem{prop}[thm]{Proposition}
\newtheorem{lem}[thm]{Lemma}
\theoremstyle{definition}

\newtheorem{defn}[thm]{Definition}

\newtheorem{rem}[thm]{Remark}
\newtheorem{ex}[thm]{Example}

\def\Journal#1#2#3#4{\emph{#1} {\bf #2}, #3 (#4)}

\def\nurel{v}
\def\nuhat{\widehat{\nurel}}
\def\ww{w}

\def\vrel#1#2{\nurel(#1,#2)}
\def\vhat#1#2{\nuhat(#1,#2)}

\def\wrel#1#2{\ww(#1,#2)}

\def\ubig{u'}
\def\Bpr{B'}
\def\BB#1#2{B(#1,#2)}

\def\I{h}
\def\et{e}
\def\EEt{E}
\def\umw{u_{\textrm{em}}}

\def\Aem{\zeta_{\textrm{em}}}
\def\AD{\zeta_{\textrm{g}}}
\def\arg{\textrm{arg}}

\def\ww{\psi}
\def\sgn{\textrm{sgn}}

\def\lra{\Leftrightarrow}
\def\ra{\rightarrow}
\def\Lra{\Leftrightarrow}

\def\Ra{\Rightarrow}
\def\be{\begin{equation}}
\def\ee{\end{equation}}
\def\beqs{\begin{eqnarray*}}
	\def\eeqs{\end{eqnarray*}}
\def\beqn{\begin{eqnarray}}
\def\eeqn{\end{eqnarray}}
\def\a{\alpha}
\def\b{\beta}

\def\d{\delta}
\def\e{\epsilon}
\def\l{\lambda}
\def\p{f}
\def\lc{\overline{\lambda}}

\def\f{f}
\def\fc{\overline{f}}

\def\Imm{\textrm{Im}}
\def\Ker{\textrm{Ker}}

\def\eb{\mathfrak{e}}
\def\bb{\mathfrak{b}}
\def\bd{\beta}
\def\v{{x}}

\def\vc{\overline{{x}}}

\def\R{\mathbb{R}}
\def\Rnn{\mathbb{R}_{\geq 0}}

\def\C{\mathbb{C}}
\def\st#1{\star\!#1}

\def\M{{M}}

\def\Sp{S^+}

\def\T{{\cal T}}
\def\up{u^\perp}
\def\Up{U^+}
\def\Cup{\C u^\bot}

\def\LV{\Lambda^2}
\def\LL{\C\Lambda^2}

\def\psiu{\psi_{u}}
\def\psie0{\psi_{e_0}}

\def\chie0{\chi_{e_0}}
\def\tr{\textrm{tr}}
\def\Fdag{F^\dag}

\def\Cop{{\rm C}}

\def\Qop{{\sf Q}}
\def\Qcop{{\sf \overline{Q}}}
\def\Iop{{\sf h}}
\def\Eop{{\sf E}}
\def\Hop{{\sf H}}
\def\Sop{{\sf S}}
\def\Top{{\sf T}}
\def\Fop{f}
\def\Gop{g}
\def\Rop{{\sf R}}

\def\vop{{\sf x}}
\def\vcop{{\sf \overline{x}}}

\def\xop{{\sf x}}
\def\yop{{\sf y}}

\def\EE{E_\textrm{R}}
\def\HH{H_\textrm{R}}
\def\QQ{Q_\textrm{R}}
\def\II{I_\textrm{R}}
\def\JJ{J_\textrm{R}}
\def\PP{\qp_\textrm{R}}
\def\tgR{\rho_{\textrm{g,R}}}

\def\Fcal{{\cal F}}
\def\Fcalc{\overline{{\cal F}}}
\def\Ucal{{\cal X}}

\def\Ccal{{\cal C}}
\def\Ccalc{\overline{{\cal C}}}
\def\Xcal{{\cal X}}

\def\Ycal{{\cal Y}}
\def\Ycalc{\overline{{\cal Y}}}
\def\IEM{I_F}
\def\E{E}
\def\B{B}
\def\H{H}
\def\Q{Q}
\def\U{X}

\def\Qc{\overline{Q}}
\def\Max{Faraday }

\def\sp{p}
\def\te{\rho_{\textrm{em}}}
\def\che{\chi}
\def\SS{\Sigma}
\def\SSb{\Sigma^\perp}

\def\us{u_\Sigma}
\def\ub{u_\bot}
\def\vs{\eb_\Sigma}
\def\vb{e_\bot}

\def\pp{\psi_\SS}

\def\aa{\gamma}
\def\aaa{\varphi}

\def\zv{z}
\def\cc{\vartheta}

\def\x{{x}}
\def\y{{y}}
\def\q{{\cal S}}
\def\qp{{\cal P}}
\def\tg{\rho_{\textrm{g}}}
\def\chg{\xi}
\def\tauc{\overline{\tau}}

\def\IX{I_X}
\def\IU{I_X}
\def\IUi{I_{X_i}}

\def\Enul{e_0}
\def\Eone{e_1}
\def\Etwo{e_2}
\def\Etr{e_3}

\def\SF{\Sigma_F}
\def\SW{\Sigma_C}

\def\tq{\theta}

\def\erad{e_{{r}}}
\def\etheta{e_{{\tq}}}
\def\ephi{e_{{\phi}}}
\def\Delc{\Delta_c}

\def\wx{w}

\def\hh{\psi}

\def\DD{D}

\def\vvec{\nu}

\def\cheE{n_E}
\def\cheH{n_B}

\def\chgE{N_E}
\def\chgH{N_H}

\def\ON{orthonormal\ }

\begin{document}

\title{Poynting vector, super-Poynting vector, and principal observers in
	electromagnetism and general relativity}

\maketitle

\author{Lode~Wylleman$^{\diamond,\dagger}$, L.~Filipe~O~Costa$^{\star}$ and Jos\'{e}~Nat\'{a}rio$^{\star}$\\
	\vspace{0.05cm}
	\\
	{\small{}$^{\diamond}$ Faculty of Science and Technology, University of Stavanger}, {\small  N-4036 Stavanger, Norway}\\
	{\small $^\dagger$ Department of Mathematical Analysis, Ghent
		University, Galglaan 2, 9000 Gent, Belgium}\\
	{\small{}$^{\star}$ CAMGSD, Departamento de Matem\'{a}tica, Instituto
		Superior T\'{e}cnico, 1049-001 Lisboa, Portugal}}\\
\rm{\small{Email}}: \href{mailto:lode.wylleman@uis.no}{\texttt{\small{lode.wylleman@uis.no}}}, \,\,\href{mailto:lfilipecosta@tecnico.ulisboa.pt}{\texttt{\small{lfilipecosta@tecnico.ulisboa.pt}}}, \,\,\href{mailto:jnatar@math.ist.utl.pt}{\texttt{\small{jnatar@math.ist.utl.pt}}}

\begin{abstract}
	In electromagnetism, the concept of Poynting vector as measured by an observer is well known. A mathematical analogue in general relativity is the super-Poynting vector of the Weyl tensor. Observers for which the (super-)Poynting vector vanishes are called {\em principal}. When, at a given point, the electromagnetic field is non-null, or the gravitational field is of Weyl-Petrov type I or D, principal observers instantaneously passing through that point always exist. We survey characterizations of such observers and study their relation to arbitrary observers. In the non-null electromagnetic case it is known that, given any observer, there is a principal observer which moves relative to the first in the direction of his Poynting vector. Replacing Poynting by super-Poynting yields a possible gravitational analogue; we show that this analogy indeed holds for any observer when the Petrov type is D, but only for a one-dimensional variety of observers when the Petrov type is I. We provide algorithms to obtain the principal observers directly from the electric and magnetic fields (in the electromagnetic case) or electric and magnetic parts of the Weyl tensor (in the gravitational case) relative to an arbitrary observer. 
	It is found that in Petrov type D doubly aligned non-null Einstein-Maxwell fields (which include all classical charged black hole solutions) the Poynting and super-Poynting vectors are aligned, at each point and for each observer, and the principal observers coincide.
	Our results are illustrated in simple examples. 
\end{abstract}

\tableofcontents

\section{Introduction}

\label{sec:introduction}

In the late 1950's, and in the framework of classical general relativity
theory, Bel published a series of influential notes \cite{Belnotes},
culminating in his PhD thesis \cite{BelPhD}. Important concepts and
results appear for the first time in these notes: the refinement of
the Petrov type classification of the Weyl tensor,\footnote{Strictly speaking, most of the original results apply to the Riemann
	tensor of a pure vacuum or Einstein spacetime, but they can be and
	have been generalized to the Weyl tensor in the presence of matter,
	and are here stated as such.} the super-energy tensor~\cite{Sen00} of the Weyl tensor which later
became known as the Bel-Robinson tensor, and new or precise definitions
of quantities relative to an observer: the electric and magnetic parts 
of the Weyl tensor (inspired by Matte's work~\cite{Matte}) and the
super-energy density and super-Poynting vector. 

The motivation for introducing the {last} two quantities is the following:
an old problem in general relativity is that although it is known
that the gravitational field and waves carry some sort of energy and
momentum (as given e.g.\ by the Landau-Lifshitz pseudotensor~\cite{LandauLifshitz}),
such quantities cannot be defined as tensors and have no local meaning
since, due to the equivalence principle, they can always be made to
locally vanish. The Bel-Robinson tensor (and its associated super-energy
scalar and super-Poynting vector) arose in a quest to devise quantities
formally analogous to the electromagnetic stress-energy tensor, in
terms of which one could define a scalar and a vector analogous to
the electromagnetic energy density and Poynting vector. The price
to pay is that these are quantities with strange dimensions and whose
physical meaning is no longer clear \cite{Sen00,AlfonsoSE,Komar}.
Nevertheless, some suggestive connections between super-energy and the
quasi-local notions of gravitational field energy density (%
e.g. Landau-Lifshitz's) {can be} found in the literature \cite{Sen00,Hawking,Teyssandier,Garecki}%
. It is moreover believed that gravitational radiation is accompanied by a super-Poynting vector~\cite{Pirani,Bel62,FerSae12,AlvarezSenovillaI,AlvarezSenovillaII,Clifton2014}; in the linear regime, this is clearly suggested by Matte's equations~\cite{CN2014}, and, recently, a relationship with the quasi-local wave energy flux has been put forth \cite{Clifton2014}; in the exact theory, it has also recently been shown~\cite{AlvarezSenovillaI} that a variant of it yields a characterization of the presence of gravitational radiation equivalent to the classical one, given by the News tensor.
This fosters the interest in studying the observers relative to which
the super-Poynting vector vanishes, and led to the proposal, inspired
by the electromagnetic analogy, of the non-existence of such observers
at a given point as a {criterion} for ``intrinsic radiation'' \cite{Bel62}.
Such research line started with the seminal work by Pirani \cite{Pirani},
who studied the eigenbivectors of the Weyl tensor (considered as an
endomorphism of self-dual bivector space, as in the original Petrov
classification), calling { the intersections of the corresponding 2-planes} `principal vectors'.
Observers with a timelike principal vector as their 4-velocity at
a point were dubbed to be instantaneously `following the gravitational
field', because in the electromagnetic {(henceforth abbreviated to EM)} analogue they correspond to
observers measuring a vanishing Poynting vector; we shall call them
{\em principal observers} here. Pirani proved that when the Petrov
type is `diagonal' at the point, i.e., algebraically general type
I or type D or trivial type O (corresponding to a vanishing Weyl tensor)
then, and only then, principal observers exist. In \cite{Bel62}
Bel pointed out the correct difference between the non-trivial Petrov
types I and D in this respect, showing that in the Petrov type I case
only one { observer}
is principal, while in the Petrov type
D case the 
{ principal observers are precisely those observers lying in the 2-plane spanned by the repeated principal null directions of the Weyl tensor}.
Bel moreover identified the principal observers as those relative
to which the super-Poynting vector vanishes, and demonstrated that
this precisely happens when the electric and magnetic parts of the
Weyl tensor (considered as endomorphisms of tangent space) 
can be simultaneously diagonalized in a {\em real} orthonormal (eigen)frame.

It is worth emphasizing that a generic spacetime is of Weyl-Petrov type I at each
point and thus allows for a unique congruence of Weyl  principal observers. On taking an orthonormal tetrad formalism where the timelike tetrad
vector is the principal one, the classification and determination of all Petrov
type I spacetimes with given additional properties becomes feasible; see
e.g.\ \cite{Wyl08} for a class of algebraically general rotating
dust spacetimes. Many examples of Petrov type D spacetimes exist~\cite{SKMHH}: the Petrov type D vacuum spacetimes and doubly aligned non-null Einstein-Maxwell fields,
which have been fully classified (with or without cosmological constant)~\cite{Kinnersley,Overview-D-alinged} and contain    
{well-known black hole solutions} such as the Schwarzschild, Reissner-Nordstr\"{o}m, Kerr, Kerr-Newman, Kerr-NUT-(A)dS, (charged or spinning) $C$ and Pleba\'{n}ski-Demia\'{n}ski
{spacetimes}; all spherically symmetric spacetimes like the Lema\^{i}tre-Tolman-Bondi
dust models, and the asymmetric Szekeres dust model, to name a few; some notable congruences of observers
in these solution families (e.g.\ the static observers in Schwarzschild, or the
Carter observers in the Kerr and Kerr-Newman cases) are actually Weyl principal.

In the present paper we will focus on the pointwise relation of general
observers to principal ones. Much of the aforementioned work in the
gravitational case has been inspired by the EM setting;
here observers with respect to which the Poynting vector vanishes
will also be called principal. For a non-zero
\Max tensor  principal observers exist precisely when
the EM field is non-null, and are those observers 
{lying in}
the 2-plane spanned by the principal null directions. It is known 
that given any observer $u^{a}$, there is 
a (unique) 
{principal observer} $\umw^a$  
which travels instantaneously relative
to $u^{a}$ in the direction of the Poynting vector $\sp^{a}$ measured
by $u^{a}$, 
with speed $\tanh(\ww)$ implicitly given by the Wheeler equation
\begin{equation}
\label{eq:MisWhe57}
\tanh(2\ww)=\sqrt{{p^{a}p_{a}}}/{\te}\,,
\end{equation}
where $\te$ is the EM energy density as measured by $u^{a}$; this will be referred to as the `EM Wheeler result'.\footnote{{
The result is already implicit in Synge's standard work} \cite{Synge}. To our knowledge, it makes its first appearance in the caption of Figure 1 of \cite{MisWhe57}, albeit with an error in the description. The result is mentioned as a problem in section 25 of \cite{LandauLifshitz} and proved in \cite{Whe77}, producing the correct formula \eqref{eq:MisWhe57}; see also Ex.\ 20.6 in \cite{Gravitation}, and \cite{Bini}. }  In his
contribution \cite{Whe77} Wheeler emphasized the ``miracle'' of
this result. He also scrutinized whether an analogous result could
hold in the gravitational case with diagonal Petrov type -- on replacing Poynting by super-Poynting vector and with a surrogate of \eqref{eq:MisWhe57}, which will be referred to as the `gravitational Wheeler analogue' -- but came to negative conclusions. In the present paper we will clarify the EM Wheeler result, prove that the gravitational Wheeler analogue {\em does} hold when the Weyl tensor is of Petrov type D at a spacetime point, and find that it is invalid for {\em general} but valid for {\em some} observers when the Petrov type is I,
{thereby refining Wheeler's negative conclusions in this case.} 

In \cite{Whe77} Wheeler also studied the possibility of obtaining analytically the boost(s) that annihilate the super-Poynting vector
{[}i.e., boost(s) relating a general observer $u^{a}$ to the principal
one(s){]} directly from the  electric and magnetic parts of the Weyl tensor as measured by an arbitrary observer, but deemed this impossible. In this paper, however, we show that this {\em can} be done, and design simple algorithms for this purpose.   

\noindent {\em Outline of the paper.} 
{}
Sec.\ \ref{sec:exec-summ} provides an executive summary, where our main results are briefly outlined.

Sec.\ \ref{sec:preliminaries} fixes
the notation and conventions and provides a technical
background.  

Sec.\ \ref{sec:EM} treats the EM case. 
First we survey characterizations, existence and locus of EM principal observers. Then we derive a version of the EM Wheeler result (theorem \ref{thm:EM}) which explicates its geometric nature and provides a simple, explicit formula for the velocity of $\umw^a$ relative to $u^a$.  
We also provide an algorithm that computes all principal observers from the electric and magnetic fields as measured by an arbitrary observer.

Sec.\ \ref{sec:Petrov-diag} deals with the gravitational case. Again we first 
survey characterizations, existence and loci of Weyl principal observers. 
Next, we derive our main result, namely the gravitational Wheeler analogue for Petrov type D (theorem \ref{thm:D}),  and point out how to deduce all principal observers from 
the electric and magnetic parts of the Weyl tensor as measured by an arbitrary observer. As an application we find that in Petrov type D doubly aligned non-null Einstein-Maxwell fields the Poynting and super-Poynting vectors are aligned, at each point and for each observer, and the principal observers coincide. Finally, the Petrov type I case is studied; we show 
that the observers for which
the gravitational Wheeler analogue {\em does} hold form a one-dimensional variety (i.e., a finite union of curves and points), which we pin down in the `degenerate'  Petrov type I subcases in the sense of \cite{McIntosh3}, 
and present an algorithm that derives the unique principal observer directly from the electric and magnetic parts of the Weyl tensor measured by an arbitrary observer. 

Sec.\ \ref{sec:Examples} illustrates our results in simple examples:
the non-null EM field of a spinning charge, {the Petrov type D Kerr-Newman
spacetimes}, and the 
Petrov type I Kasner spacetimes.

Sec.\ \ref{sec:disc} summarizes our results and comments on the
physical significance of the super-Poynting vector. 

Supporting or additional material is put in three appendices:
a geometric proof of the bijection between oriented
\ON frames in self-dual bivector and tangent space, notes on 
the Petrov classification of the Weyl tensor and `degenerate' Petrov type I, 
and a simple proof of the minimal super-energy density characterization of Weyl principal observers.

\section{Executive summary}\label{sec:exec-summ}

In electromagnetism, for a non-null Faraday tensor (i.e., in
an electromagnetic field which is not pure radiation), there are observers
for whom the Poynting vector vanishes {(and whose measured electromagnetic energy density is minimum)}. These are called {\em EM principal observers}. Given the electric and magnetic fields $E^{a}$ and $B^{a}$
as measured by some arbitrary observer (of 4-velocity) $u^{a}$, the
principal observers $u'^{a}$ are those that are boosted (with respect
to $u^{a}$) with a relative velocity $\nurel(u',u)^{a}=\nurel_{\parallel p}^{a}+\nurel_{\parallel\mathfrak{e}}^{a}$
having a \emph{fixed} component ($\nurel_{\parallel p}^{a}$) along
the Poynting vector $p^{a}=\epsilon^{a}_{\ bcd}E^{b}B^{c}u^{d}$,

\begin{center}
\begin{tabular}{lc}
\raisebox{5.5ex}{}${\displaystyle \nurel_{\parallel p}^{a}=\frac{1}{\te+|\IEM|/8\pi}\,\sp^{a}\ ;}$ & \multirow{3}{*}{~~~~~~~~~~~~~~~~~%
\begin{tabular}{l}
\raisebox{5.5ex}{}(Electromagnetism,\tabularnewline
non-null $F_{ab}$)\tabularnewline
\tabularnewline
\end{tabular}}\tabularnewline
\raisebox{5ex}{}${\displaystyle \te\equiv T_{ab}u^{a}u^{b}=\frac{1}{8\pi}(\E^{a}\E_{a}+\B^{a}\B_{a})}\ ,$ & \tabularnewline
\raisebox{6ex}{}\raisebox{2ex}{${\displaystyle \IEM\equiv\E^{a}\E_{a}-\B^{a}\B_{a}-2iE^{a}\B_{a}\ }$} & \tabularnewline
\end{tabular}
\par\end{center}

\noindent (where $T_{ab}$ is the EM energy-momentum tensor), and an \emph{arbitrary} component ($\nurel_{\parallel\mathfrak{e}}^{a}$)
along 
\[
\eb^{a}=\Re\left(\frac{f^{a}}{\sqrt{\IEM}}\right),\qquad\qquad f^{a}\equiv\E^{a}-i\B^{a}\ .
\]
The EM principal observers form then, at each point, an infinite class,
consisting of all unit vectors belonging to the distinguished timelike plane $\Sigma_F$ spanned
by the two principal null directions of the Faraday tensor, called the {\em timelike
EM principal plane}. Given $u^a$, a canonically associated EM principal observer can always be found by boosting in
the direction of $\sp^a$ (with relative velocity $\nurel_{\parallel p}^{a}$); geometrically, this is the principal observer obtained by projecting $u^a$ onto $\Sigma_F$, and the vector $\eb^a$ above  lies along 
the aligned electric and magnetic fields 
measured by this principal observer.

Here we treat the analogous gravitational problem, first posed by
Wheeler: given the electric
and magnetic parts  $E_{ab}$ and $H_{ab}$ of a non-zero Weyl tensor measured by an arbitrary
observer $u^{a}$, find the observers for which the so-called \emph{super-Poynting
vector} vanishes. Such observers are dubbed {\em Weyl principal} and
exist only for Petrov type I and D spacetimes. 

The type D case exhibits
a strong analogy with the EM counterpart: the principal observers $u'^{a}$ are those that
are boosted (with respect to $u^{a}$) with a relative velocity $\nurel(u',u)^{a}=\nurel_{\parallel\mathcal{P}}^{a}+\nurel_{\parallel\mathfrak{e}}^{a}$
having a \emph{fixed} component ($\nurel_{\parallel\mathcal{P}}^{a}$)
along the super-Poynting vector $\mathcal{P}^{a}=\frac12\epsilon^{a}_{\ bcd}E^{b}_{\ e}H^{e c}u^{d}$,

\begin{center}
\begin{tabular}{lc}
\raisebox{5.5ex}{}${\displaystyle \nurel_{\parallel\mathcal{P}}^{a}=\frac{8}{3|I|\AD(\AD+1)}\,\qp^{a}\ ;}$ & \multirow{4}{*}{~~~~~~~~~~~~~%
\begin{tabular}{l}
\raisebox{8ex}{}(Gravity,\tabularnewline
Petrov type D)\tabularnewline
\tabularnewline
\end{tabular}}\tabularnewline
\raisebox{6.5ex}{}$\AD\equiv\sqrt{\frac{1}{3}\left(\frac{8\tg}{|I|}+1\right)}\ ,$ & \tabularnewline
\raisebox{4ex}{}$\tg\equiv T_{abcd}u^{a}u^{b}u^{c}u^{d}=\frac{1}{4}(\E^{ab}\E_{ab}+\H^{ab}\H_{ab})\ ,$ & \tabularnewline
\raisebox{6ex}{}\raisebox{2ex}{${\displaystyle I=\E^{ab}\E_{ab}-\H^{ab}\H_{ab}-2i\E^{ab}\H_{ab}\ }$} & \tabularnewline
\end{tabular}
\par\end{center}

\noindent (where $T_{abcd}$ is the Bel-Robinson tensor), and an \emph{arbitrary} component ($\nurel_{\parallel\mathfrak{e}}^{a}$)
along $\eb^{a}$. The vector $\eb^{a}$ is the real part of the (up to sign unique) unit
eigenvector associated to the simple eigenvalue
of the complex tensor $\Q_{\ \ b}^{a}\equiv\E_{\ \ b}^{a}-i\H_{\ \ b}^{a}$; it is given
by
\begin{align*}
 & \eb^{a}=\Re\left(\frac{\wx^{a}}{\sqrt{\wx^{b}\wx_{b}}}\right);\qquad\qquad\wx^{a}=(\E_{\ \ b}^{a}-i\H_{\ \ b}^{a}+J\I_{\;\;b}^{a}/I)y^{b}\ ,\\
 & J=\E_{\ b}^{a}\E_{\ c}^{b}\E_{\ a}^{c}+i{\H}_{\ b}^{a}\H_{\ c}^{b}\H_{\ a}^{c}-3i\,\E_{\ b}^{a}(\E_{\ c}^{b}-i\H_{\ c}^{b})\H_{\ a}^{c}\ ,\quad \I_{ab}\equiv g_{ab}+u_{a}u_{b}\ ,
\end{align*}
where $y^{a}$ is any real spatial vector (with respect to $u^{a}$)
which is not an eigenvector associated to the repeated eigenvalue
of $\Q^a{}_{b}$; i.e. (in the generic case that $u^{a}$ is non-principal) any vector non-proportional to the super-Poynting vector $\mathcal{P}^{\alpha}$. 
The Weyl principal observers exhibit features analogous to their EM
counterparts: i) at each point, infinitely many such observers
exist, namely all unit vectors lying in the {\em timelike Weyl principal plane}
$\Sigma_C$, spanned by the two principal null directions of the Weyl
tensor; 
ii) given $u^a$, a canonically associated Weyl principal observer can always be found by boosting in
the direction of $\qp^a$ (with relative velocity $\nurel_{\parallel \qp}^{a}$), namely the one obtained by projecting $u^a$ onto $\Sigma_C$;
iii) the Weyl principal observers are those measuring minimum super-energy. 
The analogy extends to the shape of the field of the Poynting/super-Poynting
vectors and principal observers in physically analogous settings (namely
a spinning charge vs the Kerr spacetime); in Petrov type D doubly
aligned non-null Einstein-Maxwell fields (such as the Kerr-Newman
spacetime) one has $\Sigma_F=\Sigma_C$, the EM and Weyl principal observers coincide,
and the Poynting and super-Poynting vectors are aligned, at each point.

In the Petrov type I case the situation differs; at each point there is now a {\em unique} Weyl principal observer $e_0^a$, which {\em in general} is not obtained by boosting in the direction of the super-Poynting vector. It has  
relative velocity

\begin{center}
\begin{tabular}{lc}
\raisebox{5.5ex}{}${\displaystyle \nurel^{a}\equiv \nurel(e_0,u)^{a}=-\frac{1}{\gamma^{2}}\sum_{(ijk)}\Im(\v_{j}^{b}(\overline{\v_{k}})_{b})\v_{i}^{a}\ ;}$ & \multirow{3}{*}{~~~~~~~~~%
\begin{tabular}{l}
\raisebox{8ex}{}(Gravity,\tabularnewline
Petrov type I)\tabularnewline
\tabularnewline
\end{tabular}}\tabularnewline
\raisebox{5ex}{}$\gamma^{2}=\frac{1}{2}\left[\sum_{i=1}^{3}\v_{i}^{a}(\vc_{i})_{a}+1\right];\qquad{\displaystyle \v_{i}^{a}=\frac{(R^{i})_{\;\;b}^{a}y^{b}}{\sqrt{R_{cd}^{i}y^{c}y^{d}}}\ ,}$ & \tabularnewline
\raisebox{8ex}{}\raisebox{3ex}{${\displaystyle (R^{i})_{\;\;b}^{a}=\frac{\Q_{\;\;c}^{a}\Q_{\;\;c}^{a}+\l_{i}\Q_{\;\;b}^{a}+(\l_{i}^{2}-I/2)\I^a_{\;\;b}}{3\l_{i}^{2}-I/2}}$\ ,} & \tabularnewline
\end{tabular}
\par\end{center}

\noindent where $y^{a}$ is any real spatial vector which is \emph{not} an eigenvector
of $\Q_{\;\;b}^a$ with eigenvalue $\lambda_{i}$. %
Only for a one-dimensional variety of observers $u^a$  
one has $\nurel^a\propto\qp^{a}$; this variety always contains the three curves of observers obtained by arbitrarily boosting the principal observer along one of the spatial directions of the Weyl principal tetrad, but contains other observers as well, depending on the Weyl parameters; we identify these additional observers for types I$(M^\infty)$, I$(M^+)$ and I$(M^-)$ within the  extended Petrov classification by Arianrhod and McIntosh~\cite{McIntosh3}.


\section{Preliminaries}
\label{sec:preliminaries}

\subsection{Basic notation and conventions}
\label{subsec:notconv} 

\hspace{.3cm} $\boldsymbol{\S1}$. We work at a point $p$ of a spacetime $(M,g_{ab})$ with {metric} signature
$(-\,+\,+\,+)$, and use units where $8\pi G=c=1$, {with $G$ the gravitational constant and $c$ the speed of light}. For tensors abstract
index notation with small Latin letters $a,b,c,d,e,f$ is used. Abstract
indices are lowered and raised by contraction with $g_{ab}$, resp.\ $g^{ab}$
($g^{ac}g_{cb}=\delta_{b}^{a}$, with $\d^a_b$ the identity (1,1)-tensor), and round (square) brackets
around indices indicate (anti)symmetrization. $\LV$ stands for the space of real 2-forms $X_{ab}=X_{[ab]}$ at $p$. In $T_p M$, $S^{\bot}$ denotes the $g$-orthogonal
complement of a vector $S^a$ or set of vectors $S$, and {$\langle x_1^{a},x_2^{a},\ldots\rangle$
the subspace spanned by $x_1^{a},x_2^a,\ldots$}. 
A tuple $(e_{0}^{a},e_{i}^{a})$ symbolizes a restricted
\ON tetrad of $T_{p}M$ (i.e., the tetrad is properly oriented
and $e_{0}^{a}$ is future-pointing~\cite{Penrosebook1}) 
and Latin letters $i,j,k,l,m,n$ are triad indices taking values from
$1$ to $3$. In such a tetrad we take the convention $\e_{0123}=-1$
for the Levi-Civita pseudo-tensor $\e_{abcd}=\e_{[abcd]}$. When the
labels $i,j,k$ appear in one expression, e.g. $\mu_{i}=\lambda_{j}-\lambda_{k}$, then $(i,j,k)$ is 
a cyclic permutation of $(1,2,3)$; $\sum_{(ijk)}$ denotes
cyclic summation. For both abstract and triad indices the Einstein
summation convention (only) applies to two indices in opposite (upper
and lower) positions, unless stated otherwise. We abbreviate e.g.\ three-dimensional to 3d.

$\boldsymbol{\S2}$. $\C V\equiv\C\otimes V$ denotes the complexification
of a real vector space~$V$. Complex conjugation is symbolized by a bar, $\Re$ and $\Im$ indicate  real and imaginary parts; $|z|=\sqrt{\Re(z)^2+\Im(z)^2}$ denotes the modulus of $z\in\C$; we will use that the square roots of $z$ are $\pm\sqrt{z}$ with 
\begin{equation}
\label{eq:root-z}
\sqrt{z}\equiv \a+i\b,\quad \frac{1}{\sqrt{z}}=\frac{1}{|z|}(\alpha-i\beta),\quad \alpha\equiv\sqrt{\frac{|z|+\Re(z)}{2}},\quad \beta\equiv \sgn(\Im(z))\sqrt{\frac{|z|-\Re(z)}{2}},
\end{equation}
and where $\sgn:\R\ra\{-1,1\}$ 
is defined by $\sgn(x)=1$ if $x>0$ and $\sgn(x)=-1$ if $x\leq 0$.
For endomorphisms $\Fop,\Gop$ of $V$ or $\C V$ we write $\Fop\Gop\equiv \Fop\circ\Gop$ and 
inductively define $\Fop^{i+1}\equiv \Fop^i\Fop$; $\Ker(\Fop)$, $\Imm(\Fop)$ and $\tr(\Fop)$ respectively denote the kernel, image and trace of $\Fop$. $T^a_{\;\;b}$ is seen as an endomorphism of $\C T_pM$. 
The Hodge dual of a {\em bivector} $A_{ab}\in\LL$ is {denoted} $\st{A}_{ab}\equiv\frac{1}{2}\e_{abcd}A^{cd}$.
On $\C T_p M$ and $\LL$ we consider the respective metrics
\begin{equation}\label{defgG}
g:\,(v^a,w^a)\mapsto g_{ab}v^aw^b=v^aw_a,\qquad G:\,(A_{ab},B_{ab})\mapsto-\tfrac{1}{4}A^{ab}B_{ab}.
\end{equation}
A vector $v^a\in\C T_pM$ with $v^av_a=1$ is called {\em unit}, and a bivector $A_{ab}\in\LL$ with $-\frac14 A^{ab}A_{ab}=1$ {\em unitary}.

\subsection{Observers}
\label{subsec:observers} 

\hspace{.3cm} $\boldsymbol{\S1}$. Let $\Up$ denote the set of future-pointing, normalized timelike vectors $u^{a}$ at
$p$ ($u^{a}u_{a}=-1$ and $u^{0}>0$ in a restricted \ON tetrad).
An {\em observer} is identified with a worldline with {normalized} tangent vector field
$u^{a}$ (the observer's 4-velocity). Since the quantities treated
in the present paper depend only on the observer's 4-velocity and
position, we shall for short speak about the observer $u^{a}(\in\Up)$ at
$p$. The tensor $\I_{~b}^{a}$ defined by 
\begin{equation}
\label{def-hab}
\I_{ab}\equiv g_{ab}+u_{a}u_{b}
\end{equation}
represents the projector onto the (complexified) instantaneous rest space $(\C)\up$ of $u^a$ at $p$. A tensor $T^{a\ldots}{}_{b\ldots }$ that equals
$\I^{a}_{~c}\cdots \I_{b}^{~d}\cdots T^{c\ldots}{}_{d\ldots}$ is called {\em spatial}
(with respect to $u^{a}$). A spatial tensor $T_{ab}$ 
induces an endomorphism $\vvec^a\mapsto T^a_{~b}\vvec^b$ of $\Cup$ which we symbolize by $\Top$;
hence $\Iop$ is 
the identity map of $\Cup$. 
When $S_{ab}$ and $T_{ab}$ are spatial and symmetric, the vector dual to the commutator $\Sop\Top-\Top\Sop$ (with associated 4d tensor $[\Sop,\Top]^a_{\;\;b}\equiv S^a_{\;\;c}T^c_{\;\;b}-T^a_{\;\;c}S^c_{\;\;b}$) is  
$$
[\Sop,\Top]^a\equiv\e^{a}{}_{bcd}u^dS^b_{\;\;e}T^{ec}.
$$ 
For spatial vectors $x^a,y^a\in\Cup$ we write $[x,y]^a\equiv\e^{a}{}_{bcd}u^dx^by^c$ for their vector product, $\xop\yop$ for the endomorphism $\vvec^a\mapsto (\vvec^by_b)x^a$ of $\Cup$, and we will use the identity
\begin{equation}
\label{eq:id[xxyy]}
[\xop\xop,\yop\yop]^a=x^by_b\,[x,y]^a.
\end{equation}

$\boldsymbol{\S2}$. The relative motion of two observers $u_1^a,\,u_2^a\in\Up$ is described by~(see, e.g., \cite{Bini,Jantzen})
\begin{equation}\label{eq:def-upr}
u_2^a=\cosh(\ww)[u_1^a+\nurel^a]=\cosh(\ww)u_1^a+\sinh(\ww)\nuhat^a.
\end{equation}
Here $\nurel^a\equiv\vrel{u_2}{u_1}^a\in u_1^\bot$ is the velocity of $u_2^a$ relative to $u_1^a$, and $\ww\equiv\wrel{u_2}{u_1}=\wrel{u_1}{u_2}\geq0$ and 
$\cosh(\ww)= -u_2^a(u_1)_a=1/\sqrt{1-\nurel^a\nurel_a}$ 
are the associated rapidity parameter and Lorentz factor,
respectively; 
$\nuhat^a\equiv \vhat{u_2}{u_1}^a$ is the unit vector in the direction of $\nurel^a\neq 0$ if $u_2^a\neq u_1^a$ and will be formally taken to be any unit vector in $u_1^\bot$ if $u_2^a=u_1^a\Leftrightarrow \nurel^a=0\Leftrightarrow \ww=0$, such that $\vrel{u_2}{u_1}^a=\tanh(\ww)\vhat{u_2}{u_1}^a$ in any case. 
Eq.\ \eqref{eq:def-upr} defines the unique boost $\BB{u_2}{u_1}^a_{\;\;b}$
which acts trivially on $\langle u_1^a,u_2^a\rangle^\bot$ and maps $u_1^a$ to $u_2^a=\BB{u_2}{u_1}^a_{\;\;b}u_1^b$; this is the identity transformation of $T_p M$ if $u_2^a=u_1^a$, while if $u_2^a\neq u_1^a$ one has\footnote{The first equality in  \eqref{eq:transfo-nuhat} is due to $\BB{u_2}{u_1}^a_{\;\;b}$ being an orthochronous Lorentz transformation, while the second equality follows by reversing the roles of $u_1^a$ and $u_2^a$ in \eqref{eq:def-upr}; the result agrees with Eq.\ (14) in \cite{Bini} and Eq.\ (4.5) in \cite{Jantzen}.} 
\begin{equation}
\label{eq:transfo-nuhat}
\BB{u_2}{u_1}^a_{\;\;b}\vhat{u_2}{u_1}^b=\sinh(\ww)u_1^a+\cosh(\ww)\vhat{u_2}{u_1}^a=-\vhat{u_1}{u_2}^a\,.
\end{equation}

\subsection{Self-dual bivectors}
\label{subsec:sd-bivectors} 

\hspace{.3cm} $\boldsymbol{\S1}$. Bivectors ${\Xcal}_{ab}\in\LL$ that satisfy $\st{\Xcal}_{ab}=i\Xcal_{ab}$ are called {\em self-dual}; 
they form a 3d complex vector space, denoted $\Sp$. 
For any $A_{ab}\in\LL$ one has $A_{ab}^{\dag}\equiv A_{ab}-i\st{A}_{ab}\;\in\;\Sp$. 
The relations 
\begin{equation}
\label{eq:rel-Xcal-X}
X_{ab}=\Re(\Xcal_{ab})\quad\lra\quad\Xcal_{ab}=X_{ab}^{\dag}=X_{ab}-i\st{X}_{ab}
\end{equation}
define a bijection
between complex self-dual bivectors ${\Xcal}_{ab}\in\Sp$ and real bivectors $X_{ab}\in\LV$, and with this notational correspondence we put 
\begin{equation}
\label{eq:Xcal-norm}
\IX\equiv-\tfrac{1}{4}\Xcal^{ab}\Xcal_{ab}=-\tfrac{1}{2}X^{ab}X_{ab}+\tfrac{i}{2}\st{X}^{ab}X_{ab}\,.
\end{equation}
For $\Xcal_{ab},\,\Ycal_{ab}\in\Sp$ one has the identities
\begin{align}
\label{eq:XYcal-id}&\Xcal^{ab}\Ycal_{bc}+\Ycal^{ab}\Xcal_{bc}=\tfrac12 \Xcal_{gh}\Ycal^{hg}\delta^a_c \quad\Ra\quad \Xcal{}^a{}_{c}\Xcal{}^c{}_{b}=\IX\delta^a_c\,,\\
\label{eq:XcalYcalc-id}&\Xcal{}^a{}_{c}\Ycalc{}^c{}_{b}=\Ycalc{}^a{}_{c}\Xcal{}^c{}_{b}\,,\qquad \Xcal^{ab}\Ycalc_{ab}=0\,.
\end{align}
For an arbitrary but fixed observer $u^a\in\Up$ the relations 
\begin{equation}
\label{eq:def-psiuchiu}
\x^a=\Xcal^{ab}u_b \quad\lra\quad \Xcal_{ab}=X_{ab}^\dag=(2u_{[a}\x_{b]})^{\dag}=2u_{[a}\x_{b]}+i\e_{abcd}u^d\x^{c}
\end{equation}
define a fundamental isometric identification of $\Sp$ and $\Cup$ (pair of 3d complex vector space isometries, 
where the restrictions of $G$ and $g$ defined in \eqref{defgG} are taken as respective metrics)~\cite{SKMHH}:\footnote{$\Cup$ and $\Sp$ are both 3d complex vector spaces; the map $x^a\mapsto (2u_{[a}\x_{b]})^{\dag}$ is clearly an injective homomorphism and thus an isomorphism, with inverse $\Xcal_{ab}\mapsto \Xcal^{ab}u_b$; \eqref{eq:psiu-isom} follows from \eqref{eq:XYcal-id}.}
\begin{equation}
\label{eq:psiu-isom}
\x^a= \Xcal^{ab}u_{b},\,\y^a= \Ycal^{ab}u_{b}\quad \Rightarrow\quad  \x^a\y_ a=-\tfrac{1}{4}\Xcal^{ab}\Ycal_{ab},\quad \x^a\x_a= \IX\,.
\end{equation}

$\boldsymbol{\S2}$. Let $\Xcal_{ab}=X^\dag_{ab}$ be {\em unitary} ($\IU=1$). The imaginary part of \eqref{eq:Xcal-norm} gives $\st{X}^{ab}X_{ab}=0$; hence $X_{ab}$ and $\st{X}_{ab}$ are {\em simple}~\cite{Hall} and have as respective {\em blades} the orthogonal 2-planes
\begin{equation}
\label{def-Sigma}
\SS\equiv\Imm(\U^a_{\;\;b})=\Ker(\st{\U}^a_{\;\;b}),\qquad
\SSb=\Imm(\st{\U}^a_{\;\;b})=\Ker(\U^a_{\;\; b}).
\end{equation}
Combined with the real part of \eqref{eq:Xcal-norm} this leads to the existence of a 
restricted \ON tetrad $(e_0^a,e_i^a)$ such that 
\begin{equation}
\label{eq:Ucal-form}
\Ucal_{ab}=[2(e_0)_{[a}(e_1)_{b]}]^{\dag}
=\U_{ab}-i\st{\U}_{ab}
\quad\lra\quad \U^{ab}=2\Enul^{[a}\Eone^{b]},\quad\st{\U}^{ab}=-2\Etwo^{[a}\Etr^{b]}.
\end{equation}
One has $\SS=\langle e_0^a,e_1^a\rangle$ and $\SSb=\langle e_2^a,e_3^a\rangle$, with corresponding pair of projectors
\begin{equation}
\label{eq:PPbot-def}
P_{\;\;b}^{a}=\U^a_{\;\;c}\U^c_{\;\;b}=-\Enul^{a}(\Enul)_{b}+\Eone^{a}(\Eone)_{b}\,,\qquad
P_{\bot b}^{a}=-\st{\U}^a_{\;\;c}\st{\U}^c_{\;\; b}
=\Etwo^{a}(\Etwo)_{b}+\Etr^{a}(\Etr)_{b}\,.
\end{equation}
For any observer $u^a$ we have by \eqref{eq:def-psiuchiu} and \eqref{eq:psiu-isom}:
\begin{align}
\label{eq:va-def-equiv}
&\v^a
=\Ucal^{ab}u_b
\equiv\eb^a-i\bb^a\quad\lra\quad \eb^a\equiv \U^a_{\;\;b}u^b,\;\;\bb^a\equiv \st{\U}^a_{\;\;b}u^b\quad\lra\quad \U_{ab}=2u_{[a}\eb_{b]}+\e_{abcd}u^d\bb^{c},\\ 
\label{eq:va-unit}
&\v^{a}\v_{a}=\eb^{a}\eb_{a}-\bb^{a}\bb_{a}=1,\qquad\eb^{a}\bb_{a}=0.
\end{align}
Combined with \eqref{def-Sigma} it follows that $\bb^a\in\SSb$ is orthogonal to the spatial, non-zero vector $\eb^a\in\SS$, and 
\begin{equation}
\label{eq:principal-conds}
\bb^a=0\;\;(\v^a=\eb^a)\quad\Lra\quad  u^a\in\SS\, .
\end{equation}

\section{The electromagnetic case}
\label{sec:EM}

\subsection{EM principal observers}
\label{subsec:char-EM} 

In electromagnetism the \Max two-form $F_{ab}\in\LV$ 
is the governing tensor. Let $\Fcal_{ab}\equiv F_{ab}-i\st{F}_{ab}\in\Sp$ be the EM self-dual bivector. The electric and magnetic fields $\E^{a}=F^{ab}u_{b}$ and $\B^{a}=\st{F}^{ab}u_{b}$ as measured by an observer $u^{a}$ can be assembled into the complex vector $\f^a$, which by \eqref{eq:def-psiuchiu} determines $\Fcal_{ab}$:
\begin{equation}
\label{def-fa}
\f^{a}\equiv\E^{a}-i\B^{a}=\Fcal^{ab}u_{b}
\quad\Leftrightarrow\quad \Fcal_{ab}=2u_{[a}\f_{b]}+i\e_{abcd}u^d\f^{c}.
\end{equation}
Referring to \eqref{eq:Xcal-norm} and \eqref{eq:psiu-isom} the complex EM invariant $\IEM$ is defined by
\begin{align}
\label{eq:def-IEM}
\IEM&\equiv-\tfrac{1}{4}\Fcal^{ab}\Fcal_{ab}=-\tfrac{1}{2}F^{ab}F_{ab}+\tfrac{i}{2}\st{F}^{ab}F_{ab}\\
\label{eq:def-IEMb}
&=\f^{a}\f_{a}=\E^{a}\E_{a}-\B^{a}\B_{a}-2iE^{a}\B_{a}.
\end{align}
The EM field is called {null} if $\IEM=0$, else {non-null}. 

Let us review the definition, characterizations,
existence and locus of EM principal observers at a point $p$. 
The (tracefree, symmetric) EM energy-momentum tensor associated to $F_{ab}$ is given by
\begin{align}
\label{eq:Tab-EM}
T_{ab} \equiv\frac{1}{4\pi}F_{ac}F_{b}{}^{c}-\frac{1}{16\pi}F_{cd}F^{cd}g_{ab}= \frac{1}{8\pi}\Fcal_{ac}\Fcalc_{b}{}^{c}=\frac{1}{8\pi}\Fcalc_{ac}\Fcal_{b}{}^{c}
.
\end{align}
Relative to $u^a\in\Up$ the (spatial) Poynting vector $\sp^{a}$, energy density $\te$ and energy flux vector $s^a$ are 
\begin{align}
\label{eq:spa-def}
\sp^{a} & \equiv-\I^{ab}T_{bd}u^{d}=\frac{1}{4\pi}\e^{a}{}_{bcd}\E^{b}\B^{c}u^d\equiv\frac{1}{4\pi}[E,B]^a=\frac{1}{8\pi i}\e^{a}{}_{bcd}u^d\f^{b}\fc{}^{c}\equiv\frac{1}{8\pi i}[\f,\fc]^a,\\
\label{eq:te-def}
\te &\equiv T_{ab}u^{a}u^{b}=\frac{1}{8\pi}(\E^{a}\E_{a}+\B^{a}\B_{a})=\frac{1}{8\pi}\f^{a}\fc_{a},\\
\label{eq:sa-def}
s^{a} &\equiv-T^{ab}u_{b}=\sp^{a}+\te u^{a}.
\end{align}
$T^a{}_{b}$ 
satisfies $T_{~b}^{a}T_{~c}^{b}=\che^{2}\d_{c}^{a}$~\cite{Ruse} and thus has eigenvalues $\pm\che$, 
where the invariant 
\begin{equation}
\label{eq:def-che}
\che=\frac12\sqrt{T^{ab}T_{ab}}=\frac{1}{8\pi}{|\IEM|}
=\frac{1}{8\pi}|\f^a\f_a|=\frac{1}{8\pi}\sqrt{(\E^{a}\E_{a}-\B^{a}\B_{a})^{2}+4(E^{a}\B_{a})^{2}}
\end{equation} 
is the {\em proper EM energy density}~\cite{Coll}. Combined with \eqref{eq:sa-def} it follows that
\begin{equation}
\label{eq:rel-te-norm}
\te^{2}=-s^{a}s_{a}+\sp^{a}\sp_{a}=\che^{2}+\sp^{a}\sp_{a}\geq\che^2,
\end{equation}
such that $\che$ is a lower bound and in fact the infimum for $\te$ regarded as a non-negative scalar function $\Up\ra\Rnn$ (i.e., if we let $u^a$ range over $\Up$)~\cite{FerSae12,Synge}.

\begin{defn}\label{def:princobs-EM} An observer $u^{a}$ is {\em
		(EM) principal} when it measures a vanishing Poynting vector, $\sp^{a}=0$.\footnote{Such observers have been called `observers at rest' with respect to
		the EM field~\cite{ColFer}; we find however such `rest' notion somewhat
		confusing, since at each point there are infinitely many principal
		observers, which are not at rest with respect to each other, while
		(according to such notion) all being at rest with respect to the field
		(which is odd since well posed notions of relative rest are locally
		transitive \cite{BolosIntrinsic}). {Therefore} we shall not use it.
		An analogous remark applies to the same terminology used in \cite{FerSae12}
		for the gravitational case in the next section.} 
\end{defn}

By \eqref{eq:Tab-EM}-\eqref{eq:rel-te-norm}
an EM principal observer $u^a$ is characterized by any of the following conditions:\\
\vspace{-.2cm}

(i) the electric and magnetic fields are aligned ($\E^{[a}\B^{b]}=0$), i.e., linearly dependent~\cite{MisWhe57};

(ii) $\f^{a}=\sqrt{\IEM}\,\eb^{a}$ for some {\em real} unit 
vector $\eb^{a}\in\up$ \,(cf.\ p.\ 69 of \cite{LandauLifshitz});

(iii) $u^{[a}T^{b]}{}_{c}u^{c}\equiv u^{[a}F^{b]}{}_{c}F^{c}{}_{d}u^{d}
=0$, i.e., $u^a$ is an eigenvector of $T^a{}_{b}$~\cite{Pirani,Synge}
or $F^a_{\;\;c}F_{\;\;b}^{c}$;


(iv) the EM energy density $\te$ attains minimum value, namely $\che$~\cite{FerSae12,Synge, Bini,HerOrtWyll13}.\\

\vspace{-.2cm}

\noindent Note that alignment of $\E^{a}$ and $\B^a$ means $\E^{a}\propto\eb^{a}$ and $\B^{a}\propto\eb^{a}$, and thus $\f^a\propto\eb^a$, for some unit vector $\eb^{a}\in\up$, and so by \eqref{eq:def-IEMb} characterization (ii) is the complex version of (i).


We still mention two new characterizations, involving minimality of scalar functions 
$\Up\ra\Rnn$ 
and equivalent to (iv). 
First,
combining \eqref{eq:te-def} with the invariant $\E^a\E_a-\B^a\B_a=\Re(\IEM)=-\frac{1}{2}F^{ab}F_{ab}$ gives
\begin{equation}
\label{EB-norm 1}
\E^a\E_a=4\pi\te+\frac12 \Re(\IEM),\qquad \B^a\B_a=4\pi\te-\frac12 \Re(\IEM). 
\end{equation}
%
Hence, if we define the invariants 
\begin{equation}
\label{eq:def-cheEH}
\cheE\equiv\sqrt{\frac{|\IEM|+\Re(\IEM)}{2}}=|\Re(\sqrt{\IEM})|,\qquad
\cheH\equiv \sqrt{\frac{|\IEM|-\Re(\IEM)}{2}}=|\Im(\sqrt{\IEM})|
\end{equation}
(where the last equalities follow from \eqref{eq:root-z} applied to $z=\IEM$) we conclude from characterization (iv), $\che=|\IEM|/8\pi$ and \eqref{EB-norm 1} that an observer is EM principal iff one of the following similar criteria holds:\footnote{In \cite[pp.~334-335]{Synge} the invariants $\cheE$ and $\cheH$ were identified as the norms of the electric and magnetic fields acquired by {\em all} principal observers (and called therein the absolute electric and magnetic strengths) but not as the minimum norms.}\\
\vspace{-.2cm}

(v-a) the norm  $\sqrt{\E^a\E_a}$ attains minimum value, namely $\cheE$;

(v-b) the norm  $\sqrt{\B^a\B_a}$ attains minimum value, namely $\cheH$;\\
\vspace{-.2cm}

\noindent
Second, the (symmetric, spatial) tensor $\T_{ab}\equiv h_a{}^ch_b{}^dT_{cd}=\te h_{ab}-\frac{1}{4\pi}(\E_a\E_b+\B_a\B_b)$ also appears in the orthogonal splitting of $T_{ab}$ relative to $u^a$, $T_{ab}=\te u_au_b+2u_{(a}\sp_{a)}+\T_{ab}$~\cite{AlfonsoSE}.  
One has $\T^a_{\ \, a}=\te$ and $\T^{ab}\T_{ab}=\te^2+2\che^2$, so  
an EM principal observer can also be characterized by:\\
\vspace{-.2cm}

(vi) $\T^{ab}\T_{ab}$ attains minimum value, namely $3\che^2$.\\
\vspace{-.2cm}

The existence and locus of EM principal observers can be elegantly deduced from characterization (ii):
In view of \eqref{def-fa}  
such observers do not exist in the null case $\IEM=0\Leftrightarrow\che=0$ unless $F_{ab}=0$ at $p$, in which case all observers at $p$ are principal, trivially. The non-null case $\IEM\neq 0$ remains. 
Here we can normalize $\Fcal_{ab}$ to the unitary self-dual bivector $\Ucal_{ab}=\Fcal_{ab}/\sqrt{\IEM}$, where $\sqrt{\IEM}$ is one of the two square roots of $\IEM$ given by \eqref{eq:root-z} with $z=\IEM$; 
by \eqref{eq:Ucal-form} a restricted \ON tetrad  $(e_0^a,e_i^a)$ exists such that
\begin{equation}\label{eq:Fcal-non-null}
\Fcal_{ab}=\sqrt{\IEM}\Ucal_{ab},\qquad \Ucal^{ab}\equiv\U^{ab}-i\st{\U}^{ab}=2\Enul^{[a}\Eone^{b]}+2i\Etwo^{[a}\Etr^{b]}.
\end{equation}
The blade $\SS=\langle e_0^a,e_1^a\rangle$ of $X_{ab}$ is called the timelike {\em (EM) principal plane} and equips tangent space with a 2+2 structure $\SS\oplus\SS^{\bot}$; the two null directions of $\SS$ are referred to as the {\em (EM) principal null directions} (PNDs), which are the real null eigendirections (or the aligned null directions within null alignment theory~\cite{Milson04}) of $F^a_{\;\;b}$ (or $X^a_{\;\;b}$ or $\Fcal^a_{\;\;b}$), spanned by null vectors $k^a$ that satisfy $k^{[a}F^{b]}{}_{c}k^{c}=0$~\cite{Synge,Hall}.
Restricted \ON tetrads for which \eqref{eq:Fcal-non-null} holds are 
determined up to  boosts in $\SS$ and rotations in $\SSb$, i.e., given one such tetrad $(e_0^a,e_i^a)$ all other ones 
are given by
\begin{align}
\label{eq:e0e1-tetrad}&\EEt_0^a=\cosh(\aa)\et_0^a+\sinh(\aa)\et_1^a,\qquad \EEt_1^a=\sinh(\aa)\et_0^a+\cosh(\aa)\et_1^a,\qquad \aa\in\R,\\
\label{eq:e2e3-tetrad}&\EEt_2^a=\cos(\cc)\et_2^a+\sin(\cc)\et_3^a,\qquad \EEt_3^a=-\sin(\cc)\et_2^a+\cos(\cc)\et_3^a,\qquad \cc\in [0,2\pi).
\end{align}
Note that an observer $\EEt_0^a\in\SS$ and unit vector $\EEt_2^a\in\SSb$ determine a unique such tetrad by $\EEt_1^a=\U^a_{\;\;b}\EEt_0^b$ and $\EEt_3^a=\st{\U}^a_{\;\;b}\EEt_2^b$.
We could also normalize $\Fcal_{ab}$ by $-\sqrt{\IEM}$, which then yields $-\Ucal_{ab}$; thus we obtain two 2-parameter families  $(\EEt_0^a,\pm \EEt_1^a,\EEt_2^a,\pm\EEt_3^a)$ of {\em (EM) principal tetrads}.
Contraction of \eqref{eq:Fcal-non-null} with an observer $u^a$ gives
\begin{align}\label{eq:Qa-non-null}
 & \f^{a}=\sqrt{\IEM}\v^{a}\equiv \E^a-i\B^a,\qquad\v^{a}\equiv\Ucal^a_{\;\;b}u^{b}\equiv\eb^{a}-i\bb^{a}\\
\label{eq:EaBa-EM}
\lra\quad & \E^a=\Re(\sqrt{\IEM})\eb^a+\Im(\sqrt{\IEM})\bb^a,\quad \B^a=-\Im(\sqrt{\IEM})\eb^a+\Re(\sqrt{\IEM})\bb^a\,.
\end{align}
For given $u^a$  the complex vector $\v^{a}\equiv \f^a/\sqrt{\IEM}\in\Cup$ is well-defined and unit; according to characterization (ii) $u^a$ is EM principal precisely when $\v^a$ is real 
($\v^a=\eb^a$, where $\eb^a$ in (ii) and \eqref{eq:Qa-non-null} are compatible). Equation \eqref{eq:principal-conds} now shows that EM principal observers exist, and provides a characterization which may replace (ii) and describes their locus:\\
\vspace{-.3cm}

(ii)' the vector $\v^{a}\equiv \f^a/\sqrt{\IEM}$  is real 
{($\v^{a}=\eb^{a}\lra \bb^{a}=0$)}, i.e.,
$u^{a}$ lies in the timelike principal plane  $\SS$.\\
\vspace{-.2cm}

\noindent 

\noindent In summary: if an EM field is non-null at a  point, the principal observers instantaneously passing through that point are precisely those belonging to the timelike principal plane, which are the unit timelike eigenvectors of $T^a_{\;\;b}$; for these observers, and {\em only} for these, the EM energy density is minimum and the electric and magnetic vectors are aligned and have minimum norms.

\subsection{Non-null fields: general vs.\ principal observers}
\label{subsec:gen-princ-EM}


The EM Wheeler result~\cite{Whe77,LandauLifshitz,Bini} states that, given a non-principal observer $u^{a}$ in a non-null EM field, there is a (unique) principal observer $\umw^a$ which travels instantaneously relative to $u^a$ in the direction of the Poynting vector $\sp^a\neq 0$ measured by $u^a$, where the  relative speed $\tanh(\ww)$ between the two observers is implicitly given by \eqref{eq:MisWhe57}. Geometrically this means that the 2-plane $\langle u^{a},\sp^{a}\rangle$ intersects $\SS$ in a line (namely $\langle\umw^a\rangle$); thus, to the arguments Wheeler gave in \cite{Whe77}
to emphasize the ``miracle'' of the result we could add that two generic 2-planes in 4 dimensions only intersect in the origin.

Here we show that the EM Wheeler result is a simple consequence of the definition of the Poynting vector, and emphasize on the geometric nature of $\umw^a$ and $\ww$ originating from the 2+2 structure $\Sigma\oplus\Sigma^\bot$. Our analysis leads to an {\em explicit} formula for $\vrel{\us}{u}^a$ in terms of $\sp^a$ and $\te$ 
and will provide a clear parallel with the Petrov type D gravitational case studied below. 

To this end we first derive a lemma valid for any unitary self-dual bivector $\Ucal_{ab}$. We elaborate eqs.\ \eqref{eq:PPbot-def}-\eqref{eq:va-unit} and adopt the `principal' nomenclature from the non-null EM case. Given the pair of projectors \eqref{eq:PPbot-def} any non-principal observer $u^a\notin \SS$ can be uniquely decomposed along and orthogonal to $\SS$: 
\begin{align}
\label{eq:decomp-2+2}
& u^{a}=P_{\;\;b}^{a}u^{b}+P_{\bot b}^{a}u^{b}=\cosh(\pp)\us^{a}+\sinh(\pp)\ub^{a}.
\end{align}
Here $\us^{a}\in\SS$ is the observer lying along $P_{\;\;b}^{a}u^{b}$, $\pp=\wrel{u}{\us}$ is the relative rapidity of $u^a$ and $\us^a$, and 
$\ub^a=\vhat{u}{\us}^a$ is the unit vector along $P^a_{\bot b}u^{b}$; in the limit case of a principal observer ($u^a=\us^a\lra\pp=0$) we can take any unit vector in $\SSb$ for $\ub^a$. 
Eq.~\eqref{eq:decomp-2+2} defines a unique boost $\BB{u}{\us}^a_{\;\;b}$ mapping $\us^a$ to $u^a$, see Sec.~\ref{subsec:observers}. 
The inverse boost $\BB{\us}{u}^a_{\;\;b}$ is described by 

\begin{lem}\label{lem1} 
	Let $\Ucal_{ab}=\U^\dag_{ab}$ be an arbitrary unitary self-dual bivector. For any observer $u^a$ consider \eqref{eq:decomp-2+2} and define $\eb^a,\bb^a,\v^a$ as in \eqref{eq:va-def-equiv}. Then the canonically associated observer $\us^a=\cosh(\pp)[u^a+\vrel{\us}{u}^a]$ 
	lying along the projection of $u^a$ onto the principal plane $\SS$ (blade of $\U_{ab}$) is given by 
	\begin{align}\label{eq:boost-rap}
	&\cosh^2(\pp)=\eb^a\eb_a=\frac{\v^a\vc_ a+1}{2}\quad\lra\quad \v^{a}\vc_{a}=2\eb^a\eb_ a-1=\cosh(2\pp)\,,\\
	\label{eq:boost-inv}
	&\vrel{\us}{u}^a=\frac{[\eb,\bb]^a}{\eb^b\eb_b}=\frac{[\eb,\bb]^a}{\cosh^2(\pp)}\quad\lra\quad [\eb,\bb]^a=\tfrac12{\sinh(2\pp)}\vhat{\us}{u}^a\,.
	\end{align}
	Hence, when $u^a\notin\SS$ the 2-plane $\langle u^a,[\eb,\bb]^a\rangle$ coincides with the plane 	$\langle u^a,\us^a\rangle$ 
	and intersects $\SS$ in $\langle \us^a\rangle$.	
\end{lem}

To prove this lemma we extend 
$\us^{a}$ and $\ub^{a}$ to the principal tetrad 
\begin{equation}\label{EM-principaltetrad}
(\EEt_0^a,\EEt_1^a,\EEt_2^a,\EEt_3^a)=(\us^{a},\vs^{a},\ub^{a},\vb^{a})
\end{equation}
according to the note after \eqref{eq:e2e3-tetrad}, such that  
\begin{align}\label{eq:def-vsa-vba}
&\vs^{a}\equiv\U^{a}_{\;\;b}\us^{b}=\Ucal_{\;\;b}^{a}\us^{b}\equiv\v_\SS^a,\qquad\vb^{a}\equiv\st{\U}^{a}_{\;\;b}\ub^{b}=i\Ucal_{\;\;b}^{a}\ub^{b},
\quad \SS=\langle\us^{a},\vs^{a}\rangle,\; \SSb=\langle\ub^{a},\vb^{a}\rangle.
\end{align}
Observe that $\vs^{a}$ is the vector `$\eb^a$' as defined in \eqref{eq:Qa-non-null} but now relative to $\us^a$. The vectors $\eb^a$ and $\bb^a$ relative to $u^a$ are aligned to $\vs^a$ and $\vb^a$; more precisely, by substituting \eqref{eq:decomp-2+2} in the second part of \eqref{eq:va-def-equiv} and using $\U^a_{\;\;b}\ub^b=\st{\U}^a_{\;\;b}\us^b=0$ we obtain
\begin{align}\label{eq:ebbb-vsavba}
&\eb^a\equiv \U^a_{\;\;b}u^b=\cosh(\pp)\vs^a,\qquad
\bb^a\equiv \st{\U}^a_{\;\;b}u^b=\sinh(\pp)\vb^a.
\end{align}
This implies $\eb^a\eb_a=\cosh^2(\pp),\,\bb^a\bb_a=\sinh^2(\pp)$, which confirms \eqref{eq:va-unit} and together with $\v^a\vc_a=\eb^a\eb_a+\bb^a\bb_a$ gives \eqref{eq:boost-rap}. The boost 
$\BB{u}{\us}^a_{\;\;b}$ transforms \eqref{EM-principaltetrad}
to the restricted \ON tetrad $(u^a,\vs^a,-[\vs,\vb]^a,\vb^a)$, where 
\begin{equation}\label{eq:[vsvb]}
[\vs,\vb]^a\equiv \e^a{}_{bcd}\vs^b\vb^cu^d=-[\sinh(\pp)\us^a+\cosh(\pp)\ub^a]. 
\end{equation}
If $u^a\neq \us^a$ then \eqref{eq:transfo-nuhat} applied to $u_1^a=\us^a,\,u_2^a=u^a$ and $\vhat{u}{\us}^a=\ub^a$ gives 
\begin{equation}\label{eq:nuhat-[vsvb]}
\vhat{\us}{u}^a=[\vs,\vb]^a\,,
\end{equation}
and when combined with \eqref{eq:ebbb-vsavba} this implies \eqref{eq:boost-inv}. If $u^a=\us^a$ we  have $\vrel{\us}{u}^a=0,\,\pp=0$ and $\bb^a=0$ (where \eqref{eq:ebbb-vsavba} confirms \eqref{eq:principal-conds}) and may formally take any unit vector in $u^\bot$ for $\vhat{\us}{u}^a$. This proves the lemma. An alternative 
proof goes by 
combining the first part of \eqref{eq:PPbot-def} with \eqref{eq:va-def-equiv}, giving
\begin{align} 
\label{eq:Pabub-complex}&
\cosh(\pp)\us^a=
P_{\;\;b}^{a}u^{b}=\U^a_{\;\;c}\U^c_{\;\;b}u^c=\U_{~b}^{a}\eb^{b}=\eb^{b}\eb_{b}u^{a}+[\eb,\bb]^a,
\end{align}
which immediately implies \eqref{eq:boost-rap}-\eqref{eq:boost-inv}.\qed
\vspace{.2cm}

We now apply the above to non-null EM fields. 
By its definition \eqref{eq:spa-def} the Poynting vector associated to $u^a$ is a multiple of the vector product of $\E^a$ and $\B^a$ (or of $\f^a$ and $\fc^a$) and thus of 
$\eb^a$ and $\bb^a$ 
by \eqref{eq:EaBa-EM} (or by \eqref{eq:Qa-non-null} and $[\v,\vc]^a=2i[\eb,\bb]^a$); using also \eqref{eq:def-che} 
we obtain 
\begin{align}
\label{eq:spa-EM-b}
\sp^{a}\equiv \frac{1}{4\pi}[\E,\B]^a=\frac{1}{8\pi i}[\f,\fc]^a=\frac{|\IEM|}{8\pi i}[\v,\vc]^a=2\che[\eb,\bb]^a.
\end{align}
Lemma \ref{lem1} now confirms the EM Wheeler result and implies that $\umw^a=\us^a$ and $\ww=\pp$ in \eqref{eq:MisWhe57}: For $u^a\notin\Sigma$ the 2-plane $\langle u^a,\sp^a\rangle$ equals $\langle u^a,\us^a\rangle$ and intersects $\SS$ in $\langle\umw^a\rangle=\langle\us^a\rangle$;
by \eqref{eq:boost-inv}, \eqref{eq:[vsvb]}, \eqref{eq:nuhat-[vsvb]} one has
\begin{align}
\label{eq:spa-EM-c}
\sp^{a}&=2\che\cosh^2(\pp)\vrel{\us}{u}^a\\
&=\che\sinh(2\pp)\vhat{\us}{u}^a
\label{eq:spa-EM-d}
=-\che\sinh(2\pp)[\sinh(\pp)\us^a+\cosh(\pp)\ub^a].
\end{align}
Analogously, substituting \eqref{eq:Qa-non-null} in 
the final expression of 
\eqref{eq:te-def} and using \eqref{eq:def-che}, \eqref{eq:boost-rap} it follows that 
\begin{equation}\label{eq:te-EM}
\te=\che\v^a\vc_a=\che\cosh(2\pp)=2\che\cosh^2(\pp)-\che=\che+2\che\sinh^2(\pp)\,.
\end{equation}
Comparing \eqref{eq:spa-EM-c} with \eqref{eq:te-EM} gives  
$\vrel{\us}{u}^a$ explicitly in terms of $\sp^a$ and $\te$, as announced, and we obtain

\begin{thm}[Extended EM Wheeler result]\label{thm:EM} Suppose the EM field is non-null at a point $p$, and let $u^a$ be an
observer measuring a Poynting vector $\sp^a$ and energy density {\em $\te$} at $p$. Then there is a principal observer instantaneously traveling relative to $u^{a}$ in the direction of $\sp^{a}$, namely the observer $\us^a$
lying along the orthogonal projection of $u^{a}$ onto the EM principal plane $\SS$, with spatial velocity
	\begin{align}
	\label{eq:vrel-EM}
	&\vrel{\us}{u}^a=\nurel_{\parallel \sp}^a\equiv \frac{\sp^a}{\che(\textrm{{\em $\Aem$}}+1)}=\frac{\sp^a}{\textrm{{\em $\te$}}+\che}\,,
	\qquad \textrm{{\em $\Aem$}}=\frac{\textrm{{\em $\te$}}}{\che}=\cosh(2\pp)\,.
	\end{align}
Hence	
\begin{equation}\label{eq:urel-EM}
\us^a=
\cosh(\pp)[u^a+\nurel_{\parallel \sp}^a],
\qquad
\cosh(\pp)=\sqrt{\frac{\textrm{{\em $\Aem$}}+1}{2}}=\sqrt{1+\frac{\textrm{{\em $\te$}}-\che}{2\che}}\ .
\end{equation}
\end{thm}


\begin{rem}\label{rem:WheelerEM-altern}
An alternative proof of theorem \ref{thm:EM}, specific to the non-null EM case, goes by substituting \eqref{eq:Fcal-non-null}
into \eqref{eq:Tab-EM}, and invoking \eqref{eq:def-che} and $P_{\bot b}^{a}=\delta^a_b-P_{\;\;b}^{a}$, which gives (cf.\ \cite{FerSae12,Synge})
\begin{equation}\label{eq:Tab-non-null-EM}
T_{\;\;b}^{a}=\frac{1}{8\pi}\Fcal^{ac}\Fcalc_{bc}=\che(P_{\bot b}^{a}-P_{\;\;b}^{a})=\che(\delta_{b}^{a}-2P_{\;\;b}^{a})\,.
\end{equation}
Note that the 2+2 structure $\SS\oplus\SS^{\bot}$ of tangent space
embodied in the basic field $\Fcal_{ab}$ 
has carried over to the
endomorphism $T_{\;\;b}^{a}$ of $T_{p}M$, which is 
diagonalizable and has two double eigenvalues 
$-\che,\,\che$ with
respective 2d eigenspaces $\SS,\,\SS^{\bot}$ and
projectors $P_{\;\;b}^{a},\,P_{\bot b}^{a}$.
Contraction of \eqref{eq:Tab-non-null-EM} with $u^b$ and use of \eqref{eq:sa-def} 
directly leads to \eqref{eq:vrel-EM}-\eqref{eq:urel-EM}.
\end{rem}

\begin{rem}\label{rem:obsgeom-EM} 

Making the identification \eqref{EM-principaltetrad} in
\eqref{eq:e0e1-tetrad}-\eqref{eq:e2e3-tetrad} and \eqref{eq:decomp-2+2} one gets a full decomposition of any observer $u^a$ in an arbitrary but fixed principal tetrad $(\et_{0}^{a},\et_{i}^{a})$:
\begin{equation}\label{eq:decomp-further}
u^a=\cosh(\pp)[\underbrace{\cosh(\aa)\et_0^a+\sinh(\aa)\et_1^a}_{\us^a=\EEt_0^a(\aa)}]+\sinh(\pp)\underbrace{[\cos(\cc)\et_2^{a}+\sin(\cc)\et_3^{a}]}_{\ub^a=\EEt_2^a(\cc)}.
\end{equation}
This manifests the intrinsic structure of $\Up$ ({\em observer geometry}) induced by the 2+2 structure $\SS\oplus\SSb$ of tangent space, 
where an observer $u^a$ is fully determined by the projected principal observer $\us^a=\EEt_0^a(\aa)\in\SS$ (parameter $\aa\in\R$), the relative rapidity parameter $\pp=\wrel{u}{\us}\geq 0$ and, when $u^a\notin\SS\lra\pp> 0$, by the direction $\ub^a=\vhat{u}{\us}=\EEt_2^a(\cc)\in\SSb$ in which it moves relative to $\us^a$ (parameter $\cc\in[0,2\pi[$).
By \eqref{eq:spa-EM-d}, \eqref{eq:te-EM}, \eqref{eq:def-che} and \eqref{EB-norm 1}-\eqref{eq:def-cheEH} the EM energy density and the norms of the Poynting, electric and magnetic vectors only depend on $\pp$ (and not on $\aa$ nor $\cc$): 
\begin{align}\label{eq:pa-norm}
&\sqrt{\sp^a\sp_a}=\che\sinh(2\pp),
\quad \te=\che\cosh(2\pp),\\
\label{eq:EB-norm2}
&E^a E_a-\cheE^2=B^a B_a-\cheH^2=|\IEM|\,\sinh^2(\pp).
\end{align} 
From this perspective the Wheeler equation \eqref{eq:MisWhe57} with $\ww=\pp$ just expresses the ratio of the two functions \eqref{eq:pa-norm} of the single variable $2\pp$. For large values of $\pp$ all scalars \eqref{eq:pa-norm}-\eqref{eq:EB-norm2}
quadratically increase as functions of the Lorentz factor $\cosh(\pp)$ between $u^a$ and $\us^a$, where 
$\sqrt{\sp^a\sp_a}/\te\ra 1$ and $E^aE_a/B^bB_b\ra 1$ in the limit  $\pp\ra\infty$ where the observer's velocity approaches the speed of light. The value $\pp=0$ corresponds to principal observers $u^a=\us^a$, for which the minimum values are acquired (see the definition and characterizations (iv) and (v) of principal observers).

Within the observer geometry the principal observers are written as $\EEt_0^a(\aa),\,\aa\in \R$
and form the curve $\Up\cap\Sigma$. This curve 
naturally foliates $\Up$, each 2d leaf ${\cal L}(\aa)$ corresponding to a fixed $\aa\in\R$ (so consisting of all observers $u^a$ who have $\EEt_0^a(\aa)$ as their projected principal observer $\us^a$) and parametrized by $\pp$ and $\cc$. Take any $u^a\in{\cal L}(\aa)$ and consider the corresponding vector $\eb^a\neq 0$. From \eqref{eq:Qa-non-null}-\eqref{eq:EaBa-EM}, \eqref{eq:def-vsa-vba}, \eqref{eq:ebbb-vsavba} and characterizations (i) and (ii)' of EM principal observers, we see that $\eb^a$ is parallel to $\vs^a=\EEt_1^a(\aa)=\sinh(\aa)\et_0^a+\cosh(\aa)\et_1^a$ (its norm $\cosh(\pp)$ being moreover independent of $\cc$) and thus gives the direction of the aligned fields $\E_\Sigma^a=\Re(\sqrt{\IEM})\vs^a$ and $\B_\Sigma^a=-\Im(\sqrt{\IEM})\vs^a$ corresponding to $\us^a=\EEt_0^a(\aa)$; hence it lies along the unique spatial direction of $\Sigma$ orthogonal to $u^a$ or $\us^a$.  
If moreover $u^a\neq \us^a$ then, by the extended EM Wheeler result, the 2-plane $\langle u^a,\sp^a\rangle$ intersects ${\cal L}(\aa)$ in a curve 
containing $\us^a$, which separates it in two parts corresponding to $\cc=\cc(u^a)$ and $\cc\pm\pi$ and consisting of the non-principal observers who instantaneously travel relative to $\us^a$ in the same ($\ub^a=\EEt^a_2(\cc)$) or opposite ($-\ub^a$) direction as $u^a$ or, equivalently, who have Poynting vectors lying in $\langle u^a,\sp^a\rangle$.
This last equivalence is due to \eqref{eq:spa-EM-d} and \eqref{eq:decomp-further}, which moreover imply that {\em the relation between non-principal observers $u^a\in\Up\setminus\Sigma$ and measured Poynting vectors is one-to-one}: the projections of a 
given $0\neq\sp^a\in T_pM$ along and orthogonal to $\SS$ normalize to $-\us^a$ and $-\ub^a$, which yields $\aa$ and $\cc$ in the decomposition \eqref{eq:decomp-further}, while by \eqref{eq:pa-norm} the norm of $\sp^a$ yields $\pp$; since $(u^a,e^a,b^a,\sp^a)$ is a basis of $T_p M$ we also have  
\begin{equation}
\tilde{u}^a\in\langle u^a,\sp^a\rangle\lra \tilde{u}^a\v_a=0\lra \tilde{u}^a\Xcal_{ab}u^b=0\lra \tilde{u}^a\Fcal_{ab}u^b=0,
\end{equation} which for $\tilde{u}^a\notin\Sigma$ is equivalent to $u^a\in\langle \tilde{u}^a,\tilde{\sp}^a\rangle$ and else to $\tilde{u}^a=\us^a$. 
On the other hand, taking $\aa$ and $\pp>0$ fixed gives a circle of observers; gluing these circles by letting $\aa$ run through $\R$ produces the 2d level surface of observers measuring the same energy density $\te=\te(\pp)$ (or the same norm of the Poynting or electric or magnetic vector). 
\end{rem}
\vspace{.2cm}

Consider an observer $u^a$ and the $\Sigma$-projected principal observer $\us^a$. By \eqref{eq:def-vsa-vba} and \eqref{eq:ebbb-vsavba} all principal observers, i.e., {the vectors $\ubig^a\in\Up\cap\SS$
are obtained by arbitrarily boosting $\us^a$ 
along $\langle\vs^a\rangle=\langle\eb^a\rangle$, so 
\begin{equation}\label{eq:up-us-rel}
\ubig^a=\cosh(\varphi)\us^a+\sinh(\varphi)\vs^a\quad\lra\quad
\vrel{\ubig}{\us}^a=\tanh(\varphi)\vs^a=\frac{\tanh(\varphi)}{\cosh(\pp)}\eb^a,\qquad\varphi\in\R\,.
\end{equation}
In passing, note that the (unit) vector $\eb'^a$ along the aligned fields $\E'^a=\Re(\sqrt{\IEM})\eb'^a$ and $\B'^a=-\Im(\sqrt{\IEM})\eb'^a$ as measured by $\ubig^a$ is given by [see the primed versions of \eqref{eq:va-def-equiv} and \eqref{eq:EaBa-EM}, with $\bb'^a=0$, and use \eqref{eq:Ucal-form} with $e_0^a=\us^a$ and $e_1^a=\vs^a$]
\begin{equation*}
\eb'^a=X^{ab}u'_b=\cosh(\varphi)\vs^a+\sinh(\varphi)\us^a. 
\end{equation*}
By \eqref{eq:up-us-rel} the vector $\vrel{\ubig}{\us}^a$ is orthogonal to $u^a$, such that the velocities of the principal observers relative to $u^a$ are  
$\vrel{\ubig}{u}^a=\vrel{\us}{u}^a+\vrel{\ubig}{\us}^a/\cosh(\pp)$; by \eqref{eq:vrel-EM}-\eqref{eq:urel-EM} and \eqref{eq:up-us-rel} they
are thus the orthogonal sum of a fixed component along $\sp^a$, and a `free' component along $\langle\eb^a\rangle$  parametrized by $\varphi$: 
\begin{align}\label{eq:princ-all-vec-EM}
&\nurel^a\equiv\vrel{\ubig}{u}^a=\nurel_{\parallel \sp}^a+\nurel^a_{\parallel \eb},
\qquad\nurel^a_{\parallel \eb}\;=\;
\tanh(\varphi)\frac{2\eb^a}{\Aem+1},\;\varphi\in\R\;=\;\textrm{`free' vector along $\langle\eb^a\rangle$}
\,,
\end{align}
with  associated Lorentz factor\footnote{The Lorentz factor is thus minimal for $\ubig^a=\us^a\lra\varphi=0$, cf.\ the appendix of \cite{Whe77}; note that Eq.~(36) in \cite{Whe77} is equivalent to $\cosh(\beta)=\cosh(\mu)\cosh(\alpha)$, where $(\beta,\mu,\alpha)$ corresponds to our $(\psi,\varphi,\pp)$.} $\cosh(\psi)=\cosh(\varphi)\cosh(\pp)=\cosh(\varphi)\sqrt{(\Aem+1)/2}$; when $u^a$ is itself principal ($\sp^a=0$) both $\E^a$ and $\B^a$ are aligned with $\eb^{a}$.
This is illustrated in Fig.\ \ref{fig:EMDiagram}. 
\begin{figure}
	\includegraphics[width=0.98\columnwidth]{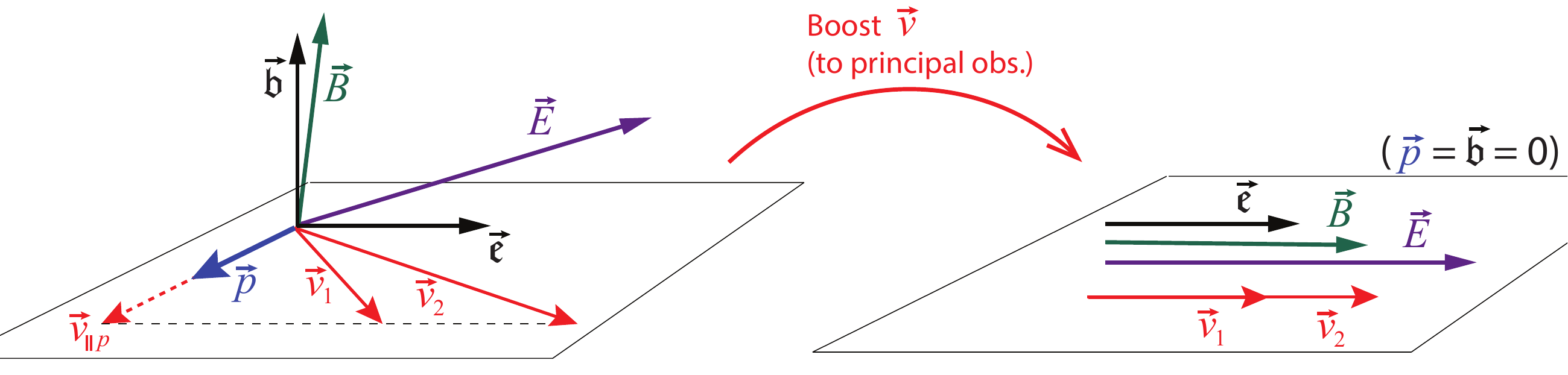}
	\protect\caption{\label{fig:EMDiagram} 
		{Velocities of the principal observers of the electromagnetic field, represented in the rest space of a generic observer  $u^a$ (measuring fields $\E^a$,$\B^a$, $\eb^a$, $\bb^a$, and a Poynting vector $\sp^a$). They move with a relative velocity $\nurel^a$ that has a fixed component $\nurel_{\parallel \sp}^a\equiv\sp^{a}/(\te+\che)$ along $\sp^a$, and a `free' component $\nurel_{\parallel \eb}^{a}$ along the orthogonal vector $\eb^{a}$. When $u^a$ is principal (right panel), the electric and magnetic fields $\E^a$ and $\B^a$ are aligned along $\eb^{a}$, and all other principal observers
			are obtained by boosting along $\langle\eb^{a}\rangle$ with arbitrary rapidity.}}
\end{figure}
The vector $\eb^a$ can be calculated by using \eqref{eq:Qa-non-null} and applying  \eqref{eq:root-z} to $z=\IEM$, giving
\begin{equation}\label{eq:calc-eb-EM}
\eb^a=\Re\left(\frac{\f^a}{\sqrt{\IEM}}\right)
=\frac{1}{|\IEM|}\left(\sqrt{\frac{|\IEM|+\Re(\IEM)}{2}}\E^a-\sgn(\Im(\IEM))\sqrt{\frac{|\IEM|-\Re(\IEM)}{2}}\B^a\right).
\end{equation}
Thus, an observer $u^a$ passing through a point $p$ and measuring 
$\f^a=\E^a-i\B^a$ may check whether the EM field is non-null at $p$, and in this case determine all principal observers directly from $\f^a$, by the following\\

{\em Algorithm for obtaining the principal observers from measured electric and magnetic fields:} 
\begin{enumerate}
	\item Compute $\IEM=\f^a\f_a$; if $\IEM\neq 0$ the EM field is non-null at $p$, and then go to step 2. 
	\item Compute $\sp^a,\,\te$ and $\che$ from \eqref{eq:spa-def},\,\eqref{eq:te-def} and \eqref{eq:def-che}; then \eqref{eq:vrel-EM} gives the velocity relative to $u^a$ of the projected principal observer $\us^a$ realizing the EM Wheeler result. 
	\item Compute $\eb^a$ from \eqref{eq:calc-eb-EM}; then  \eqref{eq:vrel-EM} and \eqref{eq:princ-all-vec-EM} give the velocities relative to $u^a$ of all principal observers $\ubig^a$.
\end{enumerate}

\begin{rem}\label{rem:PE/PM} Let $\IEM\neq 0$ and define $\sigma\in[0,\pi[$ by $e^{2i\sigma}=\IEM/|\IEM|$. By \eqref{eq:EaBa-EM} and their characterization (ii)' [and see (i)] the principal observers are precisely those for which $\cos(\sigma)\B^a+\sin(\sigma)\E^a=0$. 
By \eqref{eq:def-IEM} the condition $\Im(\IEM)=0$ means that $\E^{a}$ and $\B^{a}$ are orthogonal
for any observer ($\E^{a}\B_{a}=0$); this gives \emph{purely electric} (PE) and \emph{purely
magnetic} (PM) electromagnetic fields, characterized by $\sigma=0\lra\IEM=|\IEM|>0\lra\E^{a}\E_{a}>\B^{a}\B_{a}$
and $\sigma=\pi/2\lra\IEM=-|\IEM|<0\lra\E^{a}\E_{a}<\B^{a}\B_{a}$, respectively, and the principal observers are precisely those 
for which the magnetic (electric) field vanishes. 
In \eqref{eq:vrel-EM}-\eqref{eq:calc-eb-EM} we then have 
\begin{eqnarray}
\label{eq:PEchar-EM}
&\textrm{PE}:\qquad &\nurel_{\parallel \sp}^a=[\E,\B]^a/\E^b\E_b,\quad \eb^a=\E^a/\sqrt{\E^b\E_b-\B^b\B_b},\\
\label{eq:PMchar-EM}
&\textrm{PM}:\qquad &\nurel_{\parallel \sp}^a=[\E,\B]^a/\B^b\B_b,\quad \eb^a=\B^a/\sqrt{\B^b\B_b-\E^b\E_b},
\end{eqnarray}
in agreement with \cite{CosWylNat16}. 
Note that a real bivector that has a unitary self-dual bivector is 
`purely electric', with $\E^a=\eb^a,\,\B^a=\bb^a$ [compare 
\eqref{eq:Pabub-complex} with \eqref{eq:PEchar-EM}]; in general,
$\tilde{F}_{ab}=(\cos(\sigma)F_{ab}-\sin(\sigma)\st{F}_{ab})/\sqrt{|\IEM|}$ is such a bivector that is canonically obtained from a non-null EM field $F_{ab}$ by {\em duality rotation over $\sigma$} and normalization, and the associated electric field $\tilde{E}^a=\tilde{F}^{ab}u_b$ precisely equals the vector \eqref{eq:calc-eb-EM}.
\end{rem}

\section{The gravitational case}

\label{sec:Petrov-diag}

\subsection{Weyl principal observers}

\label{subsec:grav-basicdefs}

Consider the Weyl tensor $C_{abcd}$ at a spacetime point $p$.
One has $\tfrac{1}{2}\e_{abef}C^{ef}{}_{cd}=\tfrac{1}{2}C_{ab}{}^{ef}\e_{cdef}\equiv\st{C}_{abcd}$
and defines the self-dual Weyl tensor by $\Ccal_{abcd}\equiv C_{abcd}-i\st{C}_{abcd}$.
Given any observer $u^{a}$ the {\em electric and magnetic parts} $\E_{ab}\equiv C_{acbd}u^{c}u^{d}$ 
and $\H_{ab}\equiv \st{C}_{acbd}u^{c}u^{d}$ of the Weyl tensor relative to $u^{a}$
are spatial, traceless and symmetric, and assembled into the complex tensor
\begin{align}\label{def-Qab}
\Q_{ab}\equiv\Ccal_{acbd}u^{c}u^{d}=C_{acbd}u^cu^d-i\st{C}_{acbd}u^cu^d=\E_{ab}-i\H_{ab}.
\end{align}
Conversely, one has~\cite{SKMHH}
\begin{align}
\Ccal_{ab}{}^{cd}=4\left(u_{[\a}u^{[c}+\I_{[a}{}^{[c}\right)Q_{b]}{}^{d]}-2i\e_{abef}u^fQ^{e[c}u^{d]}-2i\e^{cdef}u_fQ_{e[a}u_{b]}\,. 
\label{eq:Cdag-reconstr}
\end{align}
This implies that the induced traceless endomorphism 
$\Qop=\Eop-i\Hop:v^a\mapsto \Q^a_{\;\;b}v^b$ of $\Cup$ 
is associated to the traceless endomorphism $\Cop:\Xcal_{ab}\mapsto-\frac{1}{4}{\Ccal}_{ab}{}^{cd}\Xcal_{cd}$
of $\Sp$ under the identification \eqref{eq:def-psiuchiu}
of $\Sp$ and $\Cup$, in the sense that if the isomorphism $\Xcal_{ab}\mapsto\Xcal^{ab}u_b$ is denoted by $\psiu$ then  
\begin{equation}
\Qop=\psiu\circ\Cop\circ\psiu^{-1}.
\label{eq:Cop-Qop-equiv}
\end{equation}
Thus, for any $u^a$, $\Qop$ has the same eigenvalues and algebraic type as $\Cop$. 
The eigenvalues  $\l_{k}\,(k=1,2,3)$ sum to zero and solve the joint characteristic equation $x^{3}-\tfrac{1}{2}I\,x-\tfrac{1}{3}J=0$, where $I$ and $J$ are the complex quadratic and cubic Weyl invariants 
\begin{eqnarray}
 I&\equiv&\tr(\Cop^{2})=\tfrac{1}{16}{\Ccal}_{ab}{}^{cd}{\Ccal}_{cd}{}^{ab}=\tfrac{1}{8}{C}_{ab}{}^{cd}(C_{cd}{}^{ab}-i\st{C}_{cd}{}^{ab})=\sum_{k=1}^3\l_{k}^{2}=-2\sum_{(ijk)}\l_j\l_k\label{eq:I-def-int}\\
 &=&\tr(\Qop^{2})=\Q_{~b}^{a}\Q_{~a}^{b}=\E_{~b}^{a}\E_{~a}^{b}-\H_{~b}^{a}\H_{~a}^{b}-2i\E_{~b}^{a}\H_{~a}^{b},\label{eq:I-def-Q}\\ J&\equiv&\tr(\Cop^{3})=-\tfrac{1}{64}{\Ccal}_{ab}{}^{cd}{\Ccal}_{cd}{}^{ef}{\Ccal}_{ef}{}^{ab}=-\tfrac{1}{16}{C}_{ab}{}^{cd}{C}_{cd}{}^{ef}({C}_{ef}{}^{ab}-i\st{C}_{ef}{}^{ab})=\sum_{k=1}^3\l_{k}^{3}=3\prod_{k=1}^{3}\l_{k}\label{eq:J-def-int}\\
 &=&\tr(\Qop^{3})=\Q_{~b}^{a}\Q_{~c}^{b}\Q_{~a}^{c}=\E_{\ b}^{a}\E_{\ c}^{b}\E_{\ a}^{c}-3\E_{\ b}^{a}\H_{\ c}^{b}\H_{\ a}^{c}+i(\H_{\ b}^{a}\H_{\ c}^{b}\H_{\ a}^{c}-3\E_{\ b}^{a}\E_{\ c}^{b}\H_{\ a}^{c}).\label{eq:J-def-Q}
\end{eqnarray}
There are six algebraic types, known as the Petrov types (see e.g.\ \cite{SKMHH}, and appendix \ref{subsec:Petrov-basic}):
\begin{itemize}
	\item Petrov type I is the `algebraically general' case $I^3\neq 6J^2$, with three simple eigenvalues. 
	\item Petrov types II and D both have $I^3=6J^2\neq 0$ and thus one double eigenvalue $\l\equiv-J/I$ and one simple eigenvalue $-2\l$, but the minimal polynomial is $(x-\l)^2(x+2\l)$ for Petrov type II while $(x-\l)(x+2\l)$ for Petrov type D, so in the latter case $Q^a_{~b}$ (relative to any $u^a$) satisfies
	\begin{equation}\label{typeD-charact}
	(\Q_{~c}^{a}-\l \I_{~c}^{a})(\Q_{~b}^{c}+2\l \I_{~b}^{c})=0,\quad\l\equiv-J/I;
	\end{equation}
	\item Petrov types III, N and O have $I=J=0$ and thus a triple eigenvalue $0$, but the minimal polynomials are $x^3,\,x^2$ and $x$, respectively; Petrov type O is thus the trivial case where $C_{abcd}=0$.
\end{itemize}
As for any endomorphism on a vector space, $\Cop$ and $\Qop$ are diagonalizable precisely when their (joint) minimal polynomial only has linear factors, i.e., when its degree equals the number of different eigenvalues; this is the case for Petrov types I, D and O but not for types II, III and N. 

Akin to the EM case we review the definition, characterizations,
existence and locus of Weyl principal observers. The basic super-energy tensor of the Weyl tensor is the (completely symmetric, tracefree) {\em Bel-Robinson tensor}~\cite{Belnotes,Sen00}~\footnote{The definition \eqref{eq:Bel-Rob-def} matches the one in e.g.\ \cite{FerSae12};
	it has a factor $1/4$ compared to the one in \cite{Sen00}, whereas
	the original Bel-Robinson tensor has a factor $1/2$~\cite{Belnotes}; as a result different conventions for super-energy density and super-Poynting vector have been used in the literature.} 
\begin{equation}\label{eq:Bel-Rob-def}
T_{abcd}  \equiv\tfrac{1}{4}{\Ccal_{a}{}^{e}}_{c}{}^{f}\Ccalc_{bedf}=\tfrac{1}{2}{C_{a}{}^{e}}_{(c}{}^{f}C_{d)fbe}-\tfrac{1}{32}C^{efgh}C_{efgh}\,g_{ab}\,g_{cd}.
\end{equation}
Following \cite{FerSae12,FerSae09} we define the invariants
\begin{equation}\label{def-chg}
\alpha\equiv\tfrac12\sqrt{T^{abcd}T_{abcd}}=\tfrac12|I|,\qquad
\chg\equiv \tfrac14\sum_{i=1}^3|\l_i|^2,\qquad
q\equiv \Im(\l_1\lc_2)=\Im(\l_2\lc_3)=\Im(\l_3\lc_1). 
\end{equation}
As shown in appendix \ref{subsec:Bel-Rob} these are related by the identity
\begin{equation}\label{eq:ids}
4\chg^2=\alpha^2+3 q^2.
\end{equation}
Relative to
an arbitrary observer $u^{a}$ 
the  (spatial) super-Poynting vector $\qp^{a}$, super-energy density
$\tg$ and super-energy flux vector $\q^{a}$ are defined by
\begin{align}
\qp^{a} & \equiv-\I^{ab}T_{bcde}u^{c}u^{d}u^{e}=\tfrac{1}{2}\e^a{}_{bcd}\E^b_{\;\;e}\H^{ec}u^d\equiv\tfrac{1}{2}[\Eop,\Hop]^a=\tfrac{1}{4i}\e^a{}_{bcd}\Q^b_{\;\;e}\Qc^{ec}u^d\equiv\tfrac{1}{4i}[\Qop,\Qcop]^a,\label{eq:qpa-def} \\
\tg & \equiv-\q^{a}u_{a}=T_{abcd}u^{a}u^{b}u^{c}u^{d}=\tfrac{1}{4}(\E^{ab}\E_{ab}+\H^{ab}\H_{ab})=\tfrac{1}{4}\Q^{ab}\Qc{}_{ab}=\tfrac14\tr(\Qcop\Qop),\label{eq:tg-def}\\
\q^{a} & \equiv-T^{abcd}u_{b}u_{c}u_{d}=\tg u^{a}+\qp^{a}.\label{eq:qa-def}
\end{align}
In \cite{FerSae12,FerSae13} 
it was proven that   
\begin{equation}\label{eq:tg2-chg2}
\qp^{a}=0\Ra\tg=\chg,
\qquad\tg^{2}\geq\tg^{2}-\qp^{a}\qp_{a}=-\q^{a}\q_{a}\geq\chg^{2}\geq \tfrac14\alpha^2 ,
\end{equation}
and that $\chg$ is the infimum for $\tg:\Up\ra\Rnn$, called the {\em proper gravitational super-energy density}.

\begin{defn}\label{def:princobs-grav} An observer $u^{a}$ is {\em
		(Weyl) principal} if the relative super-Poynting vector vanishes,
	$\qp^{a}=0$. 
\end{defn}

The following equivalent properties characterize a Weyl principal observer $u^{a}$:\\
\vspace{-.2cm}

{(i) the associated endomorphisms $\Eop$ and $\Hop$ commute}, $[\Eop,\Hop]^a=0$~\cite{Bel62}; 

(ii) $\Qop$ is diagonalizable and admits {\em real} \ON
eigenvectors $\eb_{i}^{a}\in\up$: $\Q^{ab}=\sum_{i=1}^{3}\l_{i}\eb_{i}^{a}\eb_{i}^{b}$~\cite{SKMHH};

(iii) the observer $u^{a}$ satisfies $u^{[a}T^{b]}{}_{cde}u^{c}u^{d}u^{e}\equiv u^{[a}C^{b]fg}{}_{e}C_{cfdg}u^{c}u^{d}u^{e}
=0$;

(iv) the super-energy density $\tg$ attains minimum value, namely $\chg$~\cite{FerSae12,FerSae13}.\\
\vspace{-.2cm}

\noindent 
Characterizations (i) and (iii) follow from \eqref{eq:qpa-def} and \eqref{eq:qa-def},\,\eqref{eq:Bel-Rob-def}. Since $\E_{ab}$ and $\H_{ab}$ are spatial and symmetric, commutation of $\Eop$ and $\Hop$ precisely means that $\Eop$ and $\Hop$, and thus
also $\Qop$, can be simultaneously diagonalized in a {\em real} \ON frame of $\Cup$, such that (i) is equivalent to (ii).  Finally, 
if $\tg$ attains minimum value then it should equal the infimum $\chg$, in which case the inequalities in the middle part of \eqref{eq:tg2-chg2} become equalities, implying $\qp^a=0$; this and the first part of \eqref{eq:tg2-chg2} prove characterization (iv).\footnote{Although implicit in \cite{FerSae13} it was not emphasized there that the infimum values for the functions (`super-energy scalars') appearing in characterizations (iv) and (vi) (see below) are acquired {\em precisely} by principal observers.}

Analogously to the EM case we note on additional characterizations,  involving minimality of scalar functions 
$\Up\ra\Rnn$ and equivalent to (iv). 
First, the real part of \eqref{eq:I-def-int}-\eqref{eq:I-def-Q} gives the invariant 
$\E^{ab}\E_{ab}-\H^{ab}\H_{ab}=\Re(I)=\frac{1}{8}C^{abcd}C_{abcd}$, and combined with \eqref{eq:tg-def} implies
\begin{equation}\label{EH-norm 1}
\E^{ab}\E_{ab}=2\tg+\tfrac12 \Re(I),\qquad \H^{ab}\H_{ab}=2\tg-\tfrac12 \Re(I). 
\end{equation}
Hence,  defining the invariants 
\begin{equation}\label{eq:def-chgEH}
\chgE\equiv {2\chg+\tfrac12{\Re(I)}},\qquad
\chgH\equiv {2\chg-\tfrac12{\Re(I)}}
\end{equation}
we obtain the following, new characterizations of a Weyl principal observer:\\
\vspace{-.2cm}

(v-a) 
${\E^{ab}\E_{ab}}$ 
attains  minimum value, namely $\chgE$;

(v-b) 
${\H^{ab}\H_{ab}}$ 
attains 
minimum value, namely $\chgH$.\\
\vspace{-.2cm}

\noindent 
Second, the (completely symmetric, spatial) tensors 
	$$
	t_{ab}\equiv h_a{}^ch_b{}^dT_{cdef}u^eu^f,\quad {\cal Q}_{abc}\equiv-h_a{}^ch_b{}^dh_c{}^eT_{cdef}u^f,\quad t_{abcd}\equiv h_a{}^ch_b{}^dh_c{}^eh_d{}^fT_{cdef}
	$$
also appear in the orthogonal splitting of $T_{abcd}$ relative to $u^a$,
$
T_{abcd}=\tg u_au_bu_cu_d+4\qp_{(a}u_bu_cu_{d)}+6t_{(ab}u_cu_{d)}+4{\cal Q}_{(abc}u_{d)}+t_{abcd}
$;
they satisfy $t^a{}_a=\tg,\,{\cal Q}^{ab}{}_{b}=\qp^a$ and $t^{c}{}_{cab}=t_{ab}$, can be expressed in terms of $\E_{ab},\,\H_{ab}$, and have been proven important in several contexts~\cite{AlfonsoSE,BonSen97}.
In \cite{FerSae13} it was shown that
\begin{align*}
&t^{ab}t_{ab}-\left(\tfrac16\alpha^2-\tfrac{1}{3}\chg^2\right)=\tfrac{1}{3}(\tg^2-\chg^2)+\tfrac23\qp^a\qp_a,\\
&{\cal Q}^{abc}{\cal Q}_{abc}-\left(2\chg^2-\tfrac12\alpha^2\right) =\tg^2-\chg^2+(\tg^2-\qp^a\qp_a-\chg^2),\\
& t^{abcd}t_{abcd}-\left(\alpha^2+5\chg^2\right)=\tg^2-\chg^2+4(\tg^2-\qp^a\qp_a-\chg^2)
\end{align*}
are functions $\Up\ra\Rnn$ with infimum $0$, where by \eqref{eq:tg2-chg2} the respective right hand sides are the sums of two such functions. It follows that the functions attain minimum value 0 precisely when $\tg=\chg,\,\qp^a=0$, such that a Weyl principal observer can be also characterized by either one of the following properties:\\
\vspace{-.2cm} 

(vi-a)  ${t^{ab}t_{ab}}$ takes minimum value, namely  $\frac13\chg^2+\frac16\alpha^2=\frac14(\alpha^2+q^2)$;	

(vi-b) ${{\cal Q}^{abc}{\cal Q}_{abc}}$ takes minimum value, namely  $2\chg^2-\frac12\alpha^2=\frac32 q^2$;

(vi-c) ${t^{abcd}t_{abcd}}$ takes minimum value, namely $5\chg^2+\alpha^2=(9\alpha^2+15q^2)/4$.\\
\vspace{-.2cm}	


By characterization (ii) the existence of a principal observer requires diagonalizability
of $\Cop$ and thus excludes the `non-diagonal' Petrov
types II, III and N.
If the Petrov type is O at $p$ ($C_{abcd}=0$) then every observer is principal, trivially. Hence
the non-trivial `diagonal'
Petrov types D and I remain.
Since $\Cop$ is self-adjoint due to $\Ccal_{abcd}=\Ccal_{cdab}$ and thus eigenbivectors corresponding to different eigenvalues are orthogonal,
there exists an
oriented \ON frame $(\Ucal_{ab}^{i})$ of $\Sp$ 
such that (cf.\ \cite{SKMHH}):
\begin{align}
 & \Ccal_{abcd}=\sum_{i=1}^{3}\l_{i}\Ucal_{ab}^{i}\Ucal_{cd}^{i},\qquad\Ucal_{ab}^{i}=[2(e_{0})_{[a}(e_{i})_{b]}]^{\dag}=2(e_{0})_{[a}(e_{i})_{b]}+i2(e_{j})_{[a}(e_{k})_{b]}\,.\label{eq:Cabcd-decomp-gen}
\end{align}
Here $\Ucal_{ab}^{i}$ is a unitary self-dual eigenbivector of $\Cop$
with eigenvalue $\l_{i}$; the restricted \ON tetrad $(e_{0}^{a},e_{i}^{a})$
realizing \eqref{eq:Cabcd-decomp-gen} is biunivocally related to
$(\Ucal_{ab}^{i})$, see appendix \ref{sec:orth-frames}, and is
referred to as a {\em (Weyl) principal tetrad}. Contracting \eqref{eq:Cabcd-decomp-gen}
twice with an arbitrary observer $u^{a}$ we obtain 
\begin{align}
& \Q^{ab}=\sum_{i=1}^{3}\l_{i}\v_{i}^{a}\v_{i}^{b}\quad\lra\quad \Qop=\sum_{i=1}^{3}\l_{i}\vop_i\vop_{i},\qquad\v_{i}^{a}=(\Ucal^{i})^{ab}u_{b}\equiv\eb_{i}^{a}-i\bb_{i}^{a}.\label{eq:Qab-decomp-gen}
\end{align}
The vectors $\v_{i}^{a}$ are eigenvectors of $\Qop$ with eigenvalue $\l_{i}$,
are complex in general, and by \eqref{eq:psiu-isom} form an oriented \ON frame
of $\Cup$:
\begin{align}\label{eq:vivj}
 & \v_{m}^{a}(\v_{n})_{a}=\eb_{m}^{a}(\eb_{n})_{a}-\bb_{m}^{a}(\bb_{n})_{a}=\delta_{mn},\qquad\eb_{m}^{a}(\bb_{n})_{a}+\bb_{m}^{a}(\eb_{n})_{a}=0,\qquad \v_i^a=[\v_j,\v_k]^a\,.
\end{align}
As a consequence one has, for any $y\in\Cup$:
\begin{align}\label{id:viy}
[\v_i,y]^a=(\v_j^by_b)\v_k^a-(\v_k^by_b)\v_j^a\,.
\end{align}  
Comparison of characterization (ii) of principal observers  with \eqref{eq:Qab-decomp-gen} 
tells that the principal observers are precisely those $u^{a}$
for which the vectors $\v_{i}^{a}$ can be taken to be real ($\bb_{i}^{a}=0\lra\v_{i}^{a}=\eb_{i}^{a}$),
which by \eqref{eq:principal-conds} happens precisely when $u^{a}$ belongs to the blades $\Sigma_{i}\equiv\langle e_{0}^{a},e_{i}^{a}\rangle$ of $\U_{ab}^{i}$, i.e., equals the first vector $e_0^a$ of a principal tetrad $(e_0^a,e_i^a)$ used in \eqref{eq:Cabcd-decomp-gen}. Hence, principal
observers do exist, but principal tetrads and thus principal observers may not be unique; Petrov types D and I provide the relevant distinction: 
\begin{itemize}
\item Petrov type D, characterized by one repeated and one simple eigenvalue; e.g.
\begin{equation}\label{eq:limit-D}
\l_{2}=\l_{3}=-\l_1/2\equiv \l\quad\Rightarrow\quad q=0,\quad  \chg
=\tfrac32|\l|^2=\tfrac14|I|\,.
\end{equation}
The endomorphism $\Cop$ ($\Qop$) has an up to reflection unique unitary eigenbivector  $\Ucal_{ab}^{1}\equiv\Ucal_{ab}\equiv\U_{ab}^{\dag}$ (unit eigenvector $\v_1^a\equiv\v^a$)
corresponding to the non-degenerate eigenvalue $\l_{1}=-2\l$;
the blade $\SS_1\equiv\SS$
of $\U_{ab}$ is thus uniquely defined and called the {\em (Weyl) timelike principal plane}, and its two null directions the {\em (Weyl) principal null directions} (PNDs), spanned by null vectors $k^a$ that satisfy $C_{abc[d}k_{f]}k^{b}k^{c}=0$~\cite{SKMHH,Belprinc}.
{The pair} $(\Xcal^2_{ab},\Xcal^3_{ab})$ is only determined up to a complex rotation, and so the same holds for $(\v_2^a,\v_3^a)$ in \eqref{eq:Qab-decomp-gen}. However, we can rewrite 
\eqref{eq:Qab-decomp-gen} as
\begin{align}
 & \Q^a_{~b}=\l (\I^a_{~b}-3\v^{a}\v_{b})\quad\lra\quad \Qop=\l(\Iop-3\vop\vop),\qquad \v^{a}\equiv \eb^a-i\bb^a\,.\label{eq:Qab-decomp-D} 
\end{align}
Hence characterization (ii) of principal
observers is equivalent to

\hspace{0.4cm} (ii)' the non-degenerate unit eigenvector $\v^{a}$ is real ($\bb^{a}=0\lra\v^{a}=\eb^{a}$), i.e., $u^{a}$ belongs to $\SS$.

The principal tetrads are $(\EEt_0^a,\pm \EEt_1^a, \EEt_2^a,\pm \EEt_3^a)$ with $(\EEt_0^a,\EEt_i^a)$ as in \eqref{eq:e0e1-tetrad}-\eqref{eq:e2e3-tetrad}~\cite{FerrSaezP2}. 

\item Petrov type I, characterized by all $\l_i$'s being distinct ($\l_i\neq \l_j$ for $i\neq j$). Once the eigenvalues $\l_{i}$ have been put in a certain order
the oriented frame $(\Ucal_{ab}^{i})$ in \eqref{eq:Cabcd-decomp-gen} is determined up to a simultaneous reflection of two elements, and so the same holds for the triad $(e_{i}^{a})$; hence there are 24 principal tetrads, but we will speak about `the (essentially unique) principal tetrad' $(e_0^a,e_{i}^{a})$. There are four simple Weyl PNDs, spanned by vectors $k^a$ that satisfy $C_{abc[d}k_{f]}k^{b}k^{c}=0\neq C_{abc[d}k_{f]}k^{c}$~\cite{SKMHH,Belprinc}.
Characterization (ii) of principal observers becomes:

\hspace{0.4cm}  
(ii)' the unit eigenvectors $\v_{i}^{a}$
of $\Qop$ are real ($\v_{i}^{a}=\eb_{i}^{a}=e_i^a$), i.e., $u^{a}$ is
the unique observer $e_{0}^{a}$ lying along the joint intersection
of the three timelike blades $\Sigma_{i}$.
\end{itemize}

\begin{rem} In the gravitational case the existence of Weyl principal
observers is equivalent to diagonalizability of the endomorphism
$\Cop$ (or of the endomorphism $T^{ab}{}_{cd}$ on the space of traceless symmetric tensors, see \cite{FerSae09,FerSae10} and appendix \ref{subsec:Bel-Rob}). 
In fact this parallels the EM case: by \eqref{eq:XYcal-id}-\eqref{eq:XcalYcalc-id} and \eqref{eq:Tab-EM} one has 
$\Fcal^{a}{}_{b}\Fcal^{b}{}_{c}=\IEM\d_{c}^{a}$ and $T^a{}_bT^b{}_c=\che^2\d_{c}^{a}$. 
This implies that the endomorphism $\Fcal^{a}{}_b$ of $\C T_{p}M$ is nilpotent and thus non-diagonalizable in the non-trivial null case $F_{ab}\neq 0,\,\IEM=0$, and diagonalizable in the other two cases $F_{ab}=0$ and $\IEM\neq 0$. {\em Hence diagonalizability of $\Fcal^{a}_{\;\;b}$ is equivalent to the existence of EM principal observers}, and  the same statement holds when $\Fcal^{a}_{\;\;b}$ is replaced by $F_{\;\;b}^{a}$ (cf.\ p.\ 182 of \cite{Hall}) or $T^a_{\;\;b}$.  
Contrary to the EM case, diagonalizability
of $\Cop\neq 0$, or equivalently of $\Qop\neq 0$ for any observer $u^a$, cannot be expressed entirely in terms of the invariants $I$ and $J$, since $I^3=6J^2\neq 0$ does not distinguish Petrov type II from D; however, $I^3\neq 6J^2$ distinguishes Petrov type I, which is generic and diagonal, while $I=J=0,\,C\neq 0$ corresponds to the non-diagonal Petrov types III and N. 
\end{rem}


\begin{rem}\label{rem:q=0} Consider the condition $q=0$. It is automatically satisfied in the Petrov type D case, see \eqref{eq:limit-D}, while for Petrov type I 
precisely gives 
the subcase where the invariant $M\equiv I^3/J^2-6$ is real positive or infinite ($J=0$), or equivalently 
where the four Weyl PNDs span
a 3d (instead of a 4d) vector space; see appendix \ref{subsec:Bel-Rob}. 
In general, one has $q=\Im(\l_j\lc_k)=\Im(\l_j/\l_k)$ for any $\l_k\neq 0$, such that $q=0$ corresponds to the ratio of any two non-zero eigenvalues being real, i.e., there
exist real numbers $\l_{k}'$ summing
to zero such that $\l_{k}=e^{i\sigma}\l'_{k}=\cos(\sigma)\l'_{k}+i\sin(\sigma)\l'_{k},\,k=1,2,3$, where $\sigma\in[0,\pi[$ is a {\em duality rotation index} (see \cite{FerSae04} for further discussions, and cf.~remark \ref{rem:PE/PM}). 
Then, characterizations (ii) and (vi-b) of Weyl principal observers become equivalent to~\cite{AlfonsoSE}:\\
\vspace{-.3cm}


(ii)$^*$ $\E_{ab}$ and $\H_{ab}$ are linearly dependent, $\cos(\sigma)\H_{ab}+\sin(\sigma)\E_{ab}=0$;

(vi-b)$^*$ ${\cal Q}_{abc}=0$.\\
\vspace{-.2cm}

\noindent Here (ii)$^*$ is a substitute for characterization (i) of Weyl principal observers, which resembles characterization (i) of EM principal observers even more than the general commutation condition; because of $\qp^a=Q^{ab}{}_b$ characterization (vi-b)$^*$ may be seen as a stricter form of the definition $\qp^a=0$ of Weyl principal observers. Regarding characterizations (iv) and (v), by \eqref{def-chg}-\eqref{eq:ids}, \eqref{eq:def-chgEH} and \eqref{eq:root-z} applied to $z=I$ the minimum values for $\tg$ and $\sqrt{\E^{ab}\E_{ab}},\,\sqrt{\H^{ab}\H_{ab}}$ respectively become $\chg=|I|/4$ and 
$\sqrt{\chgE}=|\Re(I)|,\,\sqrt{\chgH}=|\Im(I)|$, which formally match the corresponding values \eqref{eq:def-che} and \eqref{eq:def-cheEH} of the EM case.
Spacetimes of Petrov type D or of Petrov type I with $q=0$ at each point have been called {\em super-energy non-radiative} gravitational fields~\cite{AlfonsoSE,FerSae12}. In general one has $I/|I|=e^{2i\sigma}$. The subcase $\sigma=0$ ($\sigma=\pi/2$) corresponds to {\em purely electric (purely magnetic)} spacetimes, which by \eqref{eq:I-def-int} are characterized by all eigenvalues $\l_k$ being real (purely imaginary), or by $I=|I|>0$ ($I=-|I|<0$) {\em and $M$ being real non-negative or infinite}; by \eqref{eq:I-def-Q} any observer measures $\E^{ab}\E_{ab}>\H^{ab}\H_{ab}$ ($\E^{ab}\E_{ab}<\H^{ab}\H_{ab}$), and the principal observers are precisely those for which $\H_{ab}$ ($\E_{ab}$) vanishes; cf.~remark \ref{rem:PE/PM}, and see \cite{McIntosh1,CosWylNat16}.  
Large and important classes of purely electric spacetimes exist; 
for instance all static spacetimes, and all 
spacetimes which exhibit spherical, hyperbolic, or planar symmetry, which are automatically
of Petrov type D (or O)~\cite{StewartEllis,Goode,Lozanovski3}. See \cite{HerOrtWyll13,WyllVdB06,Lozanovski07,LozWyll11} for surveys
of the literature on purely electric or magnetic spacetimes. 

\end{rem} 

\begin{rem} In a vacuum spacetime, without sources but with a possible cosmological constant $\Lambda$ ($T_{ab}=0,\,R_{ab}=\Lambda g_{ab}$) the {\em gravitoelectric} and {\em gravitomagnetic tidal tensors} relative to an observer $u^a$ are respectively given by $(\EE)_{ab}\equiv R_{acbd}u^cu^d=\E_{ab}-\Lambda/3\,h_{ab}$ and $(\HH)_{ab}\equiv \star\!{R}_{acbd}u^cu^d=R\!\star_{acbd}u^cu^d=\H_{ab}$, and can be assembled into $(\QQ)_{ab}=(\EE)_{ab}-i(\HH)_{ab}=\Q_{ab}-\Lambda/3 h_{ab}$, where $\E_{ab},\,\H_{ab},\,\Q_{ab}$ are the relative tensors associated to the Weyl tensor defined above. 
	The endomorphism of $\Cup$ associated to $(\QQ)_{ab}$ has eigenvalues $\l_{k,\textrm{R}}=\l_k-\Lambda/3$ and invariants $\II=I+\Lambda^2/3,\,\JJ=J-\Lambda I-\Lambda^3/9$, defined as in \eqref{eq:I-def-Q},\,\eqref{eq:J-def-Q}. The corresponding super-Poynting vector and super-energy density defined as in \eqref{eq:qpa-def},\,\eqref{eq:tg-def} are simply given by $\PP^a=\qp^a$ and $\tgR=\tg+\Lambda^2/12$. For any value of $\Lambda$ it follows that {\em the corresponding Riemann principal observers are the Weyl principal observers}, with characterizations analogous to (i)-(vi).  
	Also, the spacetime is Riemann purely electric if and only if it is Weyl purely electric, and is never Riemann purely magnetic for $\Lambda\neq 0$~\cite{HerOrtWyll13,CosWylNat16}. 
\end{rem}

\subsection{Petrov type D: general vs.\ principal observers}

\label{subsec:grav-typeD}

The non-null EM and Petrov type D gravitational cases both exhibit
a distinguished unitary
self-dual bivector $\Ucal_{ab}=\U_{ab}-i\st{\U}_{ab}$, where the blade
of $\U_{ab}$ (viz.\ the timelike principal plane $\SS$) 
contains the principal observers. Concretely, $\Ucal_{ab}=\Fdag_{ab}/\sqrt{\IEM}$
in the former case, while in the latter
case $\Ucal_{ab}$ generates the eigendirection of $\Cop$
corresponding to the non-degenerate eigenvalue $-2\l$. Given
an observer $u^{a}$, the role of $\v^{a}\equiv\Ucal^{ab}u_{b}\equiv\eb^a-i\bb^a$ is
played by $(\E^a-i\B^a)/\sqrt{\IEM}$ and a unit eigenvector of $\Qop$
with non-degenerate eigenvalue $-2\l$, respectively. 

Here we strengthen the analogy by 
transferring the EM Wheeler result to the Petrov D gravitational case.
Take any observer $u^a$, and consider the final formula in \eqref{eq:qpa-def} for the associated super-Poynting vector. On comparing to \eqref{eq:spa-def} there is an important conceptual difference between the definitions of a Poynting and super-Poynting vector: the former is a {\em spatial vector product} while the latter is dual to a {\em commutator} of {\em endomorphisms corresponding to spatial 2-tensors}. For Petrov type D, however, we have the special form \eqref{eq:Qab-decomp-D} and its complex conjugate for the relevant endomorphisms $\Qop$ and $\Qcop$; since the identity map $\Iop$ of $\Cup$ commutes with any other endomorphism of $\Cup$, the dual to the commutator of $\Qop$ and $\Qcop$ is parallel to the spatial vector product of the respective unit non-degenerate eigenvectors $\v^a$ and $\vc^a$ by the crucial identity  \eqref{eq:id[xxyy]}. Using also \eqref{eq:limit-D} and
$[\v,\vc]^a=2i[\eb,\bb]^a$ we obtain
\begin{align}
\label{eq:qpa-D-1}
\qp^a&=\frac{1}{4i}[\Qop,\Qcop]^a=\frac{1}{4i}[\l(\Iop-3\vop\vop),\lc(\Iop-3\vcop\vcop)]^a
=\frac{9|\l|^2}{4i}\v^b\vc_b[\v,\vc]^a
=3\chg\v^b\vc_b [\eb,\bb]^a\,.
\end{align} 
As in the non-null EM case, we conclude by lemma \ref{lem1} that {\em the plane $\langle u^a,\qp^a\rangle$ intersects $\SS$ in $\langle \us^a\rangle$, where $\us^a$ is the principal observer along the projection of $u^a$ onto $\SS$}; moreover, we have by \eqref{eq:boost-rap}-\eqref{eq:boost-inv}:
\begin{eqnarray}
\label{eq:qpa-D-2}
\qp^a&=&3\chg\cosh(2\pp)\cosh^2(\pp)\vrel{\us}{u}^a=\frac{3\chg}{4}\sinh(4\pp)\vhat{\us}{u}^a\\
\label{eq:qpa-D-3}
&=&-\frac{3\chg}{4}\sinh(4\pp)[\sinh(\pp)\us^a+\cosh(\pp)\ub^a].
\end{eqnarray}
Notice the resemblance with \eqref{eq:spa-EM-b}. 
Analogously, the final formula in \eqref{eq:tg-def} and  $\tr(\Iop)=3,\,\tr(\vop\vop)=\tr(\vcop\vcop)=1,\,\tr(\vop\vop\vcop\vcop)=(\v^a\vc_a)^2$ produce
\begin{equation}
\label{eq:tg-D}
\tg=\frac{1}{4}\tr(\Qop\Qcop)=\frac{3|\l|^2}{4}\left[3(\v^a\vc_a)^2-1\right]=\chg+\frac{3\chg}{2}\,\sinh^2(2\pp)=\frac{\chg}{4}+\frac{3\chg}{4}\cosh(4\pp).
\end{equation} 
Note that $\tg\geq\chg$ and $\tg=\chg\Leftrightarrow \psi_\Sigma=0\Leftrightarrow \qp^a=0$, cf.\ \eqref{eq:tg2-chg2} and characterization (iv) of Weyl principal observers. 
Comparing to \eqref{eq:qpa-D-2} we arrive at our main 
\begin{thm}[Gravitational Wheeler analogue for Petrov type D]\label{thm:D} Suppose the Weyl tensor is of Petrov type
	D at a point $p$, and let $u^a$ be a non-principal observer measuring a super-Poynting vector $\qp^a$ and super-energy density $\tg$ at $p$. Then there is a principal observer instantaneously traveling relative to $u^{a}$ in the direction of $\qp^{a}$, namely the observer $\us^a$ lying along the orthogonal projection of $u^{a}$ onto the Weyl
	principal plane $\SS$, with spatial velocity
	\begin{align}
		\label{eq:vrel-D}
	&\vrel{\us}{u}^a=\nurel_{\parallel \qp}^a\equiv\frac{2\qp^a}{3\chg\AD(\AD+1)},\qquad  \AD=
	\sqrt{1+\frac{2(\tg-\chg)}{3\chg}}=\cosh(2\pp)\,.
	\end{align}
Hence 
\begin{equation}
\us^a=\cosh(\pp)[u^a+\vrel{\us}{u}^a],\qquad \cosh(\pp)=\sqrt{\frac{\AD+1}{2}}\ . \label{eq:urel-D}
\end{equation}
\end{thm}


\begin{rem}\label{rem:obsgeom-D} Remark \ref{rem:obsgeom-EM} has a clear analogue in the present context. By \eqref{eq:qpa-D-3} a non-zero super-Poynting vector corresponds to a unique non-principal observer $u^a$; in particular,
\begin{equation}
\sqrt{\qp^a\qp_a}=\frac{3\chg}{4}\sinh(4\pp)=3\chg[2\cosh^2(\pp)-1]\cosh(\pp)\sqrt{\cosh^2(\pp)-1}\,,
\end{equation}	 
and in conjunction with \eqref{eq:tg-D} this gives the following analogue of the Wheeler equation \eqref{eq:MisWhe57}:
\begin{equation}\label{eq:Wheq-D}
\tanh(4\pp)=\sqrt{\qp^a\qp_a}/(\tg-\chg/4)\,.
\end{equation}
This expresses the ratio of two functions of the single variable $4\pp$ increasing quartically (instead of quadratically in the non-null EM case) with the relative Lorentz factor $\cosh(\pp)$ of $u^a$ and $\us^a$; also $\E^{ab}\E_{ab}-\chgE^2=\H^{ab}\H_{ab}-\chgH^2=3\chg\sinh^2(2\pp)$, with the same behaviour. Equation \eqref{eq:decomp-further} remains formally the same and the corresponding observer geometry is analogous. $\Up$ is foliated by 2d leafs ${\cal L}(\aa)$, consisting of the observers $u^a$ with $\us^a=\EEt_0^a(\aa)$, and for given $u^a\in{\cal L}(\aa)$ the vector $\eb^a\propto \vs^a$ is now an eigenvector of the endomorphism $\Qop_\Sigma\equiv \Eop_\Sigma-i\Hop_\Sigma$ associated to $\us^a$ with simple eigenvalue $-2\l$, and thus of $\Eop_\Sigma$ and $\Hop_\Sigma$. Comments on the 2-plane $\langle u^a,\qp^a\rangle$ and the 2d level surface of observers with the same energy density $\tg=\tg(\pp)$ (or norm of the super-Poynting vector) are analogous, {\em mutatis mutandis}.  
\end{rem}

\begin{rem}\label{rem:qpa-eigen-D} When $u^a=\us^a$ is principal the associated endomorphism $\Qop_\Sigma$ admits {\em three real} orthogonal eigendirections [see characterization (ii) of Weyl principal observers] which are common eigendirections of $\Eop_\Sigma$ and $\Hop_\Sigma$.
When $u^a$ is non-principal, it follows from \eqref{eq:Qab-decomp-D}, \eqref{eq:qpa-D-1} and $\v_a[\v,\vc]^a=0$ that {\em $\qp^a\neq 0$ is an eigenvector of the associated endomorphism $\Qop=\Eop-i\Hop$ with degenerate eigenvalue $\l$, and thus of $\Eop$ and $\Hop$ with respective eigenvalues $\Re(\l)$ and $-\Im(\l)$}; moreover, the real eigenvectors of $\Qop$ are precisely the common (real) eigenvectors of $\Eop$ and $\Hop$; if there were two independent such eigenvectors then, since $\Q_{ab}$ and $\E_{ab},\,\H_{ab}$ are symmetric, their vector product would give a third real eigenvector and $u^a$ would be principal by characterization (ii) of Weyl principal observers, a contradiction; hence, {\em when $u^a$ is non-principal, $\langle \qp^a\rangle$ is the only real eigendirection of $\Qop$ and the only common eigendirection of $\Eop$ and $\Hop$}. 
\end{rem}

Similarly to \eqref{eq:princ-all-vec-EM}, the velocities of the principal observers $\ubig^a\in\SS$ relative to an arbitrary observer $u^a$ are the orthogonal sum of a fixed component along $\qp^a$ and a `free' component along $\langle\eb^a\rangle$:
\begin{align}\label{eq:princ-all-vec-D}
&\vrel{\ubig}{u}^a=\nurel_{\parallel \qp}^a+\nurel^a_{\parallel \eb},
\qquad\nurel^a_{\parallel \eb}\;=\;
\tanh(\varphi)\frac{2\eb^a}{\AD+1},\;\varphi\in\R\;=\;\textrm{`free' vector along $\langle\eb^a\rangle$}
\,,
\end{align}
with  associated Lorentz factor $\cosh(\varphi)\cosh(\pp)=\cosh(\varphi)\sqrt{(\AD+1)/2}$. This is illustrated in Fig.~\ref{fig:TypeDDiagram}; notice the resemblance with the EM diagram in Fig.\ \ref{fig:EMDiagram}. 
\begin{figure}
	
	\includegraphics[width=0.95\columnwidth]{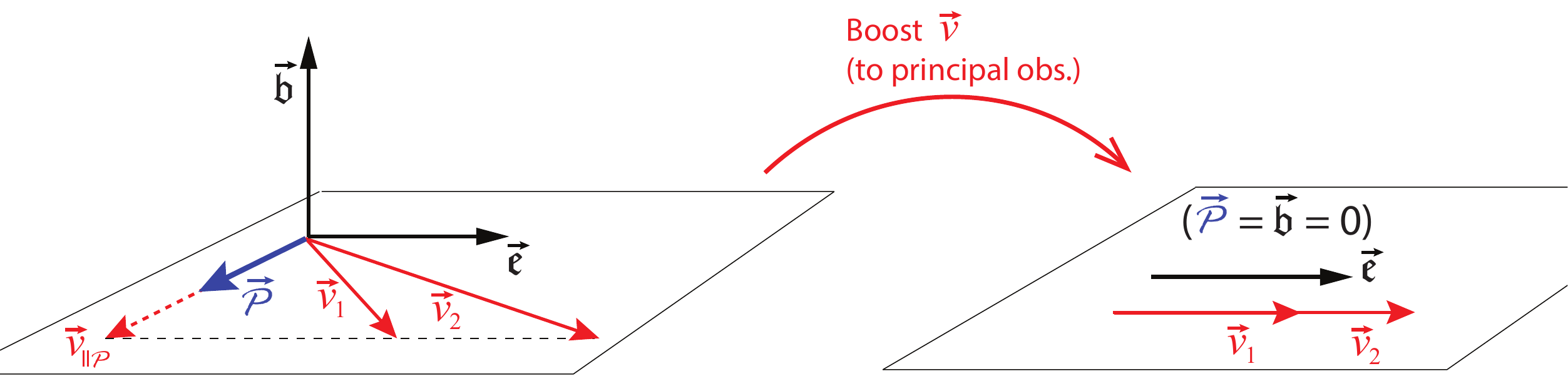}
	\protect\caption{\label{fig:TypeDDiagram}{Velocities of the Weyl principal observers, represented in the rest space of a generic observer observer $u^a$ (measuring $\eb^a$, $\bb^a$, $\Eop$, $\Hop$, and a super-Poynting vector $\qp^a$). They move with a relative velocity $\nurel^a$ that has a fixed component $\nurel_{\parallel \qp}^a$ along $\qp^{a}$, given in \eqref{eq:vrel-D}, and a `free' component $\nurel_{\parallel \eb}^{a}$ along $\eb^{a}$. When $u^a$ is principal (right panel), the non-degenerate eigendirection of $\Eop$ and $\Hop$ is  $\langle\eb^a\rangle$, and all other principal observers are obtained by boosting along this direction with arbitrary rapidity.} }
\end{figure}
To calculate $\eb^a$ directly from the electric and magnetic parts of the Weyl tensor as measured by $u^a$, one has to construct a complex eigenvector $\wx^a$ of $\Qop$ with eigenvalue $-2\l=2J/I$, normalize it to a unit vector $\v^a$, and take the real part. To construct $\wx^a$ one can apply the standard technique of solving the linear system of equations 
$(\Q_{m'i'}+2\l\delta_{m'i'})\wx^{i'}=0$, where $(e_{i'}^{a})$ is an arbitrary \ON triad of $\up$ and $\wx^a=\wx^{i'}e_{i'}^{a}$. Alternatively, one can take any $y^{a}\in\up$ that is {\em not} an eigenvector of $\Qop$ with eigenvalue $\l$ (i.e., any real vector not orthogonal to $\eb^a=\vs^a$ when $u^a=\us^a$ is principal, and any real vector not proportional to $\qp^a$ when $u^a$ is non-principal, see remark \ref{rem:qpa-eigen-D}, where for an \ON triad of $\up$ at least one of the vectors $e_{i'}^{a}$ may be taken as $y^a$) and construct $\wx^a$ from 
\begin{equation}\label{eq:comp-wx}
\wx^{a}=(\Q_{\;\;b}^{a}-\l \I^a_{~b})y^{b}\qquad [\wx^a=(\Q^{m'}_{\;\;i'}-\l\delta^{m'}_{i'})e_{m'}^a],
\end{equation}
where the result for $y^a=e_{i'}^{a}$ is indicated between square brackets. Applying \eqref{eq:root-z}  to $z=\wx^b\wx_b$ we obtain
\begin{equation}\label{eq:calc-eb-D}
\eb^a=\Re\left(\frac{\wx^a}{\sqrt{z}}\right)
=\sqrt{\frac{|z|+\Re(z)}{2|z|^2}}\,\Re(\wx^a)+\sgn(\Im(z))\sqrt{\frac{|z|-\Re(z)}{2|z|^2}}\,\Im(\wx^a),\quad z\equiv\wx^b\wx_b.
\end{equation}
Thus, an observer $u^a$ passing through a point $p$ and measuring $\Q^{ab}=(\E^{m'n'}-i\H^{m'n'})e_{m'}^ae_{n'}^b$ may check whether the gravitational field is of Petrov type D at $p$, and in this case determine all principal observers directly from $\Q^{ab}$, by the following\\

{\em Algorithm for obtaining the principal observers of a Petrov type D Weyl tensor, directly from electric and magnetic parts of the Weyl tensor measured by an arbitrary observer:} 
\begin{enumerate}
	\item Compute $I$ and $J$ from \eqref{eq:I-def-Q} and \eqref{eq:J-def-Q}; if $I\neq 0$ and $(\Q_{\;\;c}^{a}-\l \I_{~c}^{a})(\Q_{\;\;b}^{c}+2\l \I_{~b}^{c})=0,\,\l\equiv-J/I$ the gravitational field is of Petrov type D at $p$ (see appendix \ref{subsec:Petrov-basic}), and then go to step 2. 
	\item Compute $\qp^a,\,\tg,\,\chg$ from  \eqref{eq:qpa-def}, \eqref{eq:tg-def}, \eqref{eq:limit-D};  then \eqref{eq:vrel-D} gives the velocity relative to $u^a$ of the projected principal observer $\us^a$ realizing the gravitational Wheeler analogue. 
	\item Compute $\eb^a$ from \eqref{eq:comp-wx}-\eqref{eq:calc-eb-D}; then \eqref{eq:vrel-D} and \eqref{eq:princ-all-vec-D} give the relative velocities of all principal observers $\ubig^a$.
\end{enumerate}


\subsubsection{Doubly aligned non-null Einstein-Maxwell fields}\label{subsub:doubly}

Gravitational fields that have an energy-momentum tensor of the form \eqref{eq:Tab-non-null-EM} where $\Fcal_{ab}=\Fdag_{ab}$ is non-null and satisfies Maxwell's equations $\nabla^{b}\Fcal_{ab}=0$
are said to be of {\em non-null Einstein-Maxwell} type;
such fields thus exhibit an EM principal plane $\SF$ and a pair of EM PNDs at each point. 
Wherever the Weyl tensor $C_{abcd}$ is moreover of Petrov type D there is also a Weyl principal plane $\SW$ and a pair of Weyl PNDs. 

Generally one has $\SF\neq \SW$, meaning that the pairs of EM and Weyl PNDs do not coincide and, for a given observer $u^a$, the projected observers $u^a_{\SF}$ and $u^a_{\SW}$ are distinct. Let $\Xcal^F_{ab}$ and $\Xcal^C_{ab}$ denote unitary self-dual bivectors, defined up to sign by \eqref{eq:Fcal-non-null} and as the unitary eigenbivector of $\Cop$ with non-degenerate eigenvalue $-2\l=2J/I$, respectively. Since $\SF$ and $\SW$ are the blades of their real parts, and since these real parts are bijectively related to the bivectors themselves by \eqref{eq:rel-Xcal-X}, one has $\SF=\SW$ precisely when $\Xcal^F_{ab}=\pm\Xcal^C_{ab}$, which by \eqref{eq:Fcal-non-null} happens precisely when $\Fcal_{ab}$ is an eigenbivector of $\Cop$ with non-degenerate eigenvalue $-2\lambda$, $\Cop(\Fcal)_{ab}=-2\lambda\Fcal_{ab}$. This can be verified directly by any observer $u^a$, since by \eqref{eq:Qa-non-null} and \eqref{eq:Cop-Qop-equiv} this comes down to $\f^a$ being an eigenvector of $\Qop$ with eigenvalue $-2\lambda$:
\begin{equation}\label{cond-SF=SW}
\SF=\SW\qquad\Leftrightarrow\qquad \Q^a{}_{b}\f^b = -2\lambda \f^a,\quad \f^a=\E^a-i\B^a. 
\end{equation}
If this holds everywhere in the considered spacetime region (with the possible exception of a measure zero set of points where $C_{abcd}=0\lra
\l=0$) 
one speaks about Petrov type D {\em doubly aligned} non-null Einstein-Maxwell fields.
At the points where $C_{abcd}\neq 0$ the pairs of EM and Weyl PNDs coincide, and the principal observers and tetrads of the EM and gravitational fields {\em are the same}. 
For any observer $u^a$ we have $u^a_{\SF}=u^a_{\SW}\equiv \us^a$, and thus $\psi_{\SF}=\psi_{\SW}\equiv \pp$ and $\Aem=\AD$ 
in \eqref{eq:vrel-EM} and \eqref{eq:vrel-D}, and also $\eb_F^a=\eb_C^a=\eb^a$ in \eqref{eq:calc-eb-EM} and \eqref{eq:calc-eb-D}. It follows that {\em  the relative super-energy density is a quadratic function of the energy density, and the super-Poynting vector is aligned with the Poynting vector}:
\begin{equation}\label{eq:tg-Pg-doublealigned}
\tg=\frac{\chg}{2}\left[3\left(\frac{\te}{\che}\right)^2-1\right],
\qquad \qp^a
=\frac{3\chg}{2\che}\frac{\te}{\che}\,\sp^a.
\end{equation} 
All Petrov type D doubly aligned non-null Einstein-Maxwell fields (with a possible non-zero cosmological constant, and including the vacuum limits obtained by putting the EM parameters to zero) are known; the line elements are exhausted by those constructed in Refs.\ \cite{Overview-D-alinged}, comprising all well-known black hole solutions such as the Schwarzschild-Kottler, Reissner-Nordstr\"{o}m, Kerr-Newman, (charged and/or spinning) $C$ and Pleba\'{n}ski-Demia\'{n}ski
metrics. We will treat the Kerr-Newman spacetimes 
in Sec.\ \ref{sub:Kerr}.

\subsection{Petrov type I: general vs.\ principal observers}

\label{subsec:grav-typeI}

In the Petrov type I case the three eigenvalues $\l_{i}$ of $\Cop$ (equal to those for any $\Qop$) are distinct. There is a unique principal observer $e_0^a$ and an essentially unique principal tetrad $(e_0^a,e_i^a)$, where the spacelike vectors $e_i^a\in e_0^\bot$ are unit eigenvectors of the associated endomorphism $\Qop_0=\Eop_0-i\Hop_0$ of $e_0^\bot$. 
The unitary self-dual eigenbivectors $\Ucal_{ab}^{i}$ of $\Cop$ are related to the principal tetrad vectors by \eqref{eq:Cabcd-decomp-gen}. 

Consider an arbitrary observer $u^a\in \Up$. Its expansion in the    
principal tetrad $(e_{0}^{a},e_{i}^{a})$ is given by 
\begin{equation}\label{eq:u-decomp-13}
u^{a}=u^{0}e_{0}^{a}+u^{i}e_{i}^{a},\qquad u^{0}=\sqrt{1+(u^{1})^{2}+(u^{2})^{2}+(u^{3})^{2}}\,,
\end{equation}
where the components $u^{i}\in\R$ play the role of independent parameters. Equation \eqref{eq:u-decomp-13} defines the boost from $e_0^a$ to $u^a$, with {$\cosh\wrel{u}{e_0}=\cosh\wrel{e_0}{u}=u^0$} and $\vrel{u}{e_0}^a=u^ie_i^a/u^0$. The relative velocity 
\begin{equation}\label{eq:vrele0u-def}
\vrel{e_0}{u}^a=e_0^a/u^0-u^a
\end{equation}
describes the inverse boost. We are interested in an algorithm that derives $\vrel{{e_0}}{u}^a$ directly from the tensor $\Q_{ab}=\E_{ab}-\H_{ab}$ relative to $u^a$, the dependence of the super-energy-density $\tg$ on the parameters $u^i$, and to what extent a surrogate of theorem \ref{thm:D} holds.

To deal with the first point we look at the eigenvectors $\v_i^a=\eb_i^a-i\bb_i^a$ of $\Qop$. By \eqref{eq:Qab-decomp-gen} these are simply the contractions of the eigenbivectors $\Ucal^i_{ab}$ of $\Cop$ with $u^a$, and by \eqref{eq:Cabcd-decomp-gen} and \eqref{eq:u-decomp-13} we obtain their expansions in the principal tetrad: 
\begin{align}\label{eq:via-exp}
&\v_i^{a}=u^{i}e_{0}^{a}+u^{0}e_{i}^{a}+i(u^{k}e_{j}^{a}-u^{j}e_{k}^{a})\qquad\lra\qquad\eb_i^{a}=u^{i}e_{0}^{a}+u^{0}e_{i}^{a},\quad\bb_i^a=u^{j}e_{k}^{a}-u^{k}e_{j}^{a}.
\end{align}
In addition to \eqref{eq:vivj} we thus have  
\begin{align}
\label{eq:via-contr}
&\v_{i}^{a}(\overline{\v_{i}})_{a}=2(u^{0})^{2}-2(u^{i})^{2}-1=2(u^{j})^{2}+2(u^{k})^{2}+1,\quad \v_{j}^{a}(\overline{\v_{k}})_{a}=-2u^{j}u^{k}+2iu^{0}u^{i},\\
\label{eq:eibi0}
&\eb_i^a(\eb_i)_a=(u^{0})^{2}-(u^{i})^{2}=(u^{j})^{2}+(u^{k})^{2}+1=\bb_i^a(\bb_i)_a+1,\qquad\eb_i^a(\bb_i)_a=0,\\
\label{eq:eibi-contr}
&\eb_j^a(\eb_k)_a=\bb_j^a(\bb_k)_a=-u^ju^k,\qquad \eb_j^a(\bb_k)_a=-\bb_j^a(\eb_k)_a=u^0u^i. 
\end{align} 
Note that the vectors $\eb_i^a$ 
are linearly independent and thus form a (generally non-orthogonal) basis of $\up$; moreover one has
\begin{equation}\label{bia-exp-lindep}
u^0\bb_i^a=u^j\eb_k^a-u^k\eb_j^a,\qquad u^i\bb_i^a=0,
\end{equation}
which gives the expansion of the vectors $\bb_i^a$ in this basis and shows that they are linearly dependent. 
The frames $(e_{0}^{a},e_{i}^{a})$ and $(u^{a},\v_{i}^{a})$ of $\C T_{p}M$ are both oriented and orthonormal, the expansion of the latter in terms of the former being given by \eqref{eq:u-decomp-13} and \eqref{eq:via-exp}. Inverting this relation and using \eqref{bia-exp-lindep} yields\footnote{\label{foot:ei-from-e0} Given the first expression in \eqref{eq:inverse-trafo-e0} the first one in \eqref{eq:inverse-trafo-ei} is confirmed by $(e_{i})_{a}=\U^{i}_{ab}e_{0}^{b}=\Ucal^{i}_{ab}e_{0}^{b}=2(u_{[a}(\v_i)_{b]})^\dag e_0^b=2u_{[a}(\v_i)_{b]}e_0^b+i[e_0,\v_i]^a$, see \eqref{eq:def-psiuchiu} and \eqref{eq:Cabcd-decomp-gen}, and application of the identity \eqref{id:viy} to $y^a=e_0^a$.}
\begin{align}  
\label{eq:inverse-trafo-e0} &e_{0}^{a}=u^{0}u^{a}-u^{i}\v_{i}^{a}=u^{0}u^{a}-u^{i}\eb_{i}^{a},\\ 
\label{eq:inverse-trafo-ei} &e_{i}^{a}=-u^{i}u^{a}+u^{0}\v_{i}^{a}+i(u^{j}\v_{k}^{a}-u^{k}\v_{j}^{a})=-u^{i}u^{a}+u^{0}\eb_{i}^{a}+u^{j}\bb_{k}^{a}-u^{k}\bb_{j}^{a},
\end{align}
which confirms that $e_{0}^{a}$ and $e_{i}^{a}$ are real vectors. From \eqref{eq:inverse-trafo-e0} we thus obtain
\begin{equation}\label{eq:vrele0u-uform}
\vrel{e_0}{u}^a=-{u^i}\v_i^a/{u^0}=-{u^i}\eb_i^a/{u^0}.
\end{equation}
Alternatively, one can consider
the decomposition $\zv^a=\zv^0 e_0^a+\pi(\zv)^a$ of a vector $\zv^a\in T_p M$, where 
\begin{equation}
\pi(\zv)^a\equiv  [\delta^a_b+e_0^a(e_0)_b]\zv^b =\zv^i e_i^a
\end{equation}
is the projection of $\zv^a$ onto the rest space $e_0^\bot$ of the principal observer. If $\zv^a$ belongs to the rest space $\up$ of the given observer \eqref{eq:u-decomp-13} then $\zv^au_a=0$, which translates to $u^0\zv^0=\sum_{i=1}^3u^i \zv^i$ and by \eqref{eq:via-exp} yields
\begin{equation}\label{exp-wa}
\zv^a={\zv^i}\eb_i^a/{u^0}. 
\end{equation}
Applying this to $\zv^a=\vrel{e_0}{u}^a$ one has $\zv^ie_i^a=\pi(\vrel{e_0}{u})^a=-\pi(u)^a=-u^i e_i^a$ by 
\eqref{eq:vrele0u-def} and \eqref{eq:u-decomp-13}, which gives $\zv^i=-u^i$ and thus \eqref{eq:vrele0u-uform}. Using \eqref{eq:via-contr}-\eqref{eq:eibi-contr} we arrive at:
\begin{prop}\label{prop:ve0u} The velocity of $e_0^a$ relative to $u^a$ is given in terms of $x_i^a$ or $\eb_i^a,\,\bb_i^a$ by 
\begin{align}
\label{eq:vrele0u-1}\vrel{e_0}{u}^a&=-\frac{1}{2(u^0)^2}\sum_{(ijk)}\Im(\v_j^b(\overline{\v_k})_b)\v_i^a=\frac{1}{(u^0)^2}\sum_{(ijk)}\bb_{j}^{b}(\eb_{k})_{b}\eb_i^a\\ 
\label{eq:vrele0u-2}
&=\frac{1}{4i(u^0)^2}\sum_{i=1}^{3}[\v_{i},\overline{\v_{i}}]^a=\frac{1}{2(u^0)^2}\sum_{i=1}^{3}[\eb_{i},\bb_{i}]^a
\end{align}	
where the square of the associated Lorentz factor is given by 
\begin{equation}
\label{eq:u0-inverse}
2(u^{0})^2=\tfrac12\sum_{i=1}^{3}\v_{i}^{a}(\vc_{i})_{a}+\tfrac12=\sum_{i=1}^{3}\eb_{i}^{a}({\eb_{i}})_{a}-1=
\sum_{i=1}^{3}\bb_{i}^{a}(\bb_{i})_{a}+2\,.
\end{equation}
Hence the unique principal observer is $e_0^a=u^0[u^a+\vrel{e_0}{u}^a]$.
\end{prop} 
\noindent To obtain \eqref{eq:u0-inverse} one sums the first parts of \eqref{eq:via-contr} and \eqref{eq:eibi0} over $i$ and uses the second part of \eqref{eq:u-decomp-13}; the expressions \eqref{eq:vrele0u-1} 
for $\vrel{e_0}{u}^a$ 
are immediate from the above; on applying the identity \eqref{id:viy} to $y^a=\overline{\v_i}^a$ the first expressions in \eqref{eq:vrele0u-1} and \eqref{eq:vrele0u-2} are seen to be equal, and then the second expression in \eqref{eq:vrele0u-2} follows from  $[\v_i,\vc_i]^a=2i[\eb_i,\bb_i]^a$.

By proposition \ref{prop:ve0u} the problem of finding $\vrel{e_0}{u}^a$ is reduced to constructing the eigenvectors $\v_i^a$ of $\Qop$. The projector of $\Cup$
onto $\langle \v_i^a\rangle$ is given by 
\begin{equation}\label{eq:def-Ri}
\Rop^i\equiv \frac{(\Qop-\l_j\Iop)(\Qop-\l_k\Iop)}{(\l_i-\l_j)(\l_i-\l_k)}=\frac{\Qop^2+\l_i\Qop+(\l_i^2-I/2)\Iop}{3\l_i^2-I/2}\quad\lra\quad (R^i)^a_{\;\;b}=\frac{\Q^a_{\;\;c}\Q^c_{\;\;b}+\l_i\Q^a_{\;\;b}+(\l_i^2-I/2)\I^a_{~b}}{3\l_i^2-I/2},
\end{equation}
as this operator annihilates $\v_j^a,\v_k^a$ and leaves $\v_i^a$ invariant 
(see, e.g., appendix A of \cite{ColHerOrtWyll12}). Take any $y^a\in\up$ that is not an eigenvector of $\Qop$ with eigenvalue $\l_i$ [$y^a\notin\langle\v_{i}^{a}\rangle\lra (R^i)^a_{\;\;b}y^b\neq 0$]; for an arbitrary \ON tetrad $(u^{a},e_{i'}^{a})$ at least one of the vectors $e_{i'}^{a}$ may be taken as $y^{a}$. Then 
\begin{equation}\label{eq:calc-via}
\v_i^a=\frac{(R^i)^a_{\;\;b}y^b}{\sqrt{R^i_{cd}y^cy^d}}
\qquad \left[\v_i^a=\frac{(R^i)^{m'}_{\;\;i'}e_{m'}^a}{\sqrt{R^i_{i'i'}}}\right]\,,
\end{equation}
where the result for $y^a=e_{i'}^{a}$ is indicated between square brackets, and one applies \eqref{eq:root-z} to $z=R^i_{cd}y^cy^d\,[R^i_{i'i'}]$. 
We conclude that an observer $u^a$ passing through a point $p$ and measuring $\Q^{ab}=(\E^{m'n'}-i\H^{m'n'})e_{m'}^ae_{n'}^b$ may check whether the gravitational field is of Petrov type I at $p$, and in this case determine the unique Weyl principal observer directly from $\Q^{ab}$  by the following\\ 

{\em Algorithm for obtaining the principal observer of a Petrov type I Weyl tensor, directly from electric and magnetic parts of the Weyl tensor measured by an arbitrary observer:}  
\begin{enumerate}
	\item  Compute $I$ and $J$ from \eqref{eq:I-def-Q} and \eqref{eq:J-def-Q}; if $I^{3}\neq 6J^{2}$ then the Petrov type is
	I at $p$, and go to step 2. 
	\item Compute the eigenvalues $\l_{k}$ of $\Qop$ from \eqref{eq:roots}, or from \eqref{eq:L-alpha} when $q=0$. 
	\item For each $i\in\{1,2,3\}$ compute $\Rop^i$ from \eqref{eq:def-Ri}, take $y^a\in\up$ for which $(R^i)^a_{\;\;b}y^b\neq 0$, and construct the unit eigenvectors $\v_i^a\equiv \eb_i^a-i\bb_i^a$ of $\Qop$ with eigenvalue $\l_i$ from \eqref{eq:calc-via}. 
	\item The principal observer $e_0^a=u^0[u^a+\vrel{e_0}{u}^a]$ is given by \eqref{eq:vrele0u-1} or \eqref{eq:vrele0u-2}, and \eqref{eq:u0-inverse}.
\end{enumerate}
Likewise, given \eqref{eq:u0-inverse} the spatial principal tetrad vectors are found from \eqref{eq:eibi-contr} and \eqref{eq:inverse-trafo-ei}:
\begin{equation}
\label{eq:inverse-trafo-ei1}e_{i}^{a}=\frac{1}{u^0}\left[\bb_{j}^{b}(\eb_{k})_{b}u^a+(u^{0})^2\eb_i^a+\bb_{i}^{b}(\eb_{j})_{b}\bb_j^a+\bb_{i}^{b}(\eb_{k})_{b}\bb_k^a\right].
\end{equation}

\begin{rem}\label{rem:ve0u-proof-altern} In \cite{FerrSaezP1} Ferrando and S\'{a}ez gave a covariant algorithm to determine $e_0^a$ and $e_i^a$. In their approach one considers $X^i_{ab}=\Re({\cal X}^i_{ab})=2(e_{0})_{[a}(e_{i})_{b]}$ and obtains the projector $(P_{i})^a_{\;\;b}=(\U^i)^a_{\;\;c}(\U^i)^c_{\;\;b}=-\Enul^{a}(\Enul)_{b}+e_i^{a}(e_i)_{b}$ on the blade $\SS_i$, see \eqref{eq:Cabcd-decomp-gen} and \eqref{eq:PPbot-def} with $e_1^a$ replaced by $e_i^a$; summing these projectors over $i$ yields $2e_{0}^{a}(e_{0})_{b}=\delta^a_b-\sum_{i=1}^{3}(P_{i})^a_{\;\;b}$, and contraction with $u^b$ and normalization then gives $e_0^a$. On using the $i$-labeled version of \eqref{eq:Pabub-complex} this gives back the final expression in \eqref{eq:vrele0u-2}, which is thus an explicit form of the expression for $e_0^a$ in Corollary 1 of \cite{FerrSaezP1}; once $e_0^a$ is found one gets the spatial princpal tetrad vectors simply from $e_i^a=(\U^i)^a_{\;\;b}e_0^b$ (see also footnote \ref{foot:ei-from-e0}). This algorithm requires knowlegde of the eigenbivectors ${\cal X}^i_{ab}$ of the Weyl operator $\Cop$, which can be found by constructing analoga of the projectors \eqref{eq:def-Ri} from $\Cop$ (see Proposition 1 of \cite{FerrSaezP1}) and applying these to a generic self-dual bivector; however, given $\Q_{ab}=\E_{ab}-i\H_{ab}$ measured by an observer $u^a$ the operator $\Cop$ should be constructed first from \eqref{eq:Cdag-reconstr} for this purpose. Our algorithm avoids this detour and the explicit use of bivectors, and finds the principal tetrad vectors directly from $\Q_{ab}$, thereby staying entirely in the (complexified) rest space of $u^a$.  
\end{rem}


To deal with the other two points we derive expressions for the super-energy density and super-Poynting vector. Following \cite{FerSae12} we define the invariants 
\begin{equation}\label{def-ri}
r_i\equiv -[4\Re(\l_j\lc_k)+\Re(\l_k\lc_i)+\Re(\l_i\lc_j)]=2(|\l_j|^2+|\l_k|^2)-|\l_i|^2=|\l_j-\l_k|^2,
\end{equation}
which are related to the invariants $\chg$ and $q$ defined in \eqref{def-chg} by
\begin{align}
\label{rel-ri-chg}&12\chg=r_1+r_2+r_3,\\
\label{rel-ri-q}&36q^2=2r_1r_2+2r_2r_3+2r_3r_1-r_1^2-r_2^2-r_3^2.
\end{align}
The non-defining equalities in \eqref{def-ri} follow from $\l_1+\l_2+\l_3=0$; the last one was not observed in \cite{FerSae12} and makes clear that $r_i>0$ in the Petrov type I case. We will also consider the Petrov type D `limit' (or subcase), to which all equations in this section apply as well if we choose, as before, a Weyl principal tetrad $(e_0^a,e_i^a)$ such that 
\begin{equation}\label{eq:limit-D-b}
\textrm{[Petrov type D limit]}\qquad \l_2=\l_3=-\l_1/2\equiv \lambda,\qquad 
r_1=0,\qquad r_2=r_3=6\chg,\qquad q=0;
\end{equation}  
then the Weyl principal plane $\Sigma=\langle e_0^a,e_1^a\rangle$, and $u^2=u^3=0$ characterizes Weyl principal observers. 

Substituting the decomposition \eqref{eq:Qab-decomp-gen} in 
\eqref{eq:qpa-def} and \eqref{eq:tg-def} we obtain the super-energy density and super-Poynting vector in terms of the \ON triad $(\v_i^a)$ of $\Cup$:
\begin{align}
\label{tg-xi}
&\tg=\frac{1}{4}\sum_{m=1}^{3}\sum_{n=1}^3\l_{m}\overline{\l_{n}}[\v_{m}^{a}(\overline{\v_{n}})_{a}]^{2}\,,\\
\label{eq:qpa-I-a}
&\qp^a=\frac{1}{4i}\sum_{m=1}^{3}\sum_{n=1}^3\l_m\overline{\l_n}[\vop_m\vop_m,\vcop_n\vcop_n]^a=\frac{1}{4i}\sum_{(ijk)}\v_i^a(\l_j-\l_k)\sum_{n=1}^3\overline{\l_n}\v_j^b(\overline{\v_n})_b\v_k^c(\overline{\v_n})_c\,,
\end{align}
where we used \eqref{eq:id[xxyy]} and \eqref{id:viy} applied to $y^a=\overline{\v_n}^a$ for the last equality.   
Equations \eqref{eq:via-exp}-\eqref{eq:via-contr} then lead to the following expressions for 
$\tg$ (cf.\ \cite{FerSae12}) and the components of 
$\qp^a=\qp^0 e_0^a+\qp^i e_i^a$ in the principal tetrad:\footnote{The expression \eqref{eq:def-tg-Om-Phi} for $\tg$ coincides with (C.5)-(C.6) of \cite{FerSae12} except for the factor 12 in the second term of $\Omega$, which 
corrects the factor 6 in (C.5) and (C.7) of \cite{FerSae12} and gives the correct equation (C.10) of \cite{FerSae12}. Equations \eqref{eq:qa-expans1}-\eqref{eq:qa-expans2} are new. An alternative way to derive \eqref{eq:def-tg-Om-Phi}-\eqref{eq:qa-expans2} is by substituting \eqref{eq:Cabcd-decomp-gen} into \eqref{eq:Bel-Rob-def}, which yields~\cite{FerSae12,FerSae10} $T^{abcd}=\sum_{i=1}^{3}|\l_i|^2\Pi_{i}^{ab}\Pi_{i}^{cd}
+\sum_{(ijk)}\left(\l_j\overline{\l_k}\Pi_{jk}^{ab}\Pi_{jk}^{cd}+\overline{\l_j}\l_k\overline{\Pi}_{jk}^{ab}\overline{\Pi}_{jk}^{cd}\right)$
with  $\Pi_{i}^{ab}=-e_{0}^{a}e_{0}^{b}+e_{i}^{a}e_{i}^{b}-\tfrac{1}{2}g^{ab},\, \Pi_{jk}^{ab}=e_{j}{}^{(a}e_{k}{}^{b)}+ie_{0}{}^{(a}e_{i}{}^{b)}$, triply and quadruply contracting this with \eqref{eq:u-decomp-13} and using $\tg=T_{abcd}u^au^bu^cu^d$ and $\qp^a=-T^a_{\;bcd}u^bu^cu^d-\tg u^a$, see \eqref{eq:qa-def}-\eqref{eq:tg-def} and cf.\ remark \ref{rem:WheelerEM-altern}.}
\begin{align}\label{eq:def-tg-Om-Phi}
&\tg=\chg+\Omega,
\quad\Omega= A^ir_{i}+12q u^{0}u^{1}u^{2}u^{3},\quad A^{i}=(u^{0})^{2}(u^{i})^{2}-(u^{j})^{2}(u^{k})^{2},\\
&2\qp^{0}=u^{0}\left[r_{1}(u^{1})^{2}+r_{2}(u^{2})^{2}+r_{3}(u^{3})^{2}-2\Omega\right]+6qu^{1}u^{2}u^{3},\label{eq:qa-expans1}\\
&2\qp^{i}=u^{i}\left[r_{j}(u^{k})^{2}+r_{k}(u^{j})^{2}-r_{i}(u^{0})^{2}-2\Omega\right]-6qu^{0}u^{j}u^{k},\label{eq:qa-expans2}
\end{align}
Applying \eqref{exp-wa} to $w^a=\qp^a$ we obtain the expansion of $\qp^a$ in the basis $(\eb_i^a)$ of $\up$:
\begin{equation}\label{qpa-ei}
\qp^a={\qp^i}\eb_i^a/u^0. 
\end{equation}
Just as \eqref{eq:qpa-I-a} this expression makes explicit that the Poynting vector belongs to $\up$ and connects with the (observer-dependent) eigenvectors $\v_i^a$ of $\Qop$; however, it is manisfestly real and refers to the components relative to the  canonical (observer-independent) \ON frame $(e_0^a,e_i^a)$, thus synthesizing both viewpoints. 

In view of \eqref{eq:u-decomp-13} the expression \eqref{eq:def-tg-Om-Phi} give the super-energy density as a function of $(u^1,u^2,u^3)\in\R^3$. Note that this function is irrational when $q\neq0$, and polynomial  of degree 4 when $q=0$. The level surfaces of $\tg$ for Petrov type I are rather complicated, 
in contrast to those in the Petrov type D limit \eqref{eq:limit-D-b}, cf.~remarks \ref{rem:obsgeom-EM} and \ref{rem:obsgeom-D}), for which one finds back \eqref{eq:tg-D} from  \eqref{eq:def-tg-Om-Phi} by identifying \eqref{eq:decomp-further} with \eqref{eq:u-decomp-13}.
However, by \eqref{eq:tg2-chg2} and characterization (iv) of Weyl principal observers, and for all values of $q\in\R$ and of the triples $(r_1,r_2,r_3)\in\R^3_{>0}$ allowed by \eqref{rel-ri-q},
$\tg$ attains an absolute minimum value $\xi$ at $(u^1,u^2,u^3)=(0,0,0)$, corresponding to $u^a$ being the principal observer $e_0^a$ and to $\Omega=0$, while $\Omega>0$ when $(u^1,u^2,u^3)\neq (0,0,0)$. In appendix \ref{app:principal}.1 we present 
a straightforward proof of this fact. 

To what extent does a surrogate of theorem \ref{thm:D} hold? First note from \eqref{eq:qpa-I-a} that in the type D limit \eqref{eq:limit-D-b} $\qp^a$ is orthogonal to $\v_1^a\equiv \v^a\equiv\eb^a-i\bb^a$, such that $\qp^a\propto[\eb,\bb]^a\propto\vrel{\us}{u}^a$ and we retrieve the gravitational Wheeler analogue: for any $u^a$ there {\em exists a} principal observer (namely $\us^a$) lying in  $\langle u^a,\qp^a\rangle$. 
In the Petrov type I case there is only one principal observer $e_0^a$ and the ``maximal'' surrogate of theorem \ref{thm:D} would be that, for any observer $u^a$, 
the velocity of $e_0^a$ relative to $u^a$ is proportional to the super-Poynting vector of $u^a$, $\vrel{e_0}{u}^a\propto \qp^a$. However, this is not the case. Instead, it follows from \eqref{eq:vrele0u-uform} and \eqref{qpa-ei} that 
\begin{equation}\label{cond-v-propto-P}
\vrel{e_0}{u}^a=\alpha \qp^a\quad \Leftrightarrow\quad u^i=-\alpha \qp^i,\quad i=1,2,3
\quad \Leftrightarrow\quad \pi(u)^a\equiv u^0\vrel{u}{e_0}^a=-\alpha\pi(\qp)^a.
\end{equation}
Hence $\vrel{e_0}{u}^a\propto \qp^a$ precisely when $\vrel{u}{e_0}^a\propto\pi(\qp)^a$, which is a condition in the rest space $e_0^\bot$ of the principal observer. From \eqref{cond-v-propto-P} it follows that
\begin{equation}\label{ujPk}
u^j\qp^k=u^k\qp^j,\qquad i=1,2,3,
\end{equation}
where $(i,j,k)$ is a cyclic permutation of $(1,2,3)$. 
Conversely, when $\qp^a\neq 0$ equations \eqref{ujPk} imply \eqref{cond-v-propto-P}, with 
\begin{equation}\label{alpha-expr}
\alpha=\frac{-\qp^0}{u^0\qp^a\qp_a}=\frac{-u^i}{\qp^i}
\end{equation} 
for any $i$ such that $\qp^i\neq 0$, where the first equality follows from \eqref{cond-v-propto-P} and contracting \eqref{eq:vrele0u-def} with $\qp_a$. For Petrov type I the principal observer is unique and so $\qp^a=0\lra u^a=e_0^a$, such that \eqref{cond-v-propto-P} is equivalent to \eqref{ujPk}. In the Petrov type D limit \eqref{eq:limit-D-b} we regard $e_0^a$ as a given principal observer, and on solving \eqref{ujPk} we need to exclude the principal observers $u^a\neq e_0^a$, which are characterized by $u^2=u^3=0\neq u^1$ but for which $\vrel{e_0}{u}^a\neq 0=\qp^a$ and thus do not solve \eqref{cond-v-propto-P}. 
For convenience we define
\begin{equation}\label{def-ai}
a_\beta\equiv (u^\b)^2,\,\b=0,1,2,3\quad\Rightarrow\quad a_0=1+a_1+a_2+a_3,
\end{equation}
and infer from \eqref{eq:qa-expans2} that \eqref{ujPk} translates to the conditions 
\begin{align}
& u^{2}u^{3}[(a_{2}-a_{3})r_{1}+(a_{0}+a_{1})(r_{2}-r_{3})]=6u^{0}u^{1}(a_{2}-a_{3})q\,,\label{gek1}\\
& u^{3}u^{1}[(a_{1}-a_{3})r_{2}+(a_{0}+a_{2})(r_{1}-r_{3})]=6u^{0}u^{2}(a_{1}-a_{3})q\,,\label{gek2}\\
& u^{1}u^{2}[(a_{1}-a_{2})r_{3}+(a_{0}+a_{3})(r_{1}-r_{2})]=6u^{0}u^{3}(a_{1}-a_{2})q\,.\label{gek3}
\end{align}
{\em In the Petrov type D limit \eqref{eq:limit-D-b}} these equations 
reduce to $u^1u^2=u^1u^3=0$, and by the above 
we find back the property (see remark \ref{rem:obsgeom-D}) that {\em for a given principal observer $e_0^a=\EEt_0^a(\aa)$ there is a {\em 2d variety} ${\cal L}(\aa)$ of observers $u^a$ satisfying $\vrel{e_0^a}{u}^a\propto\qp^a$}, namely those for which $\us^a=e_0^a$, characterized by $u^1=0$. For Petrov type I the situation is different. 
Clearly, at most two of the equations \eqref{gek1}-\eqref{gek3} can be independent (when $u^3\neq 0$ the last equation is a consequence of the first and the second, and analogously when $u^{1}\neq0$ or $u^{2}\neq0$). A pair of non-trivial equations corresponds to the intersection ${\cal I}$ of two 2d surfaces in $\Up$. Take the square of \eqref{gek1} and \eqref{gek2}, eliminate $q^2$ by \eqref{rel-ri-q} and $a_0$ by \eqref{def-ai}. This yields two polynomial equations $K(a_i|r_j)=0$ and $L(a_i|r_j)=0$ in the variables $a_i>0$, where the $r_j$ are regarded as parameters ($i,j=1,2,3$). Computing the resultant of $K(a_i|r_j)$ and $L(a_i|r_j)$ with respect to $a_3$ leads to a non-trivial polynomial equation $M(a_1,a_2|r_j)=0$, for any values of the parameters $r_j>0$.
For each solution $(a_1,a_2)\in\R_{\geq0}^2$ of this equation there is a finite number of values $a_3\in\R_{\geq 0}$ for which $K(a_i|r_j)=0$ and $L(a_i|r_j)=0$, leading to a finite union of curves and isolated points in $\R_{\geq0}^3\ni(a_1,a_2,a_3)$, and the intersection ${\cal I}$ forms a part hereof. Note that when two of the $u^{i}$'s are zero the system \eqref{gek1}-\eqref{gek3} is identically satisfied, irrespective of $q$ and $r_i$. Hence we have

\begin{thm}\label{prop:typeI} Suppose the Weyl tensor is of Petrov
	type I at a point $p$, and let $(e_0^a,e_i^a)$ be the principal tetrad at $p$. Then 
	the observers $u^a=u^0e_0^a+u^ie_i^a$ for which \eqref{gek1}-\eqref{gek3} and thus $\vrel{e_0}{u}^a\propto \qp^a$ holds
	form a 1d variety (finite union of curves and points) in $\Up$. This variety always contains (among others) three canonical curves of observers, given by
	\begin{equation}\label{eq:threecurves}
	u^{a}=\cosh(\aaa)e_{0}^{a}+\sinh(\aaa)e_{i}^{a},\quad\aaa\in\R,\qquad i\in\{1,2,3\}
	\end{equation}
	and obtained by boosting the principal observer $e_0^a$ arbitrarily along one of the spatial tetrad vectors $e_{i}^{a}$. 
\end{thm}


\begin{rem}\label{rem:typeI-specobs} Consider the special observers \eqref{eq:threecurves}. 
For fixed $i\in\{1,2,3\}$ we have $u^0=\cosh(\aaa),\,u^i=\sinh(\aaa),\,u^j=u^k=0$, and find 
	\begin{align}\label{spec-tg-P-v}
	\tg=\chg+\tfrac14 r_i \sinh(2\aaa)^2,\qquad \qp^a=-\tfrac18 r_i\sinh(4\aaa)\eb_i^a,\qquad \vrel{e_0}{u}^a=-\tanh(\aaa)\eb_i^a 
	\end{align}
	from \eqref{eq:vrele0u-uform}, \eqref{eq:def-tg-Om-Phi} and \eqref{eq:qa-expans2}-\eqref{qpa-ei}, such that \eqref{cond-v-propto-P} is realized by
	\begin{equation}\label{spec-v-propto-P}
	\vrel{e_0}{u}^a=\frac{4\qp^a}{r_i\zeta_i(\zeta_i+1)},\qquad \zeta_i=
	\sqrt{1+\frac{4(\tg-\chg)}{r_i}}=\cosh(2\aaa)\,.
	\end{equation}  
	In the Petrov type D limit \eqref{eq:limit-D-b} the expressions \eqref{spec-tg-P-v} also apply to observers obtained by boosting any principal observer $e_0^a\in\Sigma$ along the spatial vectors of a principal tetrad $(e_0^a,e_i^a)$. In \cite{Bini} Bini et al.\ considered such observers for both Petrov types I and D, and \eqref{spec-tg-P-v} explicates their two main results: $\vrel{e_0}{u}^a\propto\qp^a$ holds, and for varying $\aaa$ the super-energy density attains its minimum at $\aaa=0$, which is a special case of characterization (iv) of a principal observer $e_0^a$; note that the first equation of \eqref{spec-tg-P-v} can be written as  $\tg=\frac{1}{24}(2r_j+2r_k-r_i)+\frac{1}{8}r_i\cosh(4\aaa)$, which is a more explicit form of Eq.\ (42) of \cite{Bini}. 
	In the Petrov type D limit \eqref{eq:limit-D-b} we can identify \eqref{eq:u-decomp-13} with \eqref{eq:decomp-further}; for $i=2\,(u^1=u^3=0)$ and $i=3\,(u^1=u^2=0)$ we have $\aaa=\pm\pp,\,\aa=0$ and $\cc=0$, resp.\ $\cc=\frac{\pi}{2}$, and retrieve from \eqref{spec-tg-P-v}-\eqref{spec-v-propto-P} that \eqref{eq:tg-D}-\eqref{eq:vrel-D} is valid, such that $\tg$ increases with $|\aaa|=\pp$; for $i=1\,(u^2=u^3=0)$, however, we have $\pp=0$: for all values of $\aaa=\aa$ the observers \eqref{eq:threecurves} with $i=1$ belong to $\Sigma$, are thus principal ($\qp^a=0$), and have the same minimal super-energy density $\tg^1(\aaa)=\chg$, in agreement with characterizations (ii)' and (iv) of principal observers. This amends the statement at the end of Sec.\ 5 of \cite{Bini} when applied to Petrov type D.\footnote{We note that Eqs.\ (39)-(42) of \cite{Bini} do not apply to boosts along $e_1^a$ in the Petrov type D limit \eqref{eq:limit-D-b}, since $\vhat{\ubig}{u}^a\equiv e_{1}^{a}$	and thus, in the notation of \cite{Bini}, $Z^{\bot\bot(\textrm{TF})}(u)=0$.} 
	
	In the Petrov type I case the special observers \eqref{eq:threecurves} may be characterized as follows. For an arbitrary observer $u^a$ given by \eqref{eq:u-decomp-13} the orthogonal eigendirections of the endomorphism $\Qop$ associated to an observer $u^a$ are generated by the three vectors $\v_i^a$ given in \eqref{eq:via-exp}. If $u^j=u^k=0$ (for some $i\in\{1,2,3\}$, with $(i,j,k)$ a cyclic permuation of $(1,2,3)$) then $\bb_i^a=0$ and so $\v_i^a=\eb_i^a$ spans a real direction in $\Cup$. Conversely, if $\v_i^a$ spans a real direction then $\eb_i^a$ and $\bb_i^a$ must be linearly dependent; by $\eb_i^a\neq 0$ and $\eb_i^a(\bb_i)_a=0$ (see \eqref{eq:eibi0}) this implies $\bb_i^a=0$, whence $\v_i^a=\eb_i^a$ and $u^j=u^k=0$ from \eqref{eq:via-exp}. Combined with \eqref{spec-tg-P-v} and by the same argument as in remark \ref{rem:qpa-eigen-D} we conclude that {\em for a non-principal observer $u^a$ the endomorphism $\Qop=\Eop-i\Hop$ has no real eigendirections except when $u^a$ is one of the observers \eqref{eq:threecurves} for some $i\in\{1,2,3\}$, in which case $\qp^a$ is an eigenvector of $\Qop$ with eigenvalue $\l_i$ (and thus of $\Eop$ and $\Hop$ with respective eigenvalues $\Re(\l_i)$ and $-\Im(\l_i)$) and  $\langle \qp^a\rangle$ is the only real eigendirection of $\Qop$ and the only common eigendirection of $\Eop$ and $\Hop$.} This relates the observers \eqref{eq:threecurves} to general observers in the Petrov type D case, cf.\ remark \ref{rem:qpa-eigen-D}. In fact, in the Petrov type D limit \eqref{eq:limit-D-b} one can write any given observer as $u^a=\cosh(\pp)\us^a+\sinh(\pp)\ub^a$ and identify $(e_0^a,e_1^a,e_2^a,e_3^a)$ with the principal tetrad $(\us^a,\vs^a,\ub^a,\vb^a)$, see \eqref{eq:decomp-2+2} and \eqref{eq:def-vsa-vba}; hence $u^1=u^3=0$ and $u^a$ takes the form \eqref{eq:threecurves} with $i=2$ and $\aaa=\pm\pp$, such that $\l_i=\l$ and one recovers remark \ref{rem:qpa-eigen-D}. 
\end{rem}

The full 1d variety of theorem \ref{prop:typeI} contains other observers than \eqref{eq:threecurves}, its shape depending on the values of $q$ and $r_i$.  
Below we work out the two cases corresponding to degenerate Petrov type I in the extended Petrov classification  
by Arianrhod and McIntosh~\cite{McIntosh3} (see appendix \ref{subsec:Bel-Rob}). To find the extra observers $u^a$ 
we require throughout that at most one of the components $u^i$ vanishes. Figures \ref{fig:typeI} and \ref{fig:illustr_r2=r3} provide illustrations.  
\begin{figure}
	\includegraphics[width=0.55\columnwidth]{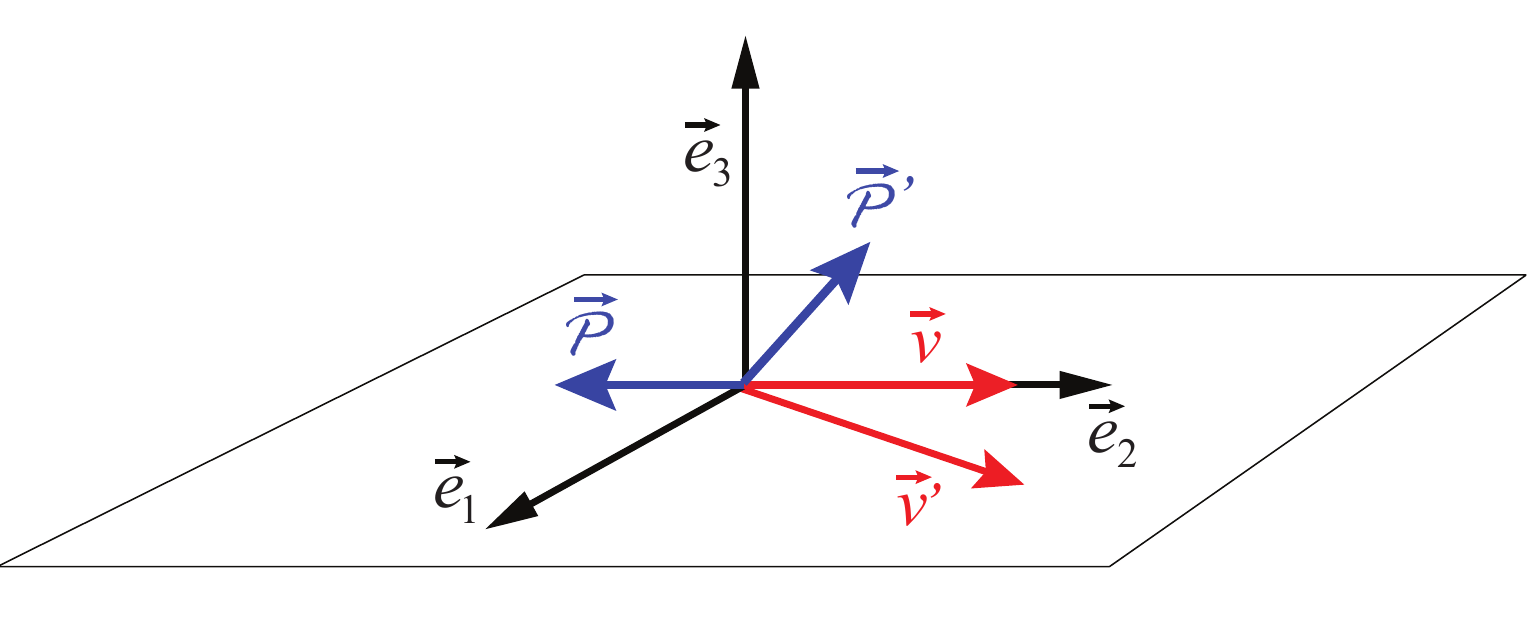}~~~~~~~\includegraphics[width=0.35\columnwidth]{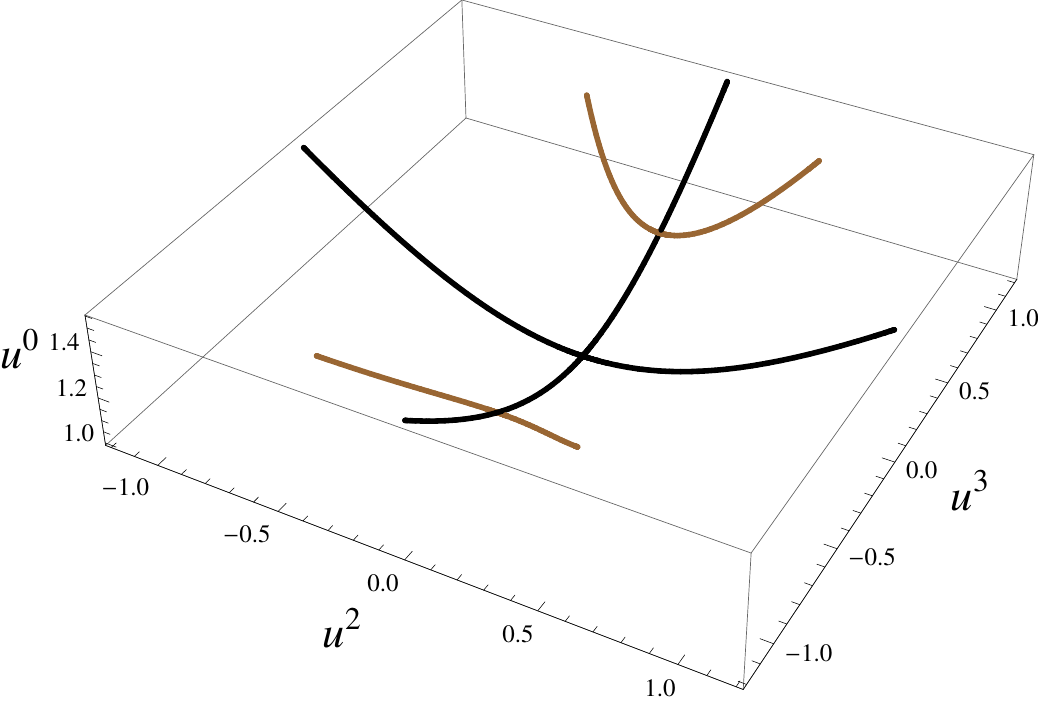}
	\protect\caption{\label{fig:typeI}{Left panel: the situation in the rest space $e_0^\bot$ of the unique principal observer $e_0^a$ in a Petrov type I spacetime. (Arrows denote projected vectors onto $e_0^\bot$.) The vectors $\vec{e}_i$ span the spatial principal directions. Observers moving (with relative velocity $\vec{\nurel}\equiv\vec{\nurel}(u,e_0)$) along one of the $\vec{e}_i$  measure a super-Poynting vector whose projection onto $e_0^\bot$, $\vec{\cal P}\equiv\pi({\cal P})^a$, is along $\vec{e}_i$; hence their boost back to the principal observer is along ${\cal P}^a$: $\vrel{e_0}{u}^a\propto {\cal P}^a$. For a generic observer $u'^a$ this does not hold: $\vrel{e_0}{u'}^a\cancel{\propto}\,\,{\cal P}'^a$. Right panel: the spacetime curves of observers $u^a$ for which $\vrel{e_0}{u}^a\propto \qp^a$, represented in the 3d diagram $e_1^\bot\lra u^1=0$, in the case where $q=0$ and $0\neq |\l_1|<|\l_{2}|<|\l_{3}|$. The black curves are the hyperbolae of observers moving along $e_2^a$ and $e_3^a$; note that the observers moving along $e_1^a$ (with $u^2=u^3=0$) cannot be represented in this diagram. The  
			two brown curves represent the `extra' observers for which $\vrel{e_0}{u}^a\propto \qp^a$ in the $q=0$ case, corresponding to Eq.~\eqref{eq:q=0}.}}
\end{figure}


\begin{ex}\label{ex:q=0} {\em The case $q=0$}. 
This covers precisely types I$(M^+)$ and I$(M^\infty)$
in the extended Petrov classification, 
see appendix \ref{subsec:Bel-Rob} and remark \ref{rem:q=0}. By proposition \ref{prop:q=0-prop} there is a unique eigenvalue with smallest modulus, say $\lambda_1$, and the Weyl PNDs span the 3d space $e_1^\bot=\langle e_0^a,e_2^a,e_3^a\rangle$. We order the principal tetrad vectors $e_2^a$ and $e_3^a$ such that $|\lambda_1|<|\lambda_2|\leq|\lambda_3|$; then the types I$(M^+)$ and I$(M^\infty)$ correspond to $\l_1\neq 0\lra|\l_2|<|\l_3|$ and $\l_1=0\lra|\l_2|=|\l_3|\lra r_2=r_3$, respectively.  
By \eqref{eq:sqrtri}
we have 
\begin{equation}\label{sqrtri}
\sqrt{r_1}=\sqrt{r_2}+\sqrt{r_3}\neq 0\qquad 
\qquad (|\l_3-\l_2|=|\l_3-\l_1|+|\l_2-\l_1|\neq 0),
\end{equation} 
whence $r_1>r_2+r_3$, and so by \eqref{def-ai} the factors between square brackets in the left hands of Eqs.~\eqref{gek2} and \eqref{gek3} exceed $(1+2a_1+2a_2)r_2>0$ and $(1+2a_1+2a_3)r_3>0$, respectively. Hence these equations are equivalent to $u^1u^2=u^1u^3=0$; by our requirement that at most one $u^i$ vanishes, and using \eqref{def-ai} and \eqref{sqrtri}, we find that 
Eqs.\  \eqref{gek1}-\eqref{gek3} reduce to
\begin{equation}\label{eq:q=0}
\begin{aligned}
&u^1=0,\qquad 2\sqrt{r_3}\,a_3-2\sqrt{r_2}\,a_2=\sqrt{r_2}-\sqrt{r_3},\\
\textrm{i.e.}\quad&u^1=0,\qquad 2\left(|\l_2|-|\l_1|\right)(u^3)^2-2\left(|\l_3|+|\l_1|\right)(u^2)^2=3|\l_1|,
\end{aligned}
\end{equation}
where we have also used \eqref{eq:sqrtri}. Thus the extra observers form two curves lying in $e_1^\bot$. In the type I$(M^+)$ case the pair $(u^2,u^3)\in\R^2$ lies on one of the two branches of a hyperbola, parametrized by $(u^2,u^3)=(b\sinh(\aaa),\pm a\cosh(\aaa)),\,\aaa \in\R$,
where $a^{-2}=\frac{2}{3}\left({|\l_{2}|}/{|\l_{1}|}-1\right),\,b^{-2}=\frac{2}{3}\left({|\l_{3}|}/{|\l_{1}|}+1\right)$; 
note that the major axis corresponds to the eigenvalue $\l_3$ with largest modulus, and $|u^2|<|u^3|$ because of $a>b\Leftrightarrow r_3<r_2$ (see \eqref{eq:sqrtri}); this case is illustrated in the right panel of Fig.~\ref{fig:typeI}. 
In the type I($M^{\infty}$) case the hyperbola degenerates to $u^3=\pm u^2$, such that the 
curves consist of the 
observers $u^{a}=\cosh(\aaa)e_{0}^{a}+\sinh(\aaa)[e_2^a\pm e_3^a]/\sqrt{2},\,\aaa\in\R$ obtained by boosting $e_0^a$ arbitrarily along the two bisectors $\langle e_2^a\pm e_3^a\rangle$ of $\langle e_2^a\rangle$ and $\langle e_3^a\rangle$. From 
\eqref{eq:def-tg-Om-Phi}-\eqref{def-ai} 
and \eqref{sqrtri}-\eqref{eq:q=0} one finds,
for both types I$(M^+)$ and I($M^{\infty}$): 
\begin{equation}\label{q0-v-propto-P}
\tg-\chg=r_2a_2+r_3a_3+\frac{(\sqrt{r_2}-\sqrt{r_3})^2}{4}=\sqrt{r_2r_3}a_0-\frac{r_1}{4}, \quad \vrel{e_0}{u}^a=\frac{2\qp^a}{\sqrt{r_2r_3}a_0}=\frac{8\qp^a}{4(\tg-\chg)+r_1}.  
\end{equation}  

Conversely, compatibility of \eqref{gek1}-\eqref{gek3} with $u^1=0\neq u^2u^3$ or a cyclic permutation hereof requires $q=0$. Hence, {\em when $q\neq 0$ the extra observers have $u^1u^2u^3\neq 0$}.   
\end{ex}

\begin{ex} {\em The case $r_2=r_3\lra|\l_2|=|\l_3|$}. This is type I$(M^-)$ in the extended Petrov classification. By \eqref{rel-ri-q} one has $36q^2=r_1(4r_2-r_1)$ and so $q={\sgn(q)}\sqrt{k(4-k)}\,r_2/6$ with $0<k\equiv r_1/r_2\leq 4$. The subcase $q=0\lra k=4$ corresponds to $\l_1=0$ and gives $a_1=0$ and $a_2=a_3$, cf.\ supra. Take now $q\neq 0$ and suppose that \eqref{gek1}-\eqref{gek3} holds with $a_2\neq a_3$ (implying $u^1u^2u^3\neq 0$). Eq.\ \eqref{gek1} then gives $u^2u^3r_1=6u^0u^1q$; taking the square hereof  yields $(a_0a_1+a_2a_3)k=4a_0a_1$, while compatibility with \eqref{gek2} requires $(a_0a_1+a_2a_3)k=a_1(a_0+a_2+a_3-a_1)$; hence $2a_0+2a_1+1=0$ by \eqref{def-ai}, a contradiction. Hence $a_2=a_3$ and it follows that, when at most one $u^i$ vanishes, the system \eqref{gek1}-\eqref{gek3} is equivalent to 
\begin{equation}\label{r2=r3-extra}
u^3=\delta u^2,\;\delta=\pm 1,\qquad \sgn(q)\sqrt{k(4-k)(1+a_1+2a_2)}(a_1-a_2)
= \delta u^1[ka_1+(3k-4)a_2+k-1]. 
\end{equation}
A detailed analysis\footnote{Taking the square of the second equation gives a polynomial equation $P(a_1|a_2,k)=0$ of degree 3 in $a_1$, where $a_2$ and $k$ are viewed as parameters, and discriminant theory can be used (see e.g.\ theorem 1 in \cite{Yang}). 
} shows that for $k\neq 1$ the second part of \eqref{r2=r3-extra} is solved by $u^1=\epsilon f(u^2)$ and additionally by $u^1=-\epsilon g(u^2)$ when $0<k<2$, where $\epsilon=\sgn(q(1-k))\delta$ and $f,\,g$ are smooth, even functions satisfying $0<f(u^2)<|u^2|<g(u^2),\,u^2\neq 0$ and $0=f(0)<g(0)$. 
For $k=1\lra r_1=r_2=r_3\lra I=0$ one has $u^1=\pm u^2=\pm u^3$, as expected by symmetry and corresponding to the non-smooth limit $f(u^2)=g(u^2)=|u^2|$ in the above; in this case one has   
\begin{equation}\label{vrel-I=0}
\vrel{e_0}{u}^a=\alpha\qp^a,\qquad 1/\alpha=2\chg u^0(a_0+a_1)[u^0+\sgn(qu^1)\delta\sqrt{3a_1}],\quad u^0=\sqrt{1+3a_1}.
\end{equation}
In the limit $k=4\lra q=0$ one finds back $u^1=0,\,u^3=\pm u^2$ (corresponding to $f=0$). Hence there are four and two smooth curves of extra observers when $0<k<2$ and $2\leq k\leq 4$, respectively; see Fig.~\ref{fig:illustr_r2=r3}. 

Conversely, compatibility of \eqref{gek1} with $a_2=a_3$ requires $r_2=r_3$. Hence, {\em when $r_2\neq r_3$ the extra observers have $u^2\neq \pm u^3$, and analogously  $r_1\neq r_2\Rightarrow u^1\neq \pm u^2$ and $r_1\neq r_3\Rightarrow u^1\neq \pm u^3$}.  
\end{ex}

\begin{figure}
	\centering\includegraphics[width=0.55\columnwidth]{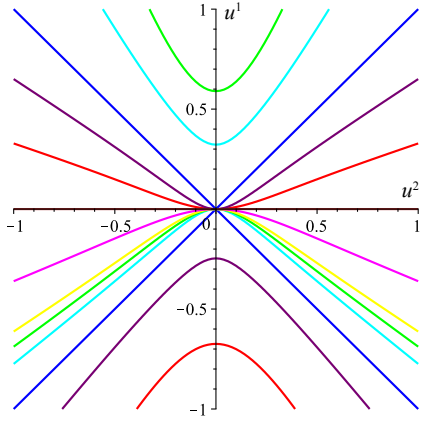}
	\vspace{-5cm}
	\protect\caption{\label{fig:illustr_r2=r3} 
		{Projections of the curves of extra observers $u^a$ satisfying $\vrel{e_0}{u}^a\propto \qp^a$ onto the $(u^2,u^1)$-plane in the degenerate Petrov type I case $r_2=r_3,\,r_1=kr_2,\,q\geq0$, corresponding to Eq.~\eqref{r2=r3-extra} with $\delta=1$, for the values $k=0.2$ (red), $k=0.6$ (purple), $k=1$ (blue), $k=1.5$ (cyan), $k=1.75$ (green), $k=2$ (yellow), $k=3$ (magenta) and $k=4\lra q=0$ (brown). Mirroring about the $u^2$-axis yields the projections of the curves corresponding to $\delta=-1$, for the same values of $k$ and $q$. For $q\leq 0$ the displays correspond to $\delta=-1$ instead, and the $u^2$-mirrors of these to $\delta=1$. In general, for $0<k<2$ and $2\leq k\leq 4$ this yields, respectively, four and two spacetime curves $u^a=\sqrt{1+(u^1)^2+2(u^2)^2}e_0^a+u^1e_1^a+u^2(e_2^a+\delta e_3^a)$ of extra observers  satisfying $\vrel{e_0}{u}^a\propto \qp^a$, different from the special observers \eqref{eq:threecurves} but including $e_0^a$ (corresponding to the origin $u^2=u^1=0$).}}
\end{figure}

\begin{rem}\label{rem:Da} Consider a principal observer $e_0^a$ and an arbitrary non-principal observer $u^a$. The condition $\vrel{e_0}{u}^a\propto \qp^a$ can also be written as 
\begin{equation}\label{eq:e0-in-piu} 
\vrel{e_0}{u}^a\propto\qp^a\quad\lra\quad e_{0}^{a}\in\langle u^{a},\qp^{a}\rangle \quad\lra\quad
\DD^a\equiv[e_0,\qp]^a\equiv \e^{a}{}_{bcd}u^d\qp^c e_0^b
\equiv\tfrac12 [\Eop,\Hop]^a_{\;\;b}e_0^b=0, 
\end{equation}
and thus boils down to the vectors $e_0^a,\,u^a,\,\qp^a$ being linearly dependent (i.e., belonging to a  2-plane), or to $e_0^a$ belonging to the kernel of $[\Eop,\Hop]^a_{\;\;b}$.
However, according to theorem \ref{prop:typeI} and the text preceeding it, \eqref{eq:e0-in-piu} does {\em not} hold for almost all observers $u^a$ in the type I case, and evenly so in the type D case if we consider $e_0^a$ as a given principal observer, leading to a non-vanishing vector $\DD^a=u^0\e^{a}{}_{bcd}u^d\vrel{e_0}{u}^b\qp^c$. In general, one can decompose $\vrel{e_0}{u}^a$ along and orthogonal to $\qp^a$; one easily finds
\begin{equation}\label{vrele0u-decomp}
\vrel{e_0}{u}^a=\nurel_{\parallel \qp}^a+\nurel^a_{\perp \qp}=-\frac{\qp^0\qp^a}{u^0\qp^b\qp_b}+\frac{\epsilon^a{}_{bcd}\qp^b\DD^cu^d}{u^0\qp^b\qp_b}
\end{equation}
[compare to \eqref{cond-v-propto-P} and \eqref{alpha-expr}]. In appendix \ref{app:principal}.2 it is proven that 
\begin{equation}\label{P0>0}
\qp^a\neq 0\quad\Rightarrow\quad -u^0\qp^0> 0,
\end{equation}
saying that {\em for a given non-principal observer $u^a$ and principal observer $e_0^a$ the relative velocity $\vrel{e_0}{u}^a$ always has a non-zero, positive component along $\qp^a$}; i.e., $\nurel_{\parallel \qp}^a=\alpha\qp^a$ with $\alpha>0$. 
Note that the formula \eqref{eq:princ-all-vec-D} in the Petrov type D case is a special instance of 
\eqref{vrele0u-decomp}, with $\nurel^a_{\perp \qp}=\nurel_{\parallel \eb}^a$ and $e_0^a=u'^a=u'^a(\varphi)$,
while \eqref{eq:vrel-D}, \eqref{q0-v-propto-P} and \eqref{vrel-I=0} confirm \eqref{P0>0}.
\end{rem}

\section{Examples}

\label{sec:Examples}

\subsection{Electromagnetic field of a spinning charge\label{sub:Electromagnetic-field-ofSC}}

The EM field generated by a charged spinning body with
charge $Q\neq 0$ and dipole moment of magnitude $\mu$ oriented along the $z$-axis,
in flat spacetime,
is described in spherical coordinates {[}$ds^{2}=-dt^{2}+dr^{2}+r^{2}(\sin(\theta)^{2}d\phi^{2}+d\theta^{2})${]}
by the vector potential $A^{a}=(Q/r){\partial}_{t}^{a}+(\mu/r^{3}){\partial}_{\phi}^{a}$. 
Consider the following \ON vector basis:
\begin{equation}
e_0^a=\partial_{t}^{a},\quad \erad^a=\partial_{r}^{a},\quad \etheta^a=\frac{1}{r}\partial_{\theta}^{a},\quad \ephi^a=\frac{1}{r\sin(\theta)}\partial_{\phi}^{a}.
\end{equation}
The electric and magnetic fields measured by the laboratory (or `static') observers 
$u_{{\rm lab}}^{a}=\partial_{t}^{a}$ are 
\begin{equation}\label{EMsphere}
E^{a}
=\frac{Q}{r^{2}}\erad^a\ ,
\qquad B^{a}
=\frac{\mu}{r^3}[2\cos(\theta)\erad^a+\sin(\theta)\etheta^a]\ .
\end{equation}
Assuming 
$r>a\equiv2\mu/Q$ (see footnote 10 of \cite{CosWylNat16})
the field is everywhere non-null. The Poynting vector 
measured by $u_{{\rm lab}}^{a}$ is 
\begin{figure}
	\includegraphics[width=0.9\columnwidth]{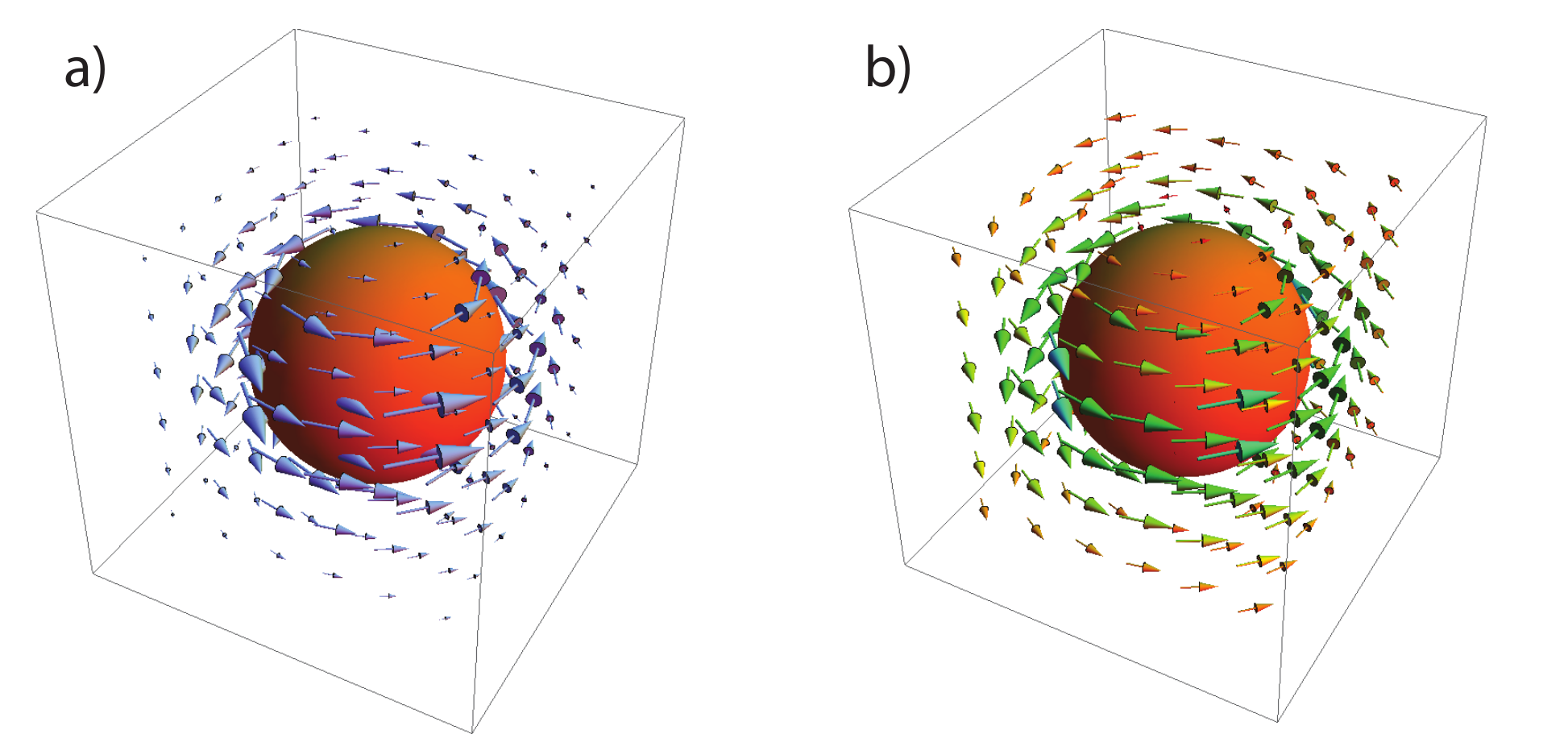}
	
	\protect\caption{\label{fig:SpinningCharge}a) Poynting vector $\sp^a=\sp^\phi\partial_{\phi}^a$
		for the electromagnetic field of a spinning charge, as measured by
		the static observers $u_{{\rm lab}}^{a}=\partial_{t}^{a}$ (the ``laboratory''
		frame); the field energy flows stationarily along circular loops parallel
		to the equatorial plane. b) The principal observers $\ubig^{a}$, for
		which the Poynting vector vanishes, have a velocity with a component
		$\nurel_{\parallel \sp}^a=\omega_{(p'=0)}\partial^a_{\phi}$ along
		$\sp^a$ given by (\ref{eq:Omega_nop}), and an arbitrary component
		along $\eb^a$. The special case of the canonical principal observers
		$\ubig^a=\us^a$, moving along circles tangent to $\sp^a$
		(i.e., $\nurel^a=\nurel_{\parallel \sp}^a$),
		is represented.}
\end{figure}
{} 
\begin{equation}\label{eq:poynting_explicit}
\sp^a
=\frac{Q\mu\sin(\theta)}{4\pi r^{5}}\ephi^a\,.
\end{equation}
Hence there is a non-vanishing flux of EM field energy;
such energy flows \emph{stationarily} around closed loops parallel
to the equatorial plane, as illustrated in Fig. \ref{fig:SpinningCharge}a.

{According to} \eqref{eq:vrel-EM} and \eqref{eq:princ-all-vec-EM} the EM principal observers $\ubig^a$
(for which $\sp'^a=0$) move relative to $u_{{\rm lab}}^{a}$
with a velocity that has a fixed component parallel to $\sp^a$
given by 
$\nurel_{\parallel \sp}^a\equiv \sp^a/(\te+\che)$, where 
\[
\te=\frac{Q^{2}r^{2}+\mu^{2}(1+3\cos^{2}(\theta))}{8\pi r^{6}};\qquad\che=\frac{|\IEM|}{8\pi},\quad \IEM=\frac{Q^{2}r^{2}-\mu^{2}(1+3\cos^{2}(\theta))-4iQr\mu\cos(\theta)}{r^{6}}\ ,
\]
which amounts to an angular velocity $\nurel_{\parallel p}^{\phi}=\ubig^{\phi}/\ubig^{t}\equiv\omega_{(p'=0)}$
given by 
\begin{align}
& \omega_{(p'=0)}=\frac{2Q\mu}{Q^{2}r^{2}+\mu^{2}(1+3\cos^{2}(\theta))+\sqrt{(Q^{2}r^{2}-\mu^{2})^{2}+2\mu^{2}(5q^{2}r^{2}+3\mu^{2})\cos^{2}(\theta)+9\mu^{2}\cos^{4}(\theta)}}\ .\label{eq:Omega_nop}
\end{align}
The relative velocity $\vrel{\ubig}{u}^{a}$ may also have an
	arbitrary component $\nurel^a_{\parallel \eb}$ parallel to the vector $\eb^{a}$, 
	as given in \eqref{eq:calc-eb-EM}. In the
special case of the canonical principal observers $\ubig^{a}=\us^{a}$
for which $\nurel^a_{\parallel \eb}=0\lra\nurel^{a}=\nurel_{\parallel \sp}^a
$ (which are observers in circular
motion), this velocity field is illustrated in Fig.\ \ref{fig:SpinningCharge}b.
Finally, in the equatorial plane $\theta=\pi/2$, which is a purely
electric region, we have $\omega_{(p'=0)}=\mu/(Qr^{2})$, and $\eb^a=\E^a r^3/\sqrt{Q^2r^2-\mu^2}$, 
yielding the velocities of the observers measuring $\Bpr^{a}=0$ obtained
in \cite{CosWylNat16}.

\subsection{Kerr-Newman spacetimes\label{sub:Kerr}}

Consider a charged rotating Kerr-Newman black hole in Boyer-Lindquist
coordinates outside the event horizon, $r>M+\sqrt{M^{2}-(a^{2}+Q^{2})}$,
where the mass $M$, angular momentum per unit mass $a$ and charge
$Q$ satisfy $M^{2}>a^{2}+Q^{2}$, and with $0<\tq<\pi$. The metric is $g_{ab}=-\Omega_{a}^{0}\Omega_{b}^{0}+\Omega_{a}^{1}\Omega_{b}^{1}+\Omega_{a}^{2}\Omega_{b}^{2}+\Omega_{a}^{3}\Omega_{b}^{3}$ with
\begin{align}
& \Omega_{a}^{0}=\frac{\sqrt{\Delta}}{\rho}(d_{a}t-a\sin^{2}(\tq)d_{a}\phi),\quad\Omega_{a}^{1}=\frac{\rho}{\sqrt{\Delta}}d_{a}r,\quad\Omega_{a}^{2}=\frac{\sin(\tq)}{\rho}[ad_a t-(r^{2}+a^{2})d_{a}\phi],\quad \Omega_{a}^{3}=\rho d_{a}\tq,
\nonumber\\
\label{eq:Kerr-Newman}
& \rho^2\equiv r^{2}+a^{2}\cos^{2}(\tq),\quad\Delta\equiv r^{2}-2Mr+a^{2}+Q^{2}.
\end{align}
This represents a gravitational field of non-null Einstein-Maxwell type, with corresponding EM potential 1-form  $A_{a}=Qr(a\sin^{2}(\theta) d_{a}\phi-d_{a}t)/\rho^{2}$ (see, e.g.,~\cite{Gravitation}).

Consider the ``laboratory'' 
observer with 4-velocity 
\begin{equation}
u^a=u_{{\rm lab}}^{a}\equiv \frac{\rho}{\sqrt{\Delc}}\partial_t^a,\qquad \Delc\equiv \Delta-a^2\sin^2(\tq)=r^{2}-2Mr+a^{2}\cos^2(\tq)+Q^{2}\,.
\end{equation}
The natural \ON tetrad associated to this observer is $(u_{{\rm lab}}^{a},\erad^a,\etheta^a,\ephi^a)$ with 
\begin{equation}\label{ON-KN}
\erad^a=\frac{\sqrt{\Delta}}{\rho}\partial_r^a,\qquad \etheta^a=\frac{1}{\rho}\partial_{\tq}^a,\qquad
\ephi^a=\frac{\sqrt{\Delc/\Delta}}{\rho\sin(\tq)}\partial_{\phi}^a+\frac{a\sin(\tq)(Q^2-2Mr)}{\rho\sqrt{\Delta\Delc}}\partial_t^a=\rho\sin(\tq)\sqrt{\frac{\Delta}{\Delc}}\nabla^a\phi,
\end{equation}
where $\nabla^{a}\phi\equiv g^{ab}d_{b}\phi$ is the gradient of the coordinate $\phi$. 
The observer measures 
electric and magnetic fields 
$E^{a}=F_{\;\; b}^{a}u_{{\rm lab}}^{b}$ and $B^{a}=\star F_{\;\;b}^{a}u_{{\rm lab}}^{b}$ given by 
\begin{align}
\label{eq:Ea-Kerr-Newman}
&\E^a
=\frac{Q}{\rho^{4}\sqrt{\Delc}}\left\{\sqrt{\Delta}[r^{2}-a^{2}\cos^{2}(\tq)]\erad^a-a^{2}r\sin(2\tq)\etheta^a\right\},\\
\label{eq:Ba-Kerr-Newman}
&\B^a
=\frac{Qa}{\rho^{4}\sqrt{\Delc}}\left\{2r\sqrt{\Delta}\cos(\tq)\erad^a+\left[r^{2}-a^{2}\cos^{2}(\tq)\right]\sin(\tq)\etheta^a\right\}\,.
\end{align}
These can be assembled into the complex vector
\begin{equation}\label{f-KN}
\f^a=\E^a-i\B^a=\frac{Q}{(r+ia\sin(\tq))^2}\v^a,\quad \v^a=\frac{\sqrt{\Delta}\erad^a-ia\sin(\tq)\etheta^a}{\sqrt{\Delc}},
\end{equation}
where $\v^a$ is a unit vector. From \eqref{eq:def-IEMb} and \eqref{eq:def-che} relevant EM invariants can be calculated:
\begin{equation}
\sqrt{\IEM}=\frac{Q}{[r+ia\cos(\tq)]^{2}}\ ,\qquad\che=\frac{|\IEM|}{8\pi}=\frac{Q^{2}}{8\pi\rho^{4}},
\end{equation}
while the EM energy density and Poynting vector relative to $u_{{\rm lab}}^{a}$ are
\begin{equation}
\te=\frac{\Delta+a^{2}\sin^{2}(\tq)}{\Delta-a^{2}\sin^{2}(\tq)}\che,\qquad 
\sp^{a}=\frac{aQ^{2}\sqrt{\Delta}\sin(\tq)}{4\pi\rho^{4}[\Delta-a^2\sin^2(\tq)]}\ephi^{a}\,.\label{eq:te-spa-KerrNewman}
\end{equation}

Concerning the Weyl tensor, the electric and magnetic parts relative to $u_{{\rm lab}}^{a}$ can be assembled into the complex tensor
$\Q_{ab}=\E_{ab}-i\H_{ab}$. In the orthonormal triad $B$ given in \eqref{ON-KN} the associated operator $\Qop$ is represented by the $3\times 3$ matrix
\begin{equation}\label{Qrep}
[\Qop]_B=\lambda\begin{bmatrix}
-2-3a^2\sin^2(\tq)/\Delc&3ia\sin(\tq)\sqrt{\Delta}&0\\
3ia\sin(\tq)\sqrt{\Delta}&1+3a^2\sin^2(\tq)/\Delc&0\\
0&0&1
\end{bmatrix},\qquad \l=\frac{M[r-ia\cos(\tq)]-Q^{2}}{\rho^{2}[r+ia\cos(\tq)]^{2}}. 
\end{equation}
This satisfies \eqref{typeD-charact}, and thus the spacetime is of Petrov type D (also in the pure Kerr limit case $Q=0$) with a double eigenvalue $\lambda$ and simple eigenvalue $-2\lambda$.\footnote{The exceptional spacetime points with $\l=0\lra C_{abcd}=0$ are given by 
$\{\tq=\frac{\pi}{2},r={Q^2}/{M}\}$; they lie on the equatorial plane and, 
with $a\geq 0$, 
outside the event horizon precisely when $a/M\leq (Q/M)^{2}-(Q/M)^{4}$.}
Clearly, $\ephi^a$ spans an eigendirection with eigenvalue $\lambda$. The eigendirection corresponding to 
$-2\lambda$ is easily found to be spanned by $\v^a$ given in \eqref{f-KN},
which is proportional to $\f^a$ when $Q\neq 0$. By virtue of \eqref{cond-SF=SW}
we have thus found back the well-known fact that the Kerr-Newman spacetime is a Petrov type D doubly aligned non-null Einstein-Maxwell field~\cite{SKMHH}. Therefore the results of Sec.~\ref{subsub:doubly} apply. By \eqref{eq:limit-D} and \eqref{Qrep} the proper gravitational super-energy density is given by 
\begin{equation}
\chg=\frac{3M^2}{2\rho^8}\left[a^2\cos^2(\tq)+\left(r-{Q}^2/{M}\right)^2\right]
\end{equation}
while the super-energy density and super-Poynting vector measured by $u_{{\rm lab}}^{a}$ are 
\begin{equation}\label{eq:tg-qpa-KerrNewman}
\frac{2\tg}{\chg}=3\left[\frac{\Delta+a^2\sin^2(\tq)}{\Delta-a^2\sin^2(\tq)}\right]^2-1\,,\quad 
\qp^a=\frac{9M^2a\sqrt{\Delta}[\Delta+a^{2}\sin^{2}(\tq)]\left[a^2\cos^2(\tq)+\left(r-{Q}/{M}\right)^2\right]}{2\rho^{8}[\Delta-a^{2}\sin^{2}(\tq)]^{2}}\ephi^{a},
\end{equation}
in agreement with \eqref{eq:tg-Pg-doublealigned}. 
Thus the Poynting and super-Poynting vectors are aligned eigenvectors of $\Qop$ with eigenvalue $\lambda$ (see remark \ref{rem:qpa-eigen-D}), which are moreover orthogonal to the surfaces of constant $\phi$ (i.e., their spatial component is purely azimuthal), just as the Poynting vector \eqref{eq:poynting_explicit} in the case of an EM spinning charge in Minkowski spacetime (but the Poynting vectors \eqref{eq:poynting_explicit} and \eqref{eq:te-spa-KerrNewman} die off as $r^{-5}$, whereas the super-Poynting vector \eqref{eq:tg-qpa-KerrNewman} dies off as $r^{-7}$). 
On the EM side this means that there is a
flow of energy around closed circular loops parallel to the equatorial
plane. Motivated by the EM analogy, some authors interpret
$\qp^{a}$ as representing a flux of ``super-energy'' \cite{Sen00,Bel62,FerSae12,AlfonsoSE};
in the spirit of such interpretation, one would say that in the Kerr and Kerr-Newman spacetime
there is a stationary flow of super-energy around closed loops parallel
to the equatorial plane; see Fig.\ \ref{fig:KN}a. 
\begin{figure}
	\includegraphics[width=0.9\textwidth]{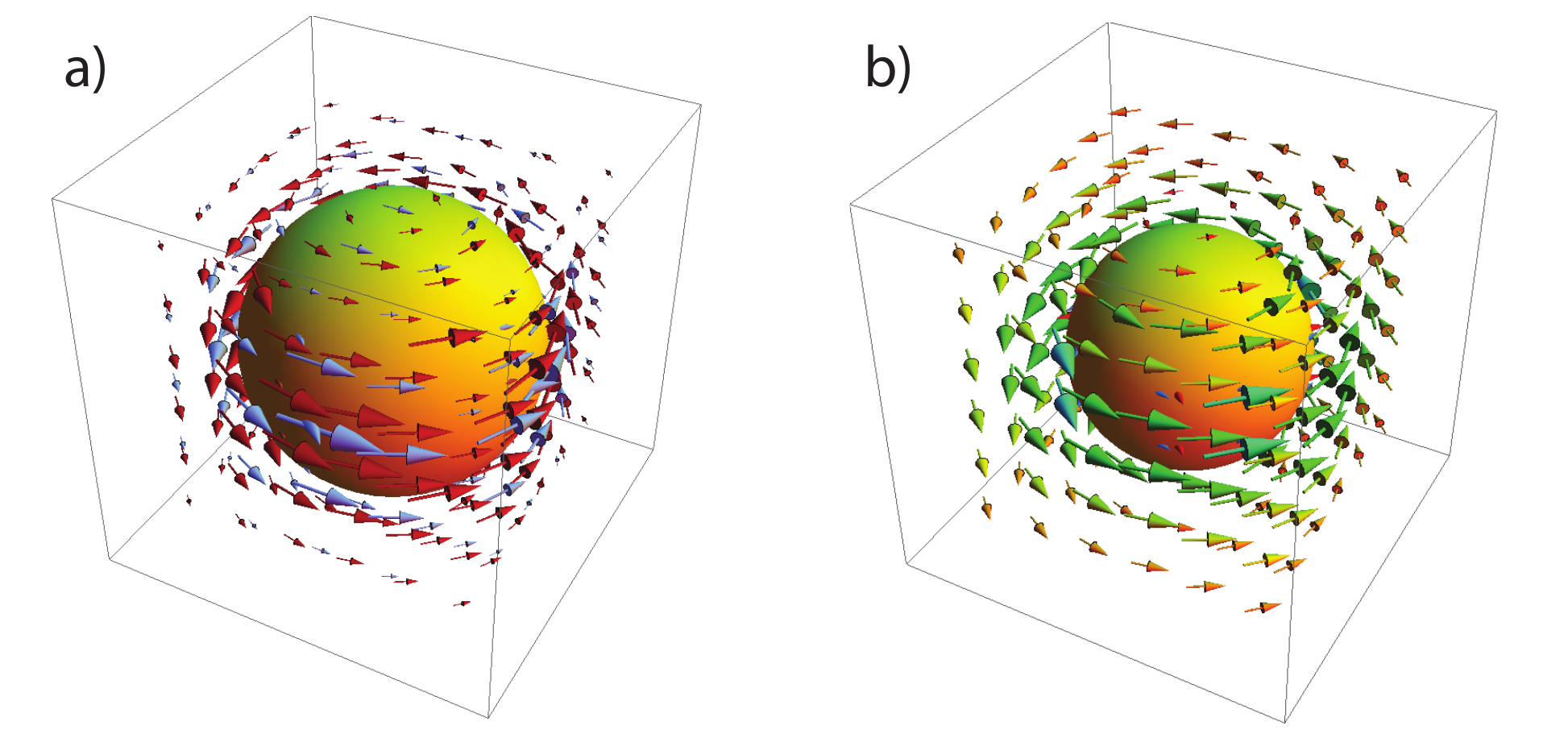}\protect\caption{\label{fig:KN}Kerr-Newman spacetime. a) Red arrows: Poynting vector
		$\sp^a$ as measured by the laboratory observers $u_{{\rm lab}}^{a}\propto\partial_{t}^{a}$,
		signaling a stationary energy flow along circular loops parallel to
		the equatorial plane; blue arrows: super-Poynting vector $\qp^a$
		as measured by $u_{{\rm lab}}^{a}$, suggesting an analogous stationary
		flow of super-energy. b) The principal observers $u'^{a}$ (for which
		{both} $\sp'^a$ and $\qp'^a$ vanish) move
		along $\sp^a\propto\qp^a$ with angular velocity \eqref{eq:Omega_noSPKN},
		and have an arbitrary velocity along $\eb^a\propto\partial_r^a$.
		The special case of the canonical principal observers $u'^{a}=\us^a$
		(Carter observers), in circular motion with relative velocity 
		parallel to $\sp^a$
		and $\qp^a$ 
		[$\vrel{u'}{u}^a=\nurel_{\parallel \sp}^a=\nurel_{\parallel \qp}^a$]
		is represented.}
\end{figure}

We want to determine the boosts to the EM and Weyl principal
observers. Since $\Sigma=\SF=\SW$ these should coincide. Indeed, from \eqref{eq:vrel-EM}-\eqref{eq:urel-EM} and \eqref{eq:vrel-D}-\eqref{eq:urel-D} we find $\Aem=\AD=1+2a^2\sin^2(\tq)/\Delc$,   
\begin{equation}\label{vusu}
\cosh(\psi_{\SF})=\cosh(\psi_{\SW})=\sqrt{\Delta/\Delc},\qquad 
\nurel_{\parallel \sp}^a
=\nurel_{\parallel \qp}^a=\frac{a\sin(\tq)}{\sqrt{\Delta}}\ephi^{a}=:\nurel_{\parallel \ephi}^a
\end{equation}
and the expression for the principal observer obtained by projecting $u_{{\rm lab}}^{a}$ onto $\Sigma$:
\begin{equation}\label{Carter}
\us^{a}=\frac{1}{\rho\sqrt{\Delta}}\left[(r^{2}+a^{2})\partial_{t}^{a}+a\partial_{\phi}^{a}\right].
\end{equation}
From \eqref{eq:calc-eb-EM}, \eqref{eq:calc-eb-D} and \eqref{f-KN} we simply have 
\begin{equation}
\eb^a=\eb_F^a=\eb_C^a=\Re(\v^a)=\sqrt{\Delta/\Delc}\erad^a,
\end{equation}
and so according to \eqref{eq:princ-all-vec-EM} or \eqref{eq:princ-all-vec-EM} the (EM or Weyl) principal observers, which are all $u'^{a}\in\SS$,
move relative to the laboratory observers
$u_{{\rm lab}}^{a}$ with relative velocity 
$
\vrel{\ubig}{u_{{\rm lab}}}^{a}=\nurel_{\parallel \ephi}^a+\nurel_{\parallel\erad}^{a}
$,
where $\nurel_{\parallel \ephi}^a$ is the \emph{fixed} component parallel
to the (super-)Poynting vector as given in \eqref{vusu} and $\nurel_{\parallel\erad}^{a}$
an \emph{arbitrary} radial component, and are obtained 
by arbitrarily boosting $\us^{a}$ along $\eb^{a}\propto\erad^a$. 
They consist thus of all observers $u'^{\alpha}$ with angular velocity 
\begin{equation}\label{eq:Omega_noSPKN}
\omega_{(\mathcal{P}'=0)}=\omega_{(p'=0)}=\frac{u'^{\phi}}{u'^{0}}=\frac{a}{(a^{2}+r^{2})}\ ,
\end{equation}
and with an arbitrary radial component $u'^{r}$. The special case $u'^{r}=0$ (that is, $u'^{a}=\us^{a}$) corresponds to a congruence of principal observers
in circular motion, called the \emph{Carter observers}~\cite{Bini,SemerakStationaryFrames}, which are plotted in Fig.\ \ref{fig:KN}b. At each point the Carter observer is the one that realizes both the EM Wheeler result (theorem \ref{thm:EM}) and the gravitational Wheeler analogue (theorem \ref{thm:D}), i.e., it is the principal observer that moves, with respect to $u_{{\rm lab}}^{a}$, in the direction of $\sp^{a}\propto\qp^{a}$. 
Comparing with the principal observers of the spinning charged body
in flat spacetime illustrated in Fig.\ \ref{fig:SpinningCharge}, and assuming
a uniform charge and mass distribution so that $\mu=(Q/2M)J=Qa/2$,
we have that, asymptotically, the angular velocities (\ref{eq:Omega_nop})
and \eqref{eq:Omega_noSPKN} match up to a factor 2.\footnote{\label{fn:SpheriodalMatching}In the equatorial plane $\theta=\tq=\pi/2$
	such matching (up to the factor 2) is actually exact since
	the metric \eqref{eq:Kerr-Newman} is written in spheroidal
	coordinates which, in the limit $M\rightarrow0$, are related to the
	spherical coordinates of Sec. \ref{sub:Electromagnetic-field-ofSC}
	by Eq.~(28) of \cite{VelocityKerr}.} In this context we note that in the work by Rosquist
\cite{VelocityKerr} it was shown that the Kerr-Newman metrics can
be obtained by a re-scaling of an orthonormal tetrad field in Minkowski
spacetime, constructed from spheroidal coordinates in differential rotation,
each spheroidal shell $r=constant$ rotating rigidly. It turns out
that the angular velocity of the shells is precisely $\omega_{(\mathcal{P}'=0)}$; therefore, the principal observers are precisely those comoving
with Rosquist's shells, plus those moving radially with respect to
them. Finally, we note that taking the limit $M\rightarrow0$ in \eqref{eq:Kerr-Newman}
one obtains an electromagnetic field in the flat Minkowski metric
written in spheroidal coordinates (cf.\ footnote \ref{fn:SpheriodalMatching}),
whose principal observers retain the same properties (such field having,
however, a complicated source whose form is not enlightening, see e.g.\ \cite{KN_Source}
and references therein).

\subsection{Kasner spacetimes\label{sub:Kasner}}

Consider the Kasner vacuum spacetimes, which are spatially homogeneous
of Bianchi type I. In suitable coordinates they admit the diagonal
line element~\cite{SKMHH}
\begin{equation}\label{eq:Kasner}
ds^{2}=-dt^{2}+t^{2\p_{1}}(dx^{1})^{2}+t^{2\p_{2}}(dx^{2})^{2}+t^{2\p_{3}}(dx^{3})^{2},
\end{equation}
where $(\p_{1},\p_{2},\p_{3})\in\R^{3}$ is a parameter triple satisfying
$\sum_{i=1}^{3}\p_{i}=\sum_{i=1}^{3}\p_{i}^{2}=1$. The Weyl eigenvalues
are $\l_{i}=-\p_{j}\p_{k}/t^{2}$, 
so the Weyl tensor is purely electric at each point, implying $q=0$ (see remark \ref{rem:q=0}). By permuting the coordinates we may assume 
	$\p_{3}\leq\p_{2}\leq\p_{1}$. The metric \eqref{eq:Kasner} represents
	Minkowski spacetime for $\p_{3}=\p_{2}=0,\,\p_{1}=1$, and
	a Petrov type D plane symmetric vacuum for $\p_{3}=-1/3,\,\p_{2}=\p_{1}=2/3$. On excluding these cases henceforth
	the Petrov type is I($M^+$),
	with $-1/3<\p_{3}<0<-\p_{3}<\p_{2}<2/3<\p_{1}<1$ and $|\p_{3}|<|\p_{2}|<|\p_{1}|$,\footnote{This can be readily inferred from $\sum_{i=1}^{3}\p_{i}=\sum_{i=1}^{3}\p_{i}^{2}=1$,
		or from the parametrization $\p_{i}=\frac{1}{3}[1+2\cos(\hh+(1-i)2\pi/3)]$,
		$0<\hh<\pi/3$, which is compatible with $\p_{3}\leq\p_{2}\leq\p_{1}$.} such that $0<|\l_{1}|<|\l_{2}|<|\l_{3}|$. 

	The essentially unique principal tetrad vectors
	lie along the coordinate vector fields at each point: $e_{0}^{a}=\partial_{t}^{a}$
	and $e_{i}^{a}=\partial_{x^{i}}^{a}/t^{\p_{i}}$. Observers of the type \eqref{eq:threecurves}, moving relative to $e_{0}^{a}$ along one of the $e_{i}^{a}$, measure, by \eqref{spec-tg-P-v}, a super-Poynting vector 
	\begin{equation*}
	\qp^a=-\frac18 t^{-4}\p_i^2(\p_j-\p_k)^2\sinh(4\aaa)[\sinh(\aaa)e_0^a+\cosh(\aaa)e_i^a]\propto\nurel^{a}(e_{0},u)\ ,
	\end{equation*}
	i.e., along the direction of $\nurel(e_{0},u)^a$; hence, their boost back to the principal observer $e_{0}^{a}$ is along $\qp^a$, see the left panel of Fig.~\ref{fig:typeI}. The other non-principal observers $u^{a}$ for which $\nurel(e_{0},u)^a\parallel\mathcal{P}^{a}$ 
	have $u^1=0$ and relative velocities parallel to $u^2e_2^a+u^3e_3^a$ with $u^{2}=[2(1-\p_{1}/\p_{3})/3]^{-1/2}\sinh(\aaa),\,
	u^{3}=\pm[2(\p_{1}/\p_{2}-1)/3]^{-1/2}\cosh(\aaa)$, see example \ref{ex:q=0} and the right panel of Fig~\ref{fig:typeI}.

Take now an observer moving along the bisector
		of $e_1^a$ and $e_2^a$: $u'^{a}=\cosh(\aaa)e_0^a+\sinh(\aaa)e_{12}^a$, where $e_{12}^2=(e_1^a+e_2^a)/\sqrt{2}$. Then the vector defined in \eqref{eq:e0-in-piu} is 
\begin{equation*}
\DD^a=\frac{3}{16}t^{-4}\p_1\p_2\p_3(\p_1-\p_2)\sinh(2\aaa)^2e_3^a\ ,
\end{equation*}
which is non-zero when the Petrov type is I$(M^+)$, again in accordance with theorem \ref{prop:typeI};\footnote{As expected, $\DD^a=0$ when the spacetime is Minkowski ($\p_{3}=\p_{2}=0,\,\p_{1}=1$) or type D ($\p_{3}=-1/3,\,\p_{2}=\p_{1}=2/3$).} hence we have $\nurel(e_{0},u')^a\nparallel\mathcal{P}'^{a}$ (i.e., the boost that takes  $u^{a}$ into the principal observer	$e_{0}^{a}$ is not along
	$\qp'^{a}$, as in the left panel of Fig. \ref{fig:typeI}).
	
	Given the electric and magnetic parts of the Weyl tensor as measured by any observer $u^{a}$ one can easily apply the algorithm at the end of Sec.\ \ref{subsec:grav-typeI}:
	one readily finds the eigenvalues $\l_{i}=-\p_{j}\p_{k}/t^{2}$ of
	the corresponding endomorphism $\Qop$, calculates the unit eigenvectors
	$\v_{i}^{a}$ by \eqref{eq:def-Ri}-\eqref{eq:calc-via}, and finds the Weyl principal
	tetrad from \eqref{eq:vrele0u-1}-\eqref{eq:inverse-trafo-ei1}.

\section{Summary and discussion}

\label{sec:disc}

In electromagnetism (EM) and general relativity theory, principal observers
are defined to be those observers for which the (super-)Poynting vector
associated to the Faraday, resp., the Weyl tensor vanishes. This precisely
happens in the EM case when the electric and magnetic fields are
aligned, and in the gravitational case when the electric and magnetic
parts of the Weyl tensor commute. Another criterion is that the (super-)energy
density attains minimum value. On identifying an observer at
a point with its 4-velocity vector, the instantaneous existence
and number of principal observers simply depends on the algebraic
type of the \Max or Weyl tensor at the point. Principal observers
precisely exist when the relevant tensor is of `diagonal' type.
In the (generic) non-null EM case and the (non-generic) Petrov type D
gravitational case the principal observers' velocities are the unit,
future-pointing timelike vectors of a distinguished 2-plane, viz.\ the
timelike principal plane, which induces a natural structure (observer
geometry) on the set of observers at the point. In the (generic) Petrov
type I gravitational case there is a unique principal observer.

We have focused on the relation of general observers to principal
ones, employing a novel calculus on the complexified rest space of a general observer. In a non-null EM field the `EM Wheeler result' holds: 
given an arbitrary observer $u^{a}$, measuring a Poynting vector $p^a$, a principal observer $\umw^{a}$ can always be reached through a boost in the direction of $p^a$.
We have clarified the underlying geometric nature of this result by showing that $\umw^{a}$ is nothing but the orthogonal projection $\us^a$ of $u^a$ onto the timelike principal plane. As our main achievement we have demonstrated that the `gravitational Wheeler analogue' (obtained by changing Poynting to super-Poynting in the statement of the EM Wheeler result) holds in a Petrov type D gravitational field, giving a deduction that emphasized the mathematical analogy between the two situations. Moreover, the (super-)energy
density depends on and increases with the rapidity relative to $\us^{a}$.
Hence the observer geometries in these two cases are simply linked to
(super-)Poynting vectors and (super-)energy densities measured by
observers. This simplicity is not shared by the Petrov type I gravitational
case, where the level sets of the super-energy density are rather
complicated, and we proved that only for a one-dimensional variety
of observers the (unique) principal observer lies in the plane spanned
by the observer and the relative super-Poynting vector. 


In all three cases (non-null EM, and Petrov type D and I gravitational cases) we have outlined algorithms to compute all principal observers from the electric and magnetic fields or electric and magnetic parts of the Weyl tensor as measured by an arbitrary observer. 

As a final note, we briefly comment on the (unavoidable) question
of the physical significance of the super-Poynting vector. Because it is
a flux of super-energy, its physical meaning, similarly to that of the super-energy itself, 
remains unclear; a \emph{solid} connection to observable effects 
has yet to be established. It is tempting however to draw a parallelism
with the situation for the Poynting vector $p^{a}$, prior to general
relativity. Although $p^{a}$ was, in some dynamical situations (such
as electromagnetic radiation), known to be a physically measurable
quantity (e.g.~by the momentum imparted on a mirror), the Poynting
vector in stationary settings such as the spinning charge in Fig.~\ref{fig:SpinningCharge}, indicating a flow of energy circulating around
in closed loops, has been questioned due to its strangeness,\footnote{Indeed $p^{a}$ looks very strange sometimes. Another example is the
case of a wire carrying a current, where outside the wire $p^{a}$
is orthogonal to the wire, pointing inwards. That implies that the
electric energy does not reach the electrons by flowing through the
wire, but instead by going from the battery into a wide area of space
around, and then inward to the wire \cite{Feynman}.} and often dismissed as immaterial, since no observable
consequences were known \cite{Feynman}. General relativity changed the picture, in
that all forms of energy and energy currents act as sources for the
gravitational field, and are therefore, in theory, measurable. In particular,
$p^{a}$ should be measurable through the frame-dragging effect it
generates \cite{BonnorDragging,HerreraGonzalezPachonRueda,HerreraBarreto}.
It should also (being an energy current) generate a ``Magnus''-like
force on spinning test bodies \cite{Magnus}. The super-Poynting vector
is likewise believed to accompany gravitational radiation~\cite{Pirani,Bel62,FerSae12,AlvarezSenovillaI,AlvarezSenovillaII,Clifton2014},
and, in stationary
settings such as those in Fig. \ref{fig:KN}, where (similarly to
its EM counterpart) it circulates around in closed loops, observable
consequences have been sought, namely an attempted link to frame-dragging\footnote{Such connection is however not well established. A known counter-example
\cite{HerreraGodel} is the G\"{o}del universe, where a rigid frame exists
relative to which $\mathcal{P}^{\alpha}$ vanishes everywhere, but
still the ``gravitomagnetic field'' $H^{a}=2\omega^{a}$ ($\omega^{a}\equiv$
vorticity), and thus frame-dragging, is present \cite{CosWylNat16}.
This shows that $\mathcal{P}^{a}$ is (at best) not the sole source
of frame-dragging. It should be noticed in this context that $\mathcal{P}^{a}$ is
built on \emph{tidal} fields (namely $H_{ab}$), whereas frame-dragging
is directly related instead to $H^{a}$, and it is important to distinguish
between the two (see \cite{CosWylNat16}).} \cite{HerreraCarrotPrisco}.

\section*{Acknowledgements}

We thank J.~M.~M.~Senovilla for useful comments and remarks. L.W.~was supported through a BOF research grant of Ghent University and through the Research Council of Norway, Toppforsk grant no. 250367: Pseudo-
Riemannian Geometry and Polynomial Curvature Invariants: Classification, Characterisation and Applications. L.F.C.~and J.N.~were supported by FCT/Portugal through projects UID/MAT/04459/2019 and UIDB/MAT/04459/2020, and grant SFRH/BDP/85664/2012.

\appendix

\section{Orthonormal frames in self-dual bivector and tangent space}

\label{sec:orth-frames}

Three unitary, orthogonal
self-dual bivectors $\Ucal_{ab}^{i}$ form an {\em \ON frame}
of $\Sp$ {[}$-\frac{1}{4}(\Ucal^{m})^{ab}\Ucal_{ab}^{n}=\d_{mn}$, $m,n=1,2,3${]}
which is moreover {\em oriented} if $\Ucal_{ab}^{i}=-i\Ucal_{ac}^{j}(\Ucal^{k})_{\;\;b}^{c}$.
The relation 
\begin{equation}\label{eq:Uiab-e0ei}
\Ucal_{ab}^{i}=[2(e_{0})_{[a}(e_{i})_{b]}]^{\dag}=2(e_{0})_{[a}(e_{i})_{b]}+2i(e_{j})_{[a}(e_{k})_{b]}\,,\qquad i=1,2,3\,
\end{equation}
determines a bijection between restricted \ON tetrads $(e_{0}^{a},e_{i}^{a})$
of $T_{p}M$ and oriented \ON frames $(\Ucal_{ab}^{i})$ of
$\Sp$. This can be seen as a consequence of the group isomorphism between the proper orthochronous
Lorentz group and the group of proper orthogonal transformations of
$\Sp$, see for instance \cite{SKMHH} or \cite{FerrSaezP1}. Here we present a simple and direct geometric proof; cf.~\cite[pp.~246-248]{Penrosebook2}. It is
easy to verify that the triple $(\Ucal_{ab}^{i})$ constructed as in
\eqref{eq:Uiab-e0ei} from a restricted \ON tetrad $(e_{0}^{a},e_{i}^{a})$ is an oriented
\ON frame of $\Sp$. Conversely, we need to show that any
oriented \ON frame $(\Ucal_{ab}^{i})$ of $\Sp$ can be written as in \eqref{eq:Uiab-e0ei}, where 
$(e_{0}^{a},e_{i}^{a})$ is moreover uniquely determined.
By \eqref{eq:Xcal-norm} and $\Ucal_{ab}^{i}=\U_{ab}^{i}-i\st{\U}_{ab}^{i}$
the orthonormality condition $-\frac{1}{4}(\Ucal^{m})^{ab}\Ucal_{ab}^{n}=\d_{mn}$
is equivalent to the three sets of conditions 
\begin{equation}
\IUi=1,\;\;i=1,2,3,\qquad(\st{\U}^{i})^{ab}\U_{ab}^{j}=0,\;\;i\neq j,\qquad({\U}^{i})^{ab}\U_{ab}^{j}=-(\st{\U}^{i})^{ab}\st{\U}_{ab}^{j}=0,\;\;i\neq j.\label{eq:Sp-ONF}
\end{equation}
The first set implies that $(\U^{i})^{ab}=2r_{i}^{[a}s_{i}^{b]},(\st{\U}^{i})^{ab}
=2{r'}_{i}^{[a}{s'}_{i}^{b]}$
for each $i$ separately, where the $(r_{i}^{a},s_{i}^{a},{r'}_{i}^{a},{s'}_{i}^{a})$
are restricted \ON frames. For $i=1,2,3$ the blades of $\U^i_{ab}$ and $\st{\U}^i_{ab}$ are denoted by $\SS_i$ and $\SS_i^\bot$, respectively. Consider now a fixed $i$. For each $j\neq i$
the second (third) set of \eqref{eq:Sp-ONF} yields $\e_{abcd}r_{i}^{a}s_{i}^{b}r_{j}^{c}s_{j}^{d}=0$
($\e_{abcd}r_{i}^{a}s_{i}^{b}r_{j}'{}^{c}s'_{j}{}^{d}=0$), which
means that $\SS_i$ and $\SS_j$ ($\SS_j^{\bot}$)
together span a 3-dimensional subspace of tangent space, and thus
intersect in a line $l_{ij}$ ($l'_{ij}$); here $l'_{ij}$ is spacelike
since the blade $\SS_j^{\bot}$ is spacelike, and thus the line
$l_{ij}$ is timelike since together with $l'_{ij}$ it spans the
timelike blade $\SS_i$. Applying this to $i=2,\,j=1$ and writing
$e_{0}^{a}$ for the unique future-pointing unit tangent vector 
along $l_{21}$ we infer $(\U^{1})^{ab}=2e_{0}^{[a}e_{1}^{b]}$ and
$(\U^{2})^{ab}=2e_{0}^{[a}e_{2}^{b]}$, where $e_{1}^{a}$ and $e_{2}^{a}$
are uniquely defined unit spacelike vectors which are both orthogonal
to $e_{0}^{a}$ and also mutually orthogonal by $({\U}^{2})^{ab}\U_{ab}^{1}=0$.
Hence we obtain a uniquely defined restricted \ON tetrad $(e_{0}^{a},e_{1}^{a},e_{2}^{a},e_{3}^{a})$.
Next consider $i=3$ and $j=1,2$. Since the timelike lines $l_{3j}$
belong to $\SS_j$ they have unit tangent vectors $u_{j}^{a}=\alpha_{j}e_{0}^{a}+\beta_{j}e_{j}^{a},\,j=1,2$,
where $\alpha_{j},\,\beta_{j}\in\mathbb{R}$ and $\a_{1}\a_{2}\neq0$.
Assume $(\b_{1},\b_{2})\neq(0,0)$; then the lines $l_{31}$ and $l_{32}$
would not coincide and thus would span $\SS_3$; hence $(\U^{3})^{ab}=Cu_{1}^{[a}u_{2}^{b]}$
for certain $C\neq0$, but then $({\U}^{3})^{ab}\U_{ab}^{j}=0,\,j=1,2$
gives $\b_{1}\a_{2}=\b_{2}\a_{1}=0$ and thus the contradiction $\a_{1}\a_{2}=0$.
Hence $\b_{1}=\b_{2}=0$ and $l_{31}=l_{32}=l_{21}$, implying $(\U^{3})^{ab}=2e_{0}{}^{[a}\vvec^{b]}$
with $\vvec^{a}$ unit spacelike and orthogonal to $e_{0}^{a}$. By $({\U}^{3})^{ab}\U_{ab}^{j}=0$, $j=1,2$,
it follows that $\vvec^{a}$ is also orthogonal to $e_{1}^{a}$ and $e_{2}^{a}$,
and thus equals $\pm e_{3}^{a}$. If we moreover require the \ON
frame $(\Ucal_{ab}^{i})$ to be oriented by $\Ucal_{ab}^{i}=-i\Ucal_{ac}^{j}(\Ucal^{k})_{\;\;b}^{c}$,
then $\vvec^{a}=e_{3}^{a}$ and we arrive at $(\Ucal^{i})^{ab}=(2e_{0}^{[a}e_{i}^{b]})^{\dag}$,
which concludes the proof.

Geometrically, $e_0^a$ is thus the observer along the intersection of the blades $\SS_i$ of the simple 2-forms $\U^i_{ab},\,i=1,2,3$, while for fixed $i$, $e_i^a=(\U^i)^a_{\;\;b}e_0^b=(\Ucal^i)^a_{\;\;b}e_0^b$ lies along the joint intersection of $\SS_i,\,\SSb_j,\,\SSb_k$. 

\section{Notes on the Weyl and Bel-Robinson tensors}
\label{sec:app}

\subsection{Petrov classification of the Weyl tensor in terms of $\Q^a_{\;\;b}$}
\label{subsec:Petrov-basic}

The Petrov classification of the Weyl tensor into the six Petrov types
can be formulated in terms of the complex tensor $\Q^a_{\;\;b}\equiv\Ccal^a{}_{cbd}u^{c}u^{d}=$
relative to any observer $u^{a}$, the projector $\I_{b}^{a}\equiv\delta_{b}^{a}+u^{a}u_{b}$ onto {the observer's complexified rest space} $\Cup$,
and the complex invariants $I=\Q_{\;\;b}^{a}\Q_{\;\;a}^{b}$ and $J=\Q_{\;\;b}^{a}\Q_{\;\;c}^{b}\Q_{\;\;a}^{c}$: 
\begin{itemize}
\item Petrov type I: $I^{3}\neq6J^{2}$; 
\item Petrov type II: $I^{3}=6J^{2}\neq0,\,(\Q_{\;\;c}^{a}-\l \I_{c}^{a})(\Q_{\;\;b}^{c}+2\l \I_{b}^{c})\neq0,\,\l\equiv-J/I$; 
\item Petrov type D: $I^{3}=6J^{2}\neq0,\,(\Q_{\;\;c}^{a}-\l \I_{c}^{a})(\Q_{\;\;b}^{c}+2\l \I_{b}^{a})=0,\,\l\equiv-J/I$; 
\item Petrov type III: $\Q_{\;\;c}^{a}\Q_{\;\;d}^{c}\Q_{\;\;b}^{d}=0\neq\Q_{\;\;c}^{a}\Q_{\;\;b}^{c}$; 
\item Petrov type N: $\Q_{\;\;c}^{a}\Q_{\;\;b}^{c}=0\neq\Q_{\;\;b}^{a}$; 
\item Petrov type O: $\Q_{\;\;b}^{a}=0$. 
\end{itemize}
Regarding the Petrov type D case,  
$(\Q_{\;\;c}^{a}-\l \I_{c}^{a})(\Q_{\;\;b}^{c}+2\l \I_{b}^{a})=0$ with $\l\neq 0$ actually implies {$I=6\l^2,\,J=-6\l^3$} and thus $I^3=6J^2$ and $\l=-J/I$. 
By \eqref{eq:Cop-Qop-equiv} the endomorphism $\Qop:\vvec^a\mapsto \Q^a_{\;\;b}\vvec^b$ of $\Cup$ 
(for any $u^a$) has the same characteristic polynomial 
\begin{equation}
c(x)=\prod_{k=1}^{3}(x-\l_{k})=x^{3}-\tfrac{1}{2}I\,x-\tfrac{1}{3}J
\label{eq:charpoly}
\end{equation}
and the same minimal polynomial polynomial $m(x)$ 
as the endomorphism $\Cop:\Xcal_{ab}\mapsto-\frac{1}{4}{\Ccal}_{ab}{}^{cd}\Xcal_{cd}$ of $\Sp$. The roots $\l_k$ of the cubic $c(x)$ are the eigenvalues of these endomorphisms. The Petrov types are characterized by the degree $d$ of $m(x)$ and the number $N$ of different eigenvalues, as summarized in
Table \ref{Table:Ptypes}. {The endomporphisms} $\Cop$ and $\Qop$ are diagonalizable
iff $N=d$, i.e.\ iff the Petrov type is I, D or O. 
Petrov types III, N and 0 have $\l=0$ as their only triple eigenvalue; Petrov types II and D have a double eigenvalue $\l=-J/I$ and a simple eigenvalue $-2\l$; the three simple eigenvalues for Petrov type I can be calculated by Cardano's formula:
\begin{equation}
\l_{k}=\B^{\frac{1}{3}}e^{\frac{2(k-1)\pi i}{3}}+\frac{I}{6}\,\B^{-\frac{1}{3}}e^{-\frac{2(k-1)\pi i}{3}},\qquad\B\equiv\frac{J}{6}+\frac{1}{6}\sqrt{\frac{-\DD}{6}},\quad\DD\equiv I^{3}-6J^{2},\quad k=1,2,3.\label{eq:roots}
\end{equation}
Here one can take any choice for the (complex) cube
and square roots, provided that in the subcase $I=0$ one chooses
$\sqrt{J^{2}}=+J$, such that always $\B\neq0$.

\begin{table}
\begin{centering}
\begin{tabular}{c|cccccc}
Petrov type & I & II & III & D & N & 0\tabularnewline
\hline 
$d$ & 3 & 3 & 3 & 2 & 2 & 1\tabularnewline
$N$ & 3 & 2 & 1 & 2 & 1 & 1 \tabularnewline
\end{tabular}
\par\end{centering}

\protect\caption{Petrov types: {$d\equiv$ degree of $m(x)$, $N\equiv$ number of different eigenvalues of $\Cop$ or $\Qop$}.}
\label{Table:Ptypes} 
\end{table}

\subsection{Bel-Robinson endomorphism and degenerate Petrov type I}
\label{subsec:Bel-Rob}

The Bel-Robinson tensor $T^{ab}{}_{cd}$ defines an endomorphism on the 9d vector space of trace-free symmetric tensors; it has eigenvalues $t_i,\,\tau_i,\,\tauc_i,\,i=1,2,3$ which can be parametrized by $t_i$ or $p_i\equiv \Re(\tau_i)$~\cite{FerSae09,FerSae10,FerSae12}:
\begin{align}\label{eq:ti-taui-def}
&\tau_i\equiv\l_j\lc_k\equiv p_i+iq,\qquad 
t_i\equiv |\l_i|^2=-(p_i+p_j)\quad\Leftrightarrow \quad p_i=\frac12(t_i-t_j-t_k),\\
&\label{eq:rel-q2}
q^2
=p_2p_3+p_3p_1+p_1p_2
=\tfrac12(t_2t_3+t_3t_1+t_1t_2)-\tfrac14(t_1^2+t_2^2+t_3^2).
\end{align}
Alternatively, one can use the invariants $r_i\equiv|\l_j-\l_k|^2,\,i=1,2,3$, coinding with (C.6) in \cite{FerSae12}: 
\begin{align}
&\label{setrpt}r_i
=t_j+t_k-2p_i
=2(t_j+t_k)-t_i
=-(4p_i+p_j+p_k)\\
\Leftrightarrow \quad
&\label{setptr}t_i=\tfrac{1}{9}(2r_j+2r_k-r_i)\quad 
\Leftrightarrow\quad 
p_i=\tfrac{1}{36}(r_j+r_k-5r_i).
\end{align}
Combining \eqref{eq:rel-q2} with \eqref{setptr} gives \eqref{rel-ri-q}. Also note that  
\begin{equation}\label{rj-rk}
t_j-t_k=p_j-p_k=(r_j-r_k)/3.
\end{equation}
By \eqref{eq:I-def-int} and \eqref{eq:ti-taui-def}, \eqref{eq:rel-q2}, \eqref{rj-rk} one obtains 
\begin{equation}
\label{eq:|I|^2}
|I|^2
=\sum_{j=1}^3\l_j^2\sum_{k=1}^3\lc_k^2
=\sum_{1\leq j,k\leq 3}(\l_j\lc_k)^2
=\sum_{i=1}^3\left(t_i^2+\tau_i^2+\tauc_i^2\right)
=4\sum_{i=1}^3 p_i^2-4q^2
=2\sum_{(ijk)}(t_j-t_k)^2.
\end{equation}
The middle expression confirms $|I|^2=T^{ab}{}_{cd}T^{cd}{}_{ab}\equiv 4\alpha^2$~\cite{FerSae09}, while the penultimate expression and $16\chg^2=(\sum_{i=1}^3t_i)^2=4\sum_{i=1}^3p_i^2+8q^2$ yield the identity \eqref{eq:ids}. 
From \eqref{eq:ti-taui-def}, \eqref{rj-rk}, \eqref{eq:|I|^2} it follows that 
\begin{equation}\label{I=0}
I=0\neq J\quad\lra\quad |\l_1|=|\l_2|=|\l_3|\neq 0\quad\lra\quad r_1=r_2=r_3\neq 0,
\end{equation}
which is the Petrov type I subcase with eigenvalues $\l_k=e^{\frac{i2\pi k}{3}}\alpha, \;\alpha\in\R$ and $q=3t_1^2/2=r_1^2/12\neq 0$.

The eigenvalue degeneracy of the Bel-Robinson endomorphism was studied in \cite{FerSae09} and easily follows from \eqref{eq:ti-taui-def}. Clearly, all eigenvalues are zero if the Petrov type is O, N or III (where $I=J=0$ and all $\l_i=0$). Excluding these cases, degeneracy occurs precisely when $q=0$ or
$t_j=t_k$ for some $j\neq k$, say 
$t_2=t_3\lra |\l_2|=|\l_3|\lra r_2=r_3$. By \eqref{eq:rel-q2}-\eqref{setptr} the intersection gives two cases:
\begin{align}
\label{II-D}I^3=6J^2\neq 0:\qquad &\l_1=-2\l_{2}=-2\l_{3}\neq 0\quad\lra\quad t_1=4t_2=4t_3\neq 0\quad\lra\quad r_1=0;\\
\label{J=0}J=0\neq I:\qquad &\l_1=0\neq\l_2=-\l_3\quad\lra\quad t_1=0\quad\lra\quad r_1=4r_2=4r_3\neq 0,
\end{align}
where the first covers types II and D and the last is the Petrov type I subcase with a vanishing Weyl eigenvalue, see \eqref{eq:J-def-int}. 
Consider the dimensionless invariant  
\begin{align}\label{eq:M-def}
M\equiv {I^3}/{J^2}-6\;\in\;\C\cup\{\infty\}.
\end{align}
The cases \eqref{I=0}, \eqref{II-D}, \eqref{J=0} respectively correspond to $M=-6,\,0,\,\infty$ and are precisely those cases with only three different Bel-Robinson eigenvalues, where other degenerate cases have six different eigenvalues. 
For $J\neq 0$, $M$
can be written as~\cite{McIntosh1}
\begin{align}\label{eq:M-expr}
M=\frac{2}{9}\mu^2,\qquad
\mu
=\frac{(\l_3-\l_2)(\l_1-\l_3)(\l_2-\l_1)}{\l_1\l_2\l_3}.
\end{align}
Multiplying numerator and denominator of $\mu$ with $\lc_1\lc_2\lc_3$ and using \eqref{eq:ti-taui-def}-\eqref{eq:rel-q2} we obtain
\begin{align}
\mu
=\frac{(\tau_2-\tauc_3)(\tau_3-\tauc_1)(\tau_1-\tauc_2)}{t_1t_2t_3}
=\frac{(t_2-t_3)(t_3-t_1)(t_1-t_2)}{t_1t_2t_3}-{i}\frac{q[4q^2+(t_1+t_2+t_3)^2]}{2t_1t_2t_3}.
\end{align}
Hence $\mu$ is real (purely imaginary) precisely when $q=0$ ($t_j=t_k$ for some $j$ and $k$); since $q=\Im(\l_j\lc_k)=\Im(\l_j/\l_k)$ whenever $\l_k\neq 0$ this proves, in a straighforward way, the known result that $M$ is real non-negative (non-positive) precisely when the ratio of any two non-zero Weyl eigenvalues is real (there are at least two Weyl eigenvalues with equal moduli). It follows that degeneracy of the Bel-Robinson eigenvalues occurs precisely when $M$ is real or infinite. 

For Petrov type I ($M\neq 0$) the corresponding cases have been denoted by I($M^+$), I$(M^{-})$ and I$(M^{\infty})$ in the extended Petrov classification of the Weyl tensor by Arianrhod and McIntosh~\cite{McIntosh3}. They considered the Penrose-Rindler disphenoid on the sphere of null directions, the vertices of which represent the four simple Weyl PNDs and which has 3 pairs of opposite edges with equal lengths $a_1,a_2,a_3$~\cite[pp.~249-252]{Penrosebook2}, and found that there is degeneracy of two kinds precisely when $M$ is real or infinite:
\begin{itemize}
	\item one has $a_j=a_k\lra t_j=t_k$ and thus there are four edges of equal length if the extended Petrov type is I$(M^{\infty})$ or I$(M^{-})$ with $M\neq -6$, and all edges are equal if $M=-6\lra I=0$;
	\item the disphenoid is planar, i.e., the PNDs span a 3d vector space precisely when $q=0$, corresponding to types I($M^+$) and I$(M^{\infty})$~\cite{McIntosh2}. 
\end{itemize}
Consider the case $q\equiv \Im(\l_j\lc_k)=0$, which is thus characterized by the existence of $\sigma\in[0,\pi[$ and {\em real} numbers $\l_{k}'$ (summing to zero) such that $\l_{k}=e^{i\sigma}\l'_{k},\,k=1,2,3$. Then
\begin{equation}
I=e^{2i\sigma}|I|,\qquad J=e^{3i\sigma}J',\,J'=3\l'_{1}\l'_{2}\l'_{3}\in\R,\qquad\l_{k}=\sqrt{I/|I|}\,\l_{k}',\qquad\M+6=\frac{|I|^{3}}{J'^{2}}\,,\label{eq:def-sigma}
\end{equation}
and rewriting \eqref{eq:charpoly} in terms of $\l'_{k},\,|I|,\,J'$ results in a trigonometric alternative to the expressions
\eqref{eq:roots}: 
\begin{align}
& \l_{k}=\sqrt{\frac{2I}{3}}\cos\left(\frac{\arccos L_{\sigma}}{3}-k\frac{2\pi}{3}\right),\quad L_{\sigma}=\frac{\textrm{sgn}(e^{-3i\sigma}J)}{\sqrt{1+\M/6}},\quad2\sigma=\arg(I)\in[0,2\pi[,\qquad k=1,2,3.\label{eq:L-alpha}
\end{align}
Since the $\l_k'$ sum to zero there is a unique $\l'_{k_0}$ with smallest absolute value, where if $|\l'_1|<|\l_2'|\leq |\l_3'|$ then either $\l_2'<\l'_1\leq 0<\l'_3$ or $\l_3'<0\leq \l_1'<\l'_2$. 
Using $|\l_k|=|\l_k'|$ and $\sqrt{r_k}=|\l_i-\l_j|=|\l_i'-\l_j'|$, and invoking theorem 2 of \cite{FerSae04}, we obtain:

\begin{prop}\label{prop:q=0-prop}
If the Petrov type is I and $q=0$ then there is a unique Weyl eigenvalue with smallest modulus, say $\l_1$, and the 3d subspace of $T_pM$ spanned by the four Weyl PNDs is $e_{1}^{\bot}=\langle e_{0}^{a},e_{2}^{a},e_{3}^{a}\rangle$, where
$(e_{0}^{a},e_{i}^{a})$ is the essentially unique Weyl principal
tetrad. Furthermore, if we arrange $|\l_3|\geq |\l_2|$ then  
type I$(M^{+})$ is characterized by $|\l_3|>|\l_2|>|\l_1|>0$ and type I$(M^{\infty})$ by $|\l_3|=|\l_2|>|\l_1|=0$, and
\begin{equation}
\label{eq:sqrtri}
|\l_3|=|\l_1|+|\l_2|,\qquad \sqrt{r_1}=\sqrt{r_2}+\sqrt{r_3}
=|\l_2|+|\l_3|>\sqrt{r_2}
=|\l_3|+|\l_1|>\sqrt{r_3}
=|\l_2|-|\l_1|\,.
\end{equation}
\end{prop}


\section{Principal observer conditions}\label{app:principal}

Let the Petrov type be I or D at $p$, let $(e_\alpha^a)\equiv(e_0^a,e_i^a)$ be a Weyl principal tetrad, and $u^a=u^\alpha e_\alpha^a=u^0e_0^a+u^ie_i^a$ an arbitrary observer. Recalling \eqref{def-ai} we write $a_\alpha\equiv(u^\alpha)^2$, 
implying $a_0=1+a_1+a_2+a_3$. We also define 
\begin{equation*}
\bd_i\equiv \l_j-\l_k
\end{equation*}
and recall $r_i=|\bd_i|^2$ and $q=\Im(\l_j\lc_k)$. Note that $\bd_3=-(\bd_1+\bd_2)$, and $(\bd_1,\bd_2)\neq (0,0)\lra(r_1,r_2)\neq (0,0)$ by the type I or D assumption. 
Below we show that the principal observers ($\qp^a=0$) are precisely those with minimal super-energy density $\tg=\chg$, and precisely those with $\qp^0=0$.\\  

{\em \ref{app:principal}.1 Minimal super-energy density}.
Consider the expression \eqref{eq:def-tg-Om-Phi} for the super-energy density:
\begin{equation*}
\tg=\chg+\Omega,\qquad \Omega=\Phi+12qA,\quad \Phi=A^ir_i,
\end{equation*}
where
\begin{equation*}
A^{i}= a_0a_i-a_ja_k,\qquad A= u^0u^1u^2u^3.
\end{equation*}
In the Petrov type I case (all $r_i>0$) the unique principal observer is the one with all $u^i=0$, while in the Petrov type D case $r_i=0\lra\alpha_i=0$ (any fixed $i$) the principal observers are those with $u^j=u^k=0$, and it follows in both cases that $\Omega=0\lra\tg=\chg$ if $u^a$ is principal. In \cite{FerSae12} Ferrando and Saez proved that $\Omega\geq 0$ as a consequence of $\Phi\geq 0$ and 
\begin{equation}\label{def-R}
R\equiv \Phi^2-(12qA)^2\geq 0.
\end{equation}
To assert the last inequality they observed that, in view of \eqref{rel-ri-q}, $R$ can be viewed as a quadratic form in the variables $r_i$:
\begin{equation}\label{eq:def-Rij}
R=R(r_1,r_2,r_3)=R^{mn}r_mr_n,\qquad R^{ii}=(A^i)^2+4A^2=(a_0a_i+a_ja_k)^2,\; R^{ij}=A^iA^j-4A^2\,(i\neq j),
\end{equation}
and that $R$
is non-negative since the principal minors of all orders are non-negative:
\begin{align*}
&\Delta^i=R^{ii}\geq 0,\quad \Delta^{ij}=4A^2\left(1+a_i+a_j\right)^2(a_i+a_j)^2\geq 0,\\
&\Delta^{123}=64A^4\left[\sum_{(ijk)}a_i^2(a_j+a_k)+\sum_{i<j}a_ia_j+2a_1a_2a_3\right]\geq 0.
\end{align*}
Based on this result applied to Weyl-like tensors and on the Bergqvist-Lankinen theorem~\cite{BerLan04} the second part of \eqref{eq:tg2-chg2} was proved in \cite{FerSae13}; as affirmed in our main text it follows that $\Omega=0\lra\tg=\chg$ holds {\em only} for principal observers. We now give an elementary proof of this fact by simply elaborating the above, where we distinguish between the cases $A\neq 0$ and $A=0$:
\begin{enumerate}
	\item $A\neq 0$ (all $u^i\neq 0\lra a_i>0$). Then $u^a$ is not a Weyl principal observer. The quadratic form $R$ is now positive definite since its leading principal minors $\Delta^1,\Delta^{12}$ and $\Delta^{123}$ are strictly positive. If $\Omega=0$ then $R=0$, whence all $r_i=|\bd_i|^2=0$, a contradiction. 
	Hence $\Omega>0$ in this case.     
	\item $A=0$. Then at least one of the $u^i$'s vanishes, say $u^3=0\lra a_3=0$. Then $\Omega=A^i|\bd_i|^2$, 
	and by 
	the triangle inequality  
	we obtain
	\begin{equation}\label{eqOmega}
	\begin{aligned}
	\Omega&=(1+a_1+a_2)\left(a_1|\bd_1|^2+a_2|\bd_2|^2\right)-a_1a_2|\bd_1+\bd_2|^2\\
	&\geq (1+a_1+a_2)\left(a_1|\bd_1|^2+a_2|\bd_2|^2\right)-a_1a_2\left[|\bd_1|+|\bd_2|\right]^2\\
	&=a_1|\bd_1|^2+a_2|\bd_2|^2+\left(a_1|\bd_1|-a_2|\bd_2|\right)^2\geq 0.
	\end{aligned}
	\end{equation}
	Hence $\Omega=0\Lra a_1\bd_1=a_2\bd_2=0$, and
	since $(\bd_1,\bd_2)\neq (0,0)$ this gives the two qualitatively different cases $\bd_1=u^2=u^3=0$ and $u^1=u^2=u^3=0$. In the former case the Petrov type is D, and in both cases  $u^a$ is a principal observer. This concludes the proof.
\end{enumerate}
{\em \ref{app:principal}.2 $\qp^0=0$}. Let $\qp^a$ be the super-Poynting vector relative to $u^a$.  By \eqref{eq:qa-expans1} we have 
\begin{equation}\label{eqa}
-u^0\qp^0= \Psi+12qB,\qquad \Psi=B^ir_i,
\end{equation}
where
\begin{equation*}
B^i=a_0(A^i-a_i/2)=a_0[a_i(a_0-1/2)-a_ja_k],\qquad B=(a_0-1/4)A.
\end{equation*}
We now show that $\qp^0=0$ implies $\qp^a=0$. In fact, we prove more strongly that \eqref{P0>0} holds, which is equivalent to
\begin{equation}\label{P0-ineq}
-u^0\qp^0\geq 0,\qquad -u^0\qp^0=0\;\lra\;\qp^a=0.
\end{equation}
To this end we first show that $\Psi\geq 0$, analogously as in appendix C of \cite{FerSae12}. By \eqref{setrpt} one calculates that
\begin{equation}
\Psi=-Q^ip_i,\qquad Q^i=(4a_0-2)a_i+\tfrac12 (a_j+a_k)+(a_j-a_k)^2\geq 0.
\end{equation} 
By \eqref{eq:ti-taui-def}-\eqref{eq:rel-q2} one has $t_3=-(p_1+p_2)$ and
$-p_3=(q^2-p_1p_2)/t_3\geq p_1p_2/(p_1+p_2)$, and so 
$$\Psi\geq -Q^1p_1-Q^2p_2+Q^3\frac{p_1p_2}{p_1+p_2}=\frac{Q}{t_3},\qquad Q=Q^1p_1^2+Q^2p_2^2+(Q^1+Q^2-Q^3)p_1p_2.$$
The quadratic form $Q$ in the variables $p_1$ and $p_2$ has principal minors that are non-negative, namely $P^1,P^2\geq 0$ and determinant 
$
\Delta_Q=\frac{9}{4}\sum_{(ijk)}a_ja_k(1+2a_j+a_k)+9a_1a_2a_3(4a_0-1)\geq 0
$. Hence $Q\geq 0$ and thus $\Psi\geq 0$, as we wanted to show. 
Similarly to \eqref{def-R} we can now define $S\equiv \Psi^2-(12qB)^2$, which by \eqref{eq:rel-q2} becomes a quadratic form in the variables $r_i$ and has components as in \eqref{eq:def-Rij} with the symbols $R,\,A$ replaced by $S,\,B$. 
The principal minors of $S$ are calculated to be
\begin{align*}
&\Delta^i=S^{ii}=(B^i)^2+4B^2\geq 0,\quad \Delta^{ij}=a_0^2B^2\left(1+2a_i+2a_j\right)^2(a_i+a_j)^2\geq 0,\\
&\Delta^{123}=4B^4\left[6A+8A\sum_{i=1}^3a_i+a_0\sum_{(ijk)}a_ja_k(1+a_j+a_k)(1+a_j+a_k)\right]\geq 0,
\end{align*}
such that $S\geq 0$. Hence $\Psi\geq |12qB|$ and by \eqref{eqa} the first part of  \eqref{P0-ineq} is proven. For the second part we argue as before:
If $A\neq 0$ then $u^a$ is not principal and $\Delta^1,\Delta^{12}$ and $\Delta^{123}$ are strictly positive, hence $S>0$ and so $-u^0\qp^0>0$; if $A=0$ with $u^3=0\lra a_3=0$ we obtain, analogous to \eqref{eqOmega}: 
	\begin{align*}
-\qp^0/u^0&=(1/2+a_1+a_2)\left(a_1|\bd_1|^2+a_2|\bd_2|^2\right)-a_1a_2|\bd_1+\bd_2|^2\\
&\geq\tfrac12\left[a_1|\bd_1|^2+a_2|\bd_2|^2\right]+\left(a_1|\bd_1|-a_2|\bd_2|\right)^2\geq 0,
\end{align*}
such that $\qp^0=0$ is only possible for a principal observer. This concludes the proof.



\begin{thebibliography}{10}
	
	
\bibitem{Belnotes} Bel L {\em Comp. R. Acad. Sci. (Paris)} \textbf{246},
3015 (1958); \textbf{247}, 1094 (1958); \textbf{247} 2096 (1958);
\textbf{248}, 1297 (1959); \textbf{248}, 2561 (1959) 

\bibitem{BelPhD} Bel L {\em Ph.D. Thesis} C.D.U et S.E.D.E.S Paris
5e 

\bibitem{Sen00} Senovilla J M M \Journal{Class. Quantum Grav.}{17}{2799}{2000} 

\bibitem{Matte} Matte A \Journal{Canadian J. Math.}{5}{1}{1953} 

\bibitem{LandauLifshitz}Landau L D, Lifshitz E M, {\em The classical
	theory of fields}, $4^{th}$ Ed. (New York: Elsevier, 1975) 

\bibitem{AlfonsoSE} Garc\'{i}a-Parrado G\'{o}mez-Lobo A {\em Class.
	Quant. Grav.} \textbf{25}, 015006 (2008) 

\bibitem{Komar}Komar A. \Journal{Phys. Rev. }{164}{1595}{1967} 

\bibitem{Hawking} Hawking S W \Journal{Astrophys. J.}{145}{544}{1966} 

\bibitem{Teyssandier}Teyssandier P, {[}gr-qc/9905080{]}

\bibitem{Garecki}Garecki J \Journal{Acta. Phys. Pol. B}{8}{159}{1977} 

\bibitem{Pirani} Pirani F \Journal{Phys. Rev.}{105}{1089}{1957}

\bibitem{Bel62} Bel L \Journal{Cah. Phys.}{16}{59}{1962};
English translation: \Journal{Gen. Rel. Grav.}{32}{2047}{2000} 

\bibitem{FerSae12} Ferrando J J and S\'{a}ez J A \Journal{Class. Quantum
	Grav.}{29}{075012}{2012} 

\bibitem{AlvarezSenovillaI} Fernández-\'{A}lvarez F and Senovilla Jos\'{e} M. M. \Journal{Phys. Rev. D}{101}{024060}{2020}

\bibitem{AlvarezSenovillaII} Fernández-\'{A}lvarez F and Senovilla Jos\'{e} M. M., [arXiv:2007.11677]

\bibitem{Clifton2014} Clifton T, Gregoris D and Rosquist K, \Journal{Class. Quantum
	Grav.}{31}{105012}{2014} 

\bibitem{CN2014}Costa L F and Nat\'{a}rio J \Journal{Gen. Rel. Grav.}{46}{1792}{2014}

\bibitem{Wyl08} Wylleman L \Journal{Class. Quamtum Grav.}{25}{172001}{2008}

\bibitem{SKMHH} Stephani H, Kramer D, MacCallum M A H, Hoenselaers
C and Herlt E, \emph{Exact Solutions to Einstein's Field Equations
(Second Edition)} (Cambridge: Cambridge University Press, 2003) 

\bibitem{Kinnersley} Kinnersley W \Journal{J. Math. Phys.}{10}{1195}{1969}

\bibitem{Overview-D-alinged} Pleba\'{n}ski J F and Hacyan S \Journal{J. Math. Phys.}{20}{1004}{1979}; Garc\'{i}a D A and Pleba\'{n}ski J F \Journal{J. Math. Phys.}{23}{1963}{1982}; Debever R, Kamran N and McLenaghan R G \Journal{Bull. Acad. Roy. Belg. Cl. Sci.}{68}{592}{1982}; idem \Journal{J. Math. Phys.}{25}{1955}{1984}; Garc\'{i}a D A \Journal{J. Math. Phys.}{25}{1951}{1984} 


\bibitem{Synge} Synge J L, \emph{Relativity: the special theory} (North-Holland Publishing Co., Amsterdam, 1956), pp.\ 334-335

\bibitem{MisWhe57} Misner C W and Wheeler J A \Journal{\em Ann.
Phys. (N.Y.)}{2}{525}{1957} 

\bibitem{Whe77} Wheeler J A \Journal{Phys. Rev. D}{16}{3384}{1977} 

\bibitem{Gravitation} Misner C W, Thorne K S and Wheeler J A, \emph{Gravitation} (W.~H.~Freeman and Co., San Francisco, 1973)

\bibitem{Bini} Bini D, Jantzen R T and Miniutti G \Journal{Int.
J. Mod. Phys. D}{11}{1439}{2002} 

\bibitem{McIntosh3} Arianrhod R and McIntosh C B G \Journal{Class.
	Quantum Grav.}{9}{1969}{1992}



\bibitem{Penrosebook1} Penrose R and Rindler W, \emph{Spinors and
Spacetime} vol 1 (Cambridge: Cambridge University Press, 1986), pp.\ 2--4 



\bibitem{Jantzen} Jantzen R T, Carini P and Bini D  \Journal{Ann. Phys.}{215}{1}{1992}, formatted with corrections on \text{arXiv:gr-qc 0106043} 


\bibitem{Hall} Hall G S, \emph{Symmetries and curvature structure
in general relativity} (Singapore: World Scientific, 2004), p.\ 174 







\bibitem{Ruse} Ruse H S \Journal{Proc. London. Math. Soc.}{41}{302}{1936}

\bibitem{Coll} Coll B \Journal{Ann. Fond. Louis de Broglie}{29}{247}{2004}

\bibitem{ColFer} Coll B and Ferrando J J \Journal{J. Math. Phys.}{30}{2918}{1989} 

\bibitem{BolosIntrinsic} V. Bol\'os, {\em Commun. Math. Phys.} \textbf{273},
217 (2007)

\bibitem{HerOrtWyll13} Hervik S, Ortaggio M and Wylleman L \Journal{Class.
	Quantum Grav.}{30}{165014}{2013}


\bibitem{Milson04} Milson R, Coley A, Pravda V and Pravdov\'{a} A {\em
Int. J. Geom. Meth. Mod. Phys.} \textbf{2} 41 (2005)


\bibitem{CosWylNat16} Costa L F, Wylleman L and Nat\'{a}rio J, \text{arXiv}:
1603.03143 



\bibitem{FerSae09} Ferrando J J and S\'{a}ez J A \Journal{Gen. Rel. Grav.}{41}{1695}{2009}

\bibitem{FerSae13} Ferrando J J and S\'{a}ez J A \Journal{Class. Quantum
Grav.}{30}{095013}{2013}







\bibitem{BonSen97} Bonilla M A G and Senovilla J M M, \Journal{Phys. Rev. Lett.}{78}{783}{1997}

\bibitem{Belprinc} Debever R, {\em Bull. Soc. Math. Belg.} \textbf{10}
112 (1958); idem, {\em Comp. R. Acad. Sci. (Paris)}
\textbf{249} 1744 (1959); Bel L {\em S\'{e}m. m\'{e}c. analytique m\'{e}c. c\'{e}leste}
\textbf{2} 1 
(1959);  Penrose R \Journal{Ann. Phys.}{10}{171}{1960}; Penrose R and Rindler W, \emph{Spinors and	Spacetime} vol 2 (Cambridge: Cambridge University Press, 1986), pp.\ 224 

\bibitem{FerrSaezP2} Ferrando J J, Morales J A and S\'{a}ez J A \Journal{Class.
	Quantum Grav.}{18}{4939}{2001}

\bibitem{FerSae10} Ferrando J J and S\'{a}ez J A \Journal{Gen. Rel. Grav.}{42}{1469}{2010}

\bibitem{FerSae04} Ferrando J J and S\'{a}ez J A \Journal{Gen. Rel.
	Grav.}{36}{2497}{2004} 

\bibitem{McIntosh1} McIntosh C B G, Arianrhod R, Wade ST and Hoenselaers
C \Journal{Class. Quantum Grav.}{11}{1555}{1994}

\bibitem{StewartEllis} Stewart J M and Ellis G F R, \Journal{J.
	Math. Phys.}{9}{1072}{1968} 

\bibitem{Goode} Goode S W and Wainwright J \Journal{Gen. Rel. Grav.}{18}{315}{1986} 

\bibitem{Lozanovski3} Lozanovski C and Carminati J \Journal{Class.
	Quantum Grav.}{20}{215}{2003} 

\bibitem{WyllVdB06} Wylleman L and Van~den Bergh N \Journal{Phys.
	Rev. D}{74}{084001}{2006} 

\bibitem{Lozanovski07} Lozanovski C \Journal{Class. Quantum Grav.}{24}{1169}{2007} 

\bibitem{LozWyll11} Lozanovski C and Wylleman L \Journal{Class.
	Quantum Grav.}{28}{075015}{2011}


\bibitem{ColHerOrtWyll12}
Coley A, Hervik S, Ortaggio M and Wylleman L \Journal{Class. Quantum Grav.}{29}{155016}{2012}


\bibitem{FerrSaezP1} Ferrando J J and S\'{a}ez J A \Journal{Class.
	Quantum Grav.}{14}{129}{1997} 




\bibitem{Yang} Yang L, Hou X R and Zeng Z,  Sci. China Ser. E \textbf{36}, 628 (1996)





\bibitem{SemerakStationaryFrames}Semer\'{a}k O {\em Gen. Rel. Grav.}
\textbf{25}, 1041 (1993)

\bibitem{VelocityKerr}Rosquist K {\em Gen. Rel. Grav.} \textbf{41}, 2619
(2009) 

\bibitem{KN_Source}Tahvildar-Zadeh A S, {[}arXiv:1410.0416{]}











%
%





\bibitem{Feynman}Feynman R, Leighton R, Sands M, \emph{The Feynman
Lectures on Physics} Vol. II (Addison-Wesley Publishing Company, 1964),
Secs. 27.5-27.6.

\bibitem{BonnorDragging}Bonnor W B, Phys. Lett. A \textbf{l58}, 23
(1991).

\bibitem{HerreraGonzalezPachonRueda}Herrera L, González G A, Pachón
L A, Rueda J. A., Class. Quantum Grav. \textbf{23}, 2395 (2006)

\bibitem{HerreraBarreto}Herrera L, Barreto W, Phys. Rev. D \textbf{86},
064014 (2012)

\bibitem{Magnus}Costa L F, Franco R, Cardoso V, Phys. Rev. D \textbf{98},
024026 (2018)

\bibitem{HerreraCarrotPrisco}Herrera L, Carrot J, Di Prisco A, Phys.
Rev. D \textbf{76} 044012 (2007)

\bibitem{HerreraGodel}Herrera L, Ibáñez J, Di Prisco A, Phys. Rev.
D \textbf{87}, 087503 (2013)

\bibitem{Penrosebook2} Penrose R and Rindler W, \emph{Spinors and
	Spacetime} vol 2 (Cambridge: Cambridge University Press, 1986)

\bibitem{McIntosh2} McIntosh C B G and Arianrhod R \Journal{Class.
	Quantum Grav.}{7}{L213}{1990}

\bibitem{BerLan04} Bergqvist G and Lankinen P \Journal{Class. Quantum Grav.}{21}{3499}{2004} 

\end{thebibliography}
\end{document}